\documentclass[fleqn,twocolumn,usenatbib,twocolappendix,numberedappendix]{openjournal}
\usepackage[table,dvipsnames]{xcolor}

\usepackage{textgreek}
\usepackage[utf8]{inputenc}
\usepackage[english]{babel}
\usepackage[fleqn]{amsmath}

\usepackage{multirow}
\usepackage{placeins}
\usepackage{booktabs}

\makeatletter
\AtBeginDocument{%
  \g@addto@macro\l@f@section{\bfseries}%
}
\makeatother

\usepackage{pgfplots}
\pgfplotsset{compat=1.18}
\usepgfplotslibrary{fillbetween}

\usetikzlibrary{arrows.meta}

\usepackage{hyperref}
\hypersetup{
    unicode, 
    colorlinks=true,
    linkcolor=linkcolor,
    citecolor=linkcolor,
    filecolor=linkcolor,
    urlcolor=linkcolor,
}
\definecolor{linkcolor}{rgb}{0.0,0.3,0.5}
\usepackage{tensind}
\tensordelimiter{?}
\DeclareGraphicsExtensions{.bmp,.png,.jpg,.pdf}
\usepackage{verbatim}
\usepackage[normalem]{ulem}
\usepackage{orcidlink}
\usepackage{soul}

\definecolor{hpurple}{HTML}{ca7b80}
\definecolor{horange}{HTML}{ffa756}
\definecolor{hred}{HTML}{d62728}
\definecolor{hgreen}{HTML}{658d6d}

\definecolor{color2}{HTML}{AE7181}
\definecolor{color1}{HTML}{6B8BA4}

\tikzset{
    lensLow/.style={draw=color1, draw opacity=0.7},
    lensHigh/.style={draw=color1, draw opacity=0.7},
    sourceStyle/.style={draw=color2, draw opacity=0.7},
}

\DeclareRobustCommand{\VAN}[3]{#2}
\let\VANthebibliography\thebibliography
\def\thebibliography{\DeclareRobustCommand{\VAN}[3]{##3}\VANthebibliography}
\DeclareUnicodeCharacter{2212}{\ensuremath{-}}

\graphicspath{ {./figures/} }

\begin{document}
\title[BigMag]{Weak lensing, big magnification: Recovering hidden lensing information from galaxy positions}

\author{Alex Alarcon $^{1}$}
\author{Carles Sánchez$^{2,3}$}

\affiliation{$^{1}$Institute of Space Sciences (ICE, CSIC), Campus UAB,
Carrer de Can Magrans, s/n, 08193 Barcelona, Spain}
\affiliation{$^{2}$Departament de F\'{\i}sica, Universitat Aut\`{o}noma de Barcelona (UAB), 08193 Bellaterra (Barcelona), Spain}
\affiliation{$^{3}$Institut de F\'{\i}sica d'Altes Energies (IFAE), The Barcelona Institute of Science and Technology, Campus UAB, 08193 Bellaterra (Barcelona), Spain}

\begin{abstract}
The combination of galaxy clustering and weak gravitational lensing is a powerful probe of cosmology, traditionally implemented through galaxy clustering and galaxy--galaxy lensing in a $2\times2$pt analysis. An alternative lensing observable is the clustering cross-correlation of foreground and background galaxy samples induced by lensing magnification, but measurements of this signal have generally achieved lower signal-to-noise than shear-based lensing. We show that, for deep background samples selected from current photometric surveys, cancellation between galaxies with positive and negative magnification responses within the same sample can substantially reduce the achievable signal. Using DES Year 6 photometry and \textsc{Balrog} synthetic-source injections, we explore high-redshift ($z\simeq0.7$--$2.5$) galaxy samples split by the sign and amplitude of their magnification response. This separation reduces signal cancellation and can substantially increase both the magnification signal-to-noise and its cosmological constraining power. We refer to the resulting combination of foreground clustering and foreground--background cross-correlations with a response-selected background sample as BigMag. Forecasts based on DES-like simulated data vectors, adopting an optimistic sample configuration and sufficiently precise calibration of the magnification responses, show that BigMag can achieve constraining power comparable to traditional $2\times2$pt while relying only on galaxy positions. The same response-selected samples also enhance inverse galaxy--galaxy lensing (IGGL), a position--shear correlation dominated by magnification terms when the position sample lies behind the shape sample. Combining BigMag and IGGL self-calibrates the magnification responses further and yields marginalized $S_8$ constraints comparable to traditional $3\times2$pt analyses, and even tighter in the more precise magnification-calibration scenarios considered here. The framework also enables the construction of near-zero-magnification-response samples, with applications to traditional $3\times2$pt and geometric lensing-ratio analyses. These results motivate further optimization and validation of galaxy selections to realize the potential of magnification as a complementary probe of cosmology.
\end{abstract}


\maketitle

\textsf{\tableofcontents}
%\tableofcontents
%\newpage

\section{Introduction}
\label{sec:intro}

Modern imaging surveys contain hundreds of millions of detected galaxies \citep{Y6Gold}, and forthcoming experiments will increase this number to several billion \citep{Ivezic2008, Akeson19_roman, Euclid_overview_paper, CSST}. Yet only a fraction of these objects typically enter cosmological analyses \citep[e.g.,][]{DESY6_3x2pt}. The requirements imposed on a galaxy sample depend strongly on the observable for which it is used. Galaxy samples used for clustering (often referred to as \emph{lens} or \emph{position} samples) are generally restricted to relatively bright galaxies at low or intermediate redshift \citep{Weaverdyck2026_Maglim}. Their observed number density can correlate with survey depth, seeing, sky brightness, stellar density, and other observing conditions. Because their autocorrelations are used to constrain galaxy bias and cosmology, these spatial variations must be accurately characterized and mitigated \citep{DESY6_systematics}. Galaxy samples used for weak lensing shear (usually called \emph{source} or \emph{shape} samples) can extend to substantially fainter objects and higher redshifts, but require reliable galaxy shape measurements, shear calibration, and redshift characterization \citep{Y6Mdet}. These requirements exclude many detected galaxies from either role.

Weak lensing \emph{magnification} offers a complementary way to exploit this broader galaxy population. By changing the observed solid angle and altering the apparent fluxes and sizes of background galaxies \citep{Bartelmann2010, PratBaconWLReview}, magnification modulates their observed number density and the probability that they enter a selected sample. This signal can be isolated by cross-correlating foreground and background populations with minimal redshift overlap: physical clustering is then strongly suppressed, leaving a correlation dominated by the magnification of background galaxies by matter traced by the foreground sample. Like galaxy-galaxy lensing, this observable probes the foreground galaxy-matter connection, but uses background galaxy positions rather than shapes. Because the background population enters only through cross-correlations, its autocorrelation need not be modeled; provided the foreground density field is adequately corrected and residual imaging fluctuations are not correlated between the samples, background systematics increase the variance without biasing the mean signal. The background selection can therefore extend to deeper, denser samples, including galaxies without usable shapes, while the foreground sample remains subject to the usual clustering-systematics requirements.

This work focuses on magnification-induced position--position correlations, which have been measured using a range of background populations \citep{2005ApJ...633..589S,2009A&A...507..683H,2012MNRAS.426.2489M,Garcia_Fernandez_2018,2014MNRAS.442.2680G}. Although well established, such measurements have generally achieved lower signal-to-noise than shape-based weak-lensing measurements \citep{Bernstein09,Bartelmann2010,2010MNRAS.401.2093V,Joachimi2010,2014MNRAS.437.2471D,2022MNRAS.513.1210M,2025Galax..13...89C}.
One important and avoidable source of this reduced statistical power, recognised since the earliest quasar–galaxy analyses \citep{2002A&A...386..784M}, is cancellation within conventionally selected background samples. A selected population responds through two competing effects: expansion of the observed area dilutes its surface density, while magnification of galaxy fluxes and sizes changes the number of objects satisfying the selection \citep{2021MNRAS.504.1452V, y3-2x2ptmagnification, Legnani2026}. The net response can therefore be positive, negative, or close to zero, and a conventional sample may combine populations with opposite responses whose lensing signals partially cancel. Separating these populations preserves the oppositely signed contributions and also helps distinguish magnification from residual intrinsic clustering: the magnification signal reverses sign between the samples, whereas clustering caused by redshift overlap does not \citep[e.g.][]{2009A&A...507..683H}.

Crucially, the magnification response is a property that can be estimated before measuring the cosmological cross-correlation. Catalogue-level perturbations have been used to propagate artificial magnification through the complex selection of spectroscopic samples and through flux-dependent photometric-redshift assignment \citep{Wenzl_2024,2025PhRvD.112j3514Q}. Image-level synthetic-source injection methods, such as \textsc{Balrog}, extend this logic through detection and measurement: simulated galaxies are magnified by a known amount, injected into real survey images, and processed through the same detection, measurement, and selection pipeline as the data \citep{Balrog_Y1,balrog,Y6Balrog}. The change in the number of selected objects then provides a direct calibration of the sample-selection response \citep{y3-2x2ptmagnification,Legnani2026}. A self-organizing map (SOM) \citep{Myles:2021:SOMPZ,DESY6lenspz,DESY6sourcepz} offers a natural representation of this information: cells containing galaxies with similar observed colors and magnitudes can be assigned both a redshift distribution and a magnification response. The SOM can then be used to construct high-redshift samples whose responses have a controlled sign and amplitude while maintaining the required redshift separation from the foreground sample.

%In this work, we use this information to split high-redshift galaxy samples into populations with positive and negative net magnification responses. 
In this work, we use injection-calibrated responses not only to calibrate a given selection but to define it. We split high-redshift galaxy samples into populations with positive and negative net magnification responses in the multi-dimensional colour–magnitude space of the SOM. Treating these populations as separate density fields prevents their signals from cancelling and provides cross-correlations with opposite signs around the same foreground structures. We call the combination of foreground clustering and its cross-correlation with these dense, response-optimized background samples BigMag. Its structure closely parallels that of a traditional $2\times2$pt analysis \citep{DESy3_2x2pt_maglim, DESy3_2x2pt_redmagic}. In both cases, the foreground autocorrelation constrains the combination of galaxy bias and matter clustering, while a lensing cross-correlation breaks the degeneracy between galaxy bias and the cosmological fluctuation amplitude. Traditional $2\times2$pt uses galaxy--galaxy lensing for this purpose; BigMag instead uses magnification-induced correlations between galaxy positions. The amplitude calibration provided by the magnification coefficient consequently plays a role analogous (although not identical) to the multiplicative shear calibration in galaxy--galaxy lensing.

The same response-selected high-redshift samples also enhance inverse galaxy--galaxy lensing (IGGL) \citep{IGGL}. In the IGGL configuration, the galaxy-position sample lies behind the source-shape sample. The conventional density--shear contribution is then strongly suppressed, and the position--shear correlation is dominated by the correlation between the magnification of the high-redshift position sample and the shear of the foreground source sample, together with the corresponding intrinsic-alignment contribution. BigMag and IGGL depend on the same magnification coefficients but probe them through different correlations and redshift kernels. Their combination can therefore partially self-calibrate the magnification responses while simultaneously constraining cosmology. It also provides a valuable consistency test because the two observables have different dependencies on galaxy bias, shear calibration, intrinsic alignments, and observational systematics.

We demonstrate this approach using DES Year 6 photometry, \textsc{Balrog} synthetic-source injections, and SOM-based redshift calibration to construct galaxy selections with controlled redshift distributions and magnification responses. We then use DES-informed number densities, redshift distributions, and response coefficients to generate simulated clustering and weak-lensing data vectors. These forecasts quantify the gain from separating positive- and negative-response populations, compare BigMag with traditional $2\times2$pt, and study the additional information supplied by IGGL. We also compare the combined magnification observables with cosmic shear and standard $3\times2$pt analyses. Finally, we investigate the calibration precision required for the magnification coefficients, the sensitivity of the measurement to residual redshift overlap, and the information available from magnification ratios on scales excluded from the fiducial cosmological analysis.

The paper is organized as follows. Section~\ref{sec:theory}
introduces the magnification contribution to galaxy-density
correlations, motivates response-based sample selection, and
discusses IGGL and sensitivity to observational systematics.
Section~\ref{sec:data_sims} presents the DES Year 6 and
\textsc{Balrog} data products, explores magnitude- and SOM-based
selections, and describes the simulated data vectors.
Section~\ref{sec:results} presents the cosmological forecasts
for BigMag and IGGL and their comparison and combination with
traditional weak-lensing probes. We then discuss the implications
and requirements for applying the method to data and summarize
our conclusions. The appendices provide further tests of
magnification-response selection and calibration, assess the
impact of redshift overlap, explore small-scale magnification
ratios, and derive the treatment of a redshift-dependent
magnification response.

\section{Theory and motivation}
\label{sec:theory}

This section develops the theoretical framework for the magnification observables used in this work. We describe how intrinsic clustering and weak-lensing magnification contribute to galaxy density correlations, motivating background-galaxy selections based on their magnification response. We then discuss magnification in galaxy–galaxy lensing, focusing on the inverse configuration, and assess the impact of observational systematics.

\subsection{Galaxy density and magnification bias}
\label{sec:gal_density_mag}

In large-scale structure surveys, the observed angular distribution of galaxies is used to trace the underlying matter density field through galaxy clustering measurements. In the absence of lensing effects, the intrinsic galaxy density contrast in tomographic bin $i$ can be written as the line-of-sight projection of the three-dimensional galaxy density field,
\begin{equation}
\label{eq:delta_int_proj}
\delta_{g,i}^\mathrm{int}(\hat{\mathbf n})
=
\int d\chi\, W_g^i(\chi)\,\delta_g^{3\mathrm D}(\hat{\mathbf n}\chi,\chi),
\end{equation}
where $\chi$ is the comoving distance and $W_g^i(\chi)$ denotes the normalized redshift selection function of the galaxy sample.

In practice, the observed projected galaxy density contrast, $\delta_g^\mathrm{obs}(\hat{\mathbf n})$, receives contributions from both the intrinsic clustering of galaxies, described by Eq.~(\ref{eq:delta_int_proj}), and gravitational lensing by foreground matter along the line of sight. Lensing remaps the apparent angular positions of background galaxies and alters their observed sizes and fluxes while conserving surface brightness. The strength of this effect is quantified by the magnification factor $\mu$, which depends on the weak-lensing convergence $\kappa$ and shear $\gamma$ \citep{2001PhR...340..291B},
\begin{equation}
\label{eq:mu_def}
\mu
=
\frac{1}{(1-\kappa)^2-|\gamma|^2}
\simeq \frac{1}{(1-2\kappa)}
\simeq
1+2\kappa,
\end{equation}
where the last equality holds in the weak-lensing regime, $\kappa\ll1$ and $|\gamma|\ll1$.

Magnification affects the observed galaxy number density through two competing effects: the stretching of the solid angle, which dilutes the observed density by a factor $\mu^{-1}$, and the increase in galaxy total flux and image area by a factor $\mu$, which alters the probability that galaxies satisfy a given selection. The balance between these competing effects is commonly referred to as \emph{magnification bias}, and depends on the properties of the galaxy population and the selection criteria. The resulting contribution to the observed galaxy density contrast can be written as
\begin{equation}
\label{eq:delta_g_split}
\delta_g^\mathrm{obs}(\hat{\mathbf n})
=
\delta_g^\mathrm{int}(\hat{\mathbf n})
+
\delta_g^\mathrm{mag}(\hat{\mathbf n},\kappa).
\end{equation}

The magnification-induced overdensity can be expressed in terms of the observed galaxy number density $n(\hat{\mathbf n},\kappa)$ relative to the unlensed case $n(\hat{\mathbf n},0)$ \citep{Bernstein09, Joachimi2010}:
\begin{equation}
\label{eq:delta_mag_ratio}
\delta_g^\mathrm{mag}(\hat{\mathbf n},\kappa)
\equiv
\frac{n(\hat{\mathbf n},\kappa)}{n(\hat{\mathbf n},0)}-1.
\end{equation}

In practice, the galaxy samples in imaging surveys are selected using criteria applied to observed (lensed) properties (like galaxy fluxes or colors), which we represent here by a vector $\mathbf{F'}$. Following \citet{y3-2x2ptmagnification, Legnani2026}, we can write the observed number density of galaxies at a given position $\hat{\mathbf n}$ as an integral over the distribution of intrinsic (unlensed) galaxy properties, $\mathbf{F}$:
\begin{equation}
\label{eq:n_kappa_def}
n(\hat{\mathbf n},\kappa)
=
\frac{1}{\Delta\Omega(\kappa)}
\int d\mathbf F\; S(\mathbf F')\,N(\mathbf F,\hat{\mathbf n}),
\end{equation}
where $S(\mathbf F')$ denotes the selection function applied to the lensed properties $\mathbf F'=\mathbf F'(\mathbf F, \kappa)$, and $N(\mathbf F,\hat{\mathbf n})$ represents the absolute number of galaxies with intrinsic properties $\mathbf F$. The factor $\Delta\Omega(\kappa)$ corresponds to the lensed solid angle (sky area) element. In the weak-lensing regime, the observed solid angle element is stretched by the magnification factor $\mu$ of Eq.~(\ref{eq:mu_def}), and therefore the lensed sky area element transforms as $\Delta\Omega(\kappa)\simeq \mu \Delta\Omega(0) = \Delta\Omega(0)/(1-2\kappa)$. Using this relation, Eq.~(\ref{eq:n_kappa_def}) can be rewritten as
\begin{equation}
\label{eq:n_kappa_area}
n(\hat{\mathbf n},\kappa)
=
\frac{1-2\kappa}{\Delta\Omega(0)}
\int d\mathbf F\; S(\mathbf F')\,N(\mathbf F,\hat{\mathbf n}).
\end{equation}

Expanding the selection function to first order in $\kappa$ around $\kappa = 0$,
\begin{equation}
\label{eq:S_expand}
S(\mathbf F')
\simeq
S(\mathbf F)
+
\kappa
\left.\frac{\partial S}{\partial\kappa}\right|_{\kappa=0},
\end{equation}
and neglecting terms of order $\kappa^2$, we obtain
\begin{align}
\label{eq:n_expand}
n(\hat{\mathbf n},\kappa)
&\simeq
\frac{1-2\kappa}{\Delta\Omega(0)}
\int d\mathbf F\;
S(\mathbf F)\,N(\mathbf F,\hat{\mathbf n})
\nonumber\\
&\quad
+
\frac{\kappa}{\Delta\Omega(0)}
\int d\mathbf F\;
\left.\frac{\partial S}{\partial\kappa}\right|_{\kappa=0}
N(\mathbf F,\hat{\mathbf n})
\nonumber\\
&=
(1-2\kappa)\,n(\hat{\mathbf n},0)
\nonumber\\
&\quad
+
\frac{\kappa}{\Delta\Omega(0)}
\int d\mathbf F\;
\left.\frac{\partial S}{\partial\kappa}\right|_{\kappa=0}
N(\mathbf F,\hat{\mathbf n}),
\end{align}
where the unlensed number density is
\begin{equation}
\label{eq:n0_def}
n(\hat{\mathbf n},0)
=
\frac{1}{\Delta\Omega(0)}
\int d\mathbf F\;
S(\mathbf F)\,N(\mathbf F,\hat{\mathbf n}).
\end{equation}

The first term in Eq.~(\ref{eq:n_expand}) corresponds to the galaxy number density in the absence of lensing, $n(\hat{\mathbf n},0)$, multiplied by a factor $(1-2\kappa)$ coming from the change in area element. The second term accounts for the response of the number of selected galaxies per unit of unlensed area due to changes in $\kappa$. Substituting Eq.~(\ref{eq:n_expand}) into Eq.~(\ref{eq:delta_mag_ratio}), the magnification-induced overdensity can be written, to leading order in $\kappa$, as
\begin{align}
\label{eq:delta_mag_linear}
\delta_g^\mathrm{mag}(\hat{\mathbf n})
&\simeq
-2\,\kappa(\hat{\mathbf n})
\nonumber\\
&
+
\frac{\kappa(\hat{\mathbf n})}{n(\hat{\mathbf n},0)\,\Delta\Omega(0)}
\int d\mathbf F\;
\left.\frac{\partial S}{\partial\kappa}\right|_{\kappa=0}
N(\mathbf F,\hat{\mathbf n})
\nonumber\\
&\equiv
\kappa(\hat{\mathbf n})
\left[
C_\mathrm{area}
+
C_\mathrm{sample}\right] \equiv
C_\mathrm{mag}\,\kappa(\hat{\mathbf n})
,
\end{align}
where the area contribution is $C_\mathrm{area}=-2$ and the second term, $C_\mathrm{sample}$, captures the response of the selected galaxy number density to lensing-induced changes in the observed galaxy properties (and the total factor is the sum of these, $C_\mathrm{mag}$). 

Equation~(\ref{eq:delta_mag_linear}) shows explicitly that magnification affects the observed galaxy density through two competing contributions. The first term, $C_\mathrm{area}=-2$, arises from the stretching of the sky area and always acts to dilute the observed number density. The second term, $C_\mathrm{sample}$, reflects the response of the galaxy selection to lensing-induced changes in observed fluxes and sizes. Because magnification increases the apparent flux and size of background galaxies while simultaneously diluting their surface density, the net sign and amplitude of the magnification-induced overdensity depend on the balance between these effects. For flux-limited samples, this balance is governed by the slope of the cumulative luminosity function \citep[e.g.][]{Joachimi2010, Garcia_Fernandez_2018}: steep slopes enhance the flux-boosting term and can lead to a positive magnification response, whereas shallow slopes result in a net suppression of the observed density. Consequently, the overall magnification coefficient $C_\mathrm{mag}=C_\mathrm{area}+C_\mathrm{sample}$ can be positive, negative, or close to zero, depending on the properties of the selected galaxy sample (we will explore this further in this paper, especially in \S\ref{sec:selections_alpha}).

\subsection{Modeling galaxy clustering}
\label{sec:model}

Now we turn to the modeling of the galaxy clustering observable. To this end, we need to consider the observed galaxy density in the sky, $\delta_g^\text{obs}$, defined for different galaxy selections (e.g., distinct redshift bins or samples at similar redshifts but selected by other properties). For this observed galaxy density of a given bin $i$, we need to include the effects of the intrinsic galaxy density and the change in the galaxy density due to weak lensing magnification from mass between the galaxy and the observer, as in Eq.~(\ref{eq:delta_g_split})\footnote{Later on, in  \S\ref{sec:systematics}, we will also consider contributions from systematic effects in the data. }:
\begin{equation}
\label{eq:delta_g}
    \delta_{g,i}^\text{obs} =  \delta_{g,i}^\text{int} + \delta_{g,i}^\text{mag} = b_i\delta_{m,i} + C_\mathrm{mag,\emph{i}} \; \kappa_{i}, 
\end{equation}
where in the second equality we assume the linear galaxy bias assumption valid on large scales, $\delta_{g,i}^\mathrm{int} = b_i \, \delta_{m,i}$, and the developments of Eq.~(\ref{eq:delta_mag_linear}). We are interested in the two-point statistics of this observed galaxy density field between two samples, which we label $i$ and $j$. Using Eq.~(\ref{eq:delta_g}), the angular cross-correlation of the observed fields can be written as

\begin{equation}
\label{eq:gg_obs_terms}
\begin{aligned}
\left\langle \delta_{g,i}^\mathrm{obs} \, \delta_{g,j}^\mathrm{obs} \right\rangle
=&\;
\underbrace{
b_i b_j \left\langle \delta_{m,i} \, \delta_{m,j} \right\rangle
}_{\text{intrinsic clustering}}
\\[4pt]
&+
\underbrace{
b_jC_\mathrm{mag,\emph{i}}\left\langle \kappa_{i} \, \delta_{m,j} \right\rangle
}_{\text{magnification of sample } i}
\\[4pt]
&+
\underbrace{
b_i C_\mathrm{mag,\emph{j}}\left\langle \kappa_{j} \, \delta_{m,i} \right\rangle
}_{\text{magnification of sample } j}
\\[4pt]
&+
\underbrace{
C_\mathrm{mag,\emph{i}} \; C_\mathrm{mag,\emph{j}}
\left\langle \kappa_{i} \, \kappa_{j} \right\rangle
}_{\text{magnification--magnification}}
\end{aligned}
\end{equation}

\begin{figure}
\centering
\begin{tikzpicture}

\definecolor{color2}{HTML}{AE7181}
\definecolor{color1}{HTML}{6B8BA4}

\begin{axis}[
    width=1.05\linewidth,
    height=0.55\linewidth,
    xmin=0, xmax=2.3,
    ymin=0, ymax=1.05,
    axis lines=left,
    xlabel={\normalsize Redshift $z$},
    ylabel={\normalsize $N(z)$},
    ytick=\empty,
    xtick=\empty,
    axis line style={line width=0.8pt},
    legend style={
        draw=none,
        fill=none,
        at={(0.97,0.97)},
        anchor=north east
    },
    legend image code/.code={
        \draw[#1, line width=1.2pt] (0cm,0cm) -- (0.6cm,0cm);
    },
]

% Baseline
\addplot[name path global=baseline, domain=0:2.3, forget plot] {0};

% Gaussian 1 (Bin i)
\addplot[
    name path global=g1,
    domain=0.2:0.8,
    samples=300,
    line width=1.2pt,
    draw=color1
]
{exp(-((x-0.50)^2)/(2*(0.05^2)))};
%\addlegendentry{Bin $i$}

\addplot[draw=none, fill=color1, fill opacity=0.4,forget plot]
    fill between[of=g1 and baseline];

% Gaussian 2 (Bin j)
\addplot[
    name path global=g2,
    domain=0.8:2.1,
    samples=300,
    line width=1.2pt,
    draw=color2
]
{exp(-((x-1.45)^2)/(2*(0.18^2)))};
%\addlegendentry{Bin $j$}

\addplot[draw=none, fill=color2, fill opacity=0.4]
    fill between[of=g2 and baseline];

\addlegendimage{line width=1.2pt, draw=blue}
\addlegendentry{Bin $i$}
\addlegendimage{line width=1.2pt, draw=color1}
\addlegendentry{Bin $j$}
\end{axis}
\end{tikzpicture}

\caption{Illustrative redshift distributions of two galaxy bins occupying non-overlapping redshift ranges, enabling isolation of the magnification-induced galaxy cross-correlation signal.}
\label{fig:nzs_simple}
\end{figure}
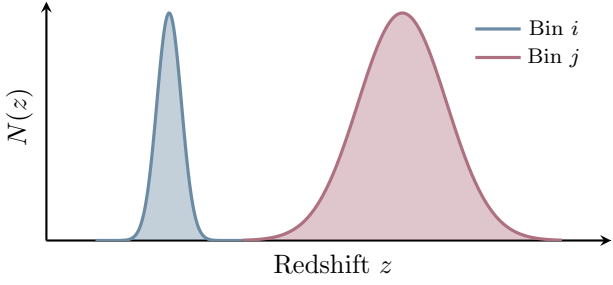

We now consider the case in which the two samples occupy non-overlapping redshift ranges, with sample $i$ at lower redshift and sample $j$ at higher redshift (see illustration in Figure \ref{fig:nzs_simple}). In this configuration, several simplifications occur. First, the matter density fields are uncorrelated, $\langle \delta_{m,i} \delta_{m,j} \rangle \simeq 0$. Second, the convergence field at the redshift of the foreground sample does not correlate with the intrinsic density of the background sample, $\langle \kappa_{i} \, \delta_{m,j} \rangle \simeq 0$. Finally, the convergence--convergence term is typically subdominant on the angular scales of interest.

Under these assumptions, the observed cross-correlation reduces to
\begin{equation}
\label{eq:gg_cross_mag}
\left\langle \delta_{g,i}^\mathrm{obs} \, \delta_{g,j}^\mathrm{obs} \right\rangle
\simeq
b_i C_\mathrm{mag,\emph{j}}\left\langle \kappa_{j} \, \delta_{m,i} \right\rangle.
\end{equation}

This remaining term describes the correlation between the matter density field traced by foreground galaxies and the convergence field sourced by the same large-scale structure along the line of sight to the background sample. In this sense, Eq.~(\ref{eq:gg_cross_mag}) highlights the three key ingredients required to extract cosmological information from magnification-induced galaxy cross-correlations:
\begin{enumerate}
    \item The galaxy bias of the foreground sample, $b_i$; 
    \item The magnification response of the background sample, also called magnification coefficient, $C_\mathrm{mag,\emph{j}}$;
    \item The matter--convergence correlation, $\langle \kappa_{j} \, \delta_{m,i} \rangle$, which contains the cosmological information given the redshift distribution of each galaxy sample.
\end{enumerate}

Two important notes are in order here. On the one hand, the galaxy bias parameter, $b_i$, is a property of the foreground (low-redshift) sample, it will generally be positive ($b_i\geq1$ for typical galaxy samples) and cannot be known \emph{a priori} but should be constrained using the galaxy auto-correlation of that foreground sample. On the other hand, the magnification parameter $C_\mathrm{mag,\emph{j}}$ is a property of the background (high-redshift) sample and can be estimated independently using sample-selection response techniques, in particular synthetic source injection (SSI) algorithms, such as Balrog in DES \citep{Balrog_Y1,balrog}. Crucially, the magnification coefficient can be positive, negative, or zero depending on the value of $C_\mathrm{sample,\emph{j}}$, according to Eq.~(\ref{eq:delta_mag_linear}). Therefore, given that galaxy bias is positive,  the sign of $C_\mathrm{mag,\emph{j}}$ (which can be characterized \emph{a priori}) will determine the sign of magnification-induced galaxy cross-correlations.

To leading order in the convergence, the observed autocorrelation of foreground
bin $i$ can be written as
\begin{equation}
\label{eq:gg_auto}
\left\langle
\delta_{g,i}^{\mathrm{obs}}
\delta_{g,i}^{\mathrm{obs}}
\right\rangle
=
b_i^2
\left\langle
\delta_{m,i}\delta_{m,i}
\right\rangle
+
2b_i C_{\mathrm{mag},i}
\left\langle
\delta_{m,i}\kappa_i
\right\rangle
+
\mathcal{O}(\kappa^2).
\end{equation}

For typical low-redshift lens samples and on the angular scales considered in this work,
the magnification contribution to the auto-correlation is subdominant with respect to
the intrinsic clustering term.
The auto-correlation can therefore be approximated as
\begin{equation}
\label{eq:gg_auto_bias}
\left\langle \delta_{g,i}^\mathrm{obs} \, \delta_{g,i}^\mathrm{obs} \right\rangle
\simeq
b_i^2 \left\langle \delta_{m,i} \, \delta_{m,i} \right\rangle ,
\end{equation}
allowing the galaxy bias of the foreground sample to be constrained directly from
its clustering signal.

Taken together, Eqs.~(\ref{eq:gg_cross_mag}) and (\ref{eq:gg_auto_bias}) show that cosmological information can be obtained\footnote{These simplifications are introduced for qualitative and pedagogical clarity; the theoretical predictions used in subsequent sections include the full set of intrinsic-clustering and magnification contributions.} from the joint analysis of the auto-correlation of a foreground galaxy sample and its cross-correlation with a background sample in a non-overlapping redshift bin. In this configuration, the auto-correlation primarily constrains the galaxy bias of the foreground sample, while the cross-correlation isolates the magnification-induced signal that probes the matter distribution along the line of sight and depends linearly on the magnification coefficient of the background galaxies. 

Two key implications follow from this framework, which form the main motivation for the analysis presented in this paper:
\begin{itemize}
\item The magnification coefficient of background galaxies, which sets the amplitude (and sign) of the cross-correlation signal, can be estimated \emph{a priori} for galaxy samples using synthetic source injection (SSI) techniques, such as \textsc{Balrog}. This enables the design and optimisation of sample selections that maximise the magnification signal, an aspect that is particularly important in the presence of mixed populations that would otherwise lead to partial cancellation between negative and positive magnification contributions. We explore this strategy in detail in \S\ref{sec:selections_alpha}.
\item Because the analysis does not require the auto-correlation of the background sample, the cosmological inference is significantly less sensitive to observational systematics affecting that sample. This relaxes the requirements on systematics mitigation for the background galaxies, allowing for more aggressive selections and the inclusion of a larger number of objects. The impact of observational systematics in this context is discussed in \S\ref{sec:systematics}.
\end{itemize}

\subsection{Optimizing galaxy selections for magnification}
\label{sec:selections_alpha}

\begin{figure}
\centering
\begin{tikzpicture}

\definecolor{color3}{HTML}{363737}
\definecolor{color1}{HTML}{6B8BA4}
\definecolor{color2}{HTML}{AE7181}

\begin{axis}[
    width=1.1\linewidth,
    height=0.55\linewidth,
    xmin=-0.5, xmax=2.3,
    ymin=0, ymax=1.05,
    axis lines=left,
    xlabel={\normalsize Magnification coeff.\ $C_\mathrm{sample}$},
    ylabel={\normalsize $N(C_\mathrm{sample})$},
    ytick=\empty,
    xtick=\empty,
    axis line style={line width=0.8pt},
    clip=false, % allow labels outside the axes if needed
]

% Baseline (y=0)
\addplot[name path global=baseline, domain=-0.5:2.3, forget plot] {0};

% Gaussian curve (peak normalized to 1)
\addplot[
    name path global=g1,
    domain=-0.5:2.3,
    samples=400,
    line width=1.2pt,
    draw=color3,
    opacity=0.75
]
{exp(-((x-0.8)^2)/(2*(0.4^2)))};

% --- Fill LEFT of x=1 ---
\path[name path=clipL] (axis cs:-0.5,0) rectangle (axis cs:1,1.2);
\addplot[draw=none, fill=color1, fill opacity=0.35, forget plot]
    fill between[of=g1 and baseline, soft clip={domain=-0.5:1}];

% --- Fill RIGHT of x=1 ---
\path[name path=clipR] (axis cs:1,0) rectangle (axis cs:2.3,1.2);
\addplot[draw=none, fill=color2, fill opacity=0.35, forget plot]
    fill between[of=g1 and baseline, soft clip={domain=1:2.3}];

% Vertical dashed line at alpha=1
\addplot[
    forget plot,
    domain=0:1.05,
    samples=2,
    dashed,
    line width=1.0pt,
    draw=color3,
    opacity=0.9
] ({1},{x});

% Label for alpha=1 line
\node[anchor=south, font=\normalsize]
    at (axis cs:1,1.02)
    {$C_\mathrm{sample} = 2$};
\node[anchor=south, font=\small]
    at (axis cs:1,1.16)
    {null $\times$-corr.};

\draw[<-, line width=1pt, color1!70!black]
(axis cs:0.6,0.37) -- (axis cs:0.96,0.37);
% Labels for left/right regions
\node[anchor=north, font=\small, text=color1!70!black]
    at (axis cs:0.69,0.32)
    {negative};
\node[anchor=north, font=\small, text=color1!70!black]
    at (axis cs:0.69,0.2)
    {$\times$-corr.};

\draw[->, line width=1pt, color2!70!black]
(axis cs:1.03,0.37) -- (axis cs:1.26,0.37);
\node[anchor=north, font=\small, text=color2!70!black]
    at (axis cs:1.22,0.32)
    {positive};

    \node[anchor=north, font=\small, text=color2!70!black]
    at (axis cs:1.23,0.2)
    {$\times$-corr.};

\end{axis}
\end{tikzpicture}

\caption{
Illustrative distribution of the sample part of the magnification coefficient, $C_\mathrm{sample}$, for a set of galaxies (see Eq.\ref{eq:delta_mag_linear}).
Galaxies with $C_\mathrm{sample}<2$ produce a negative magnification-induced cross-correlation ($\times$-corr.),
those with $C_\mathrm{sample}=2$ yield a null signal, and galaxies with $C_\mathrm{sample}>2$
produce a positive magnification-induced cross-correlation.
}
\vspace{0.5em}
\label{fig:alpha_distribution}
\end{figure}
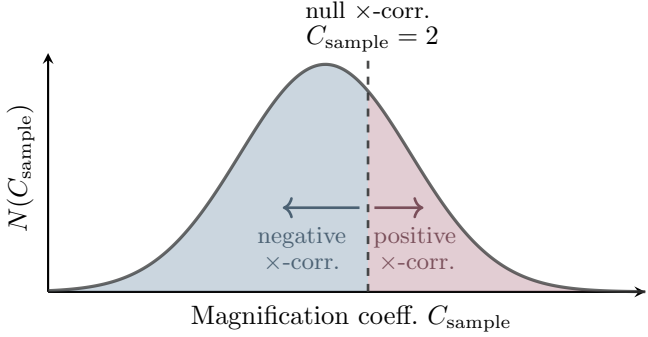

\begin{figure*}
    \centering
    \includegraphics[width=0.95\linewidth]{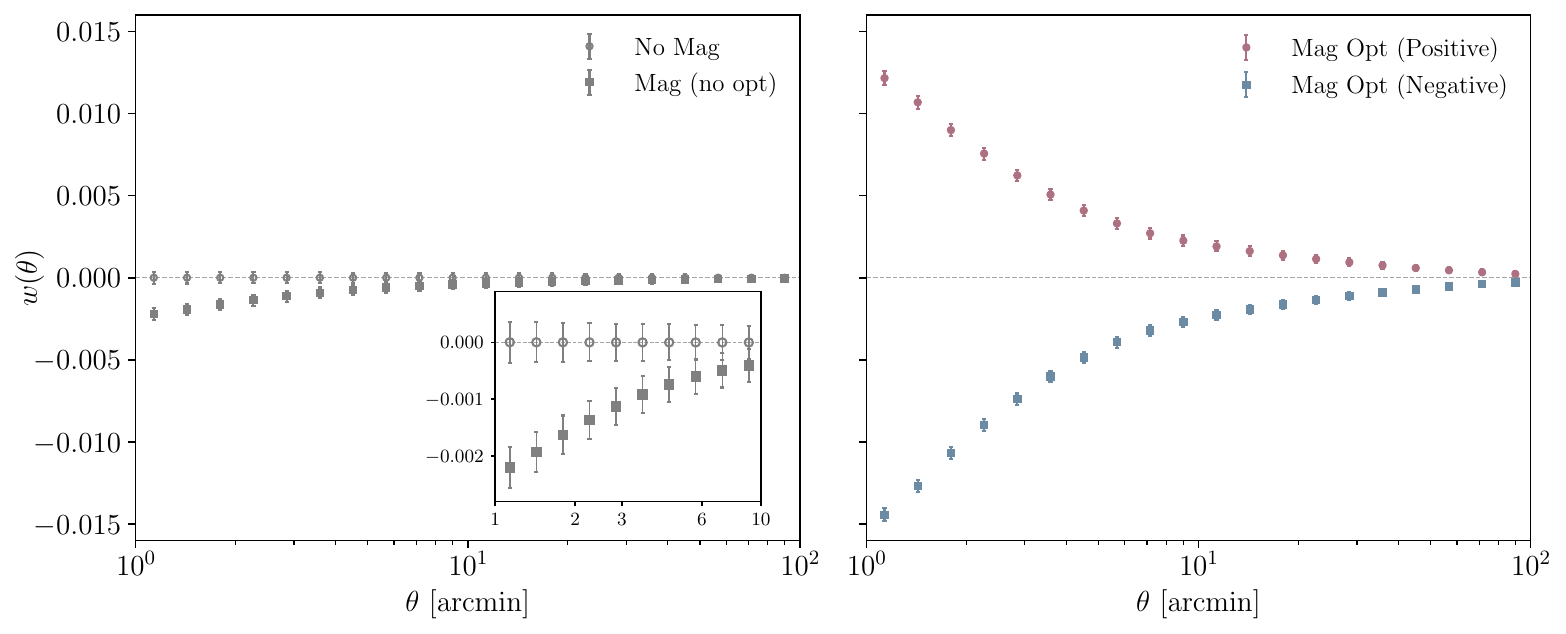}
    \caption{
Impact of magnification-selected galaxy samples on simulated
foreground--background clustering cross-correlations. The foreground
and background samples have mean redshifts $\langle z\rangle=0.34$
and $1.70$, respectively, with DES Y6-informed number densities.
The foreground magnification response is set to zero throughout.
Left: the unsplit background configuration without magnification
($C_{\rm mag}=0$, open circles) and with its adopted response
($C_{\rm mag}=-0.29$, grey squares). The latter yields
$S/N=10.2$ over the displayed angular range; the inset highlights
the small-scale signal.
Right: configurations with positive and negative background
responses, $C_{\rm mag}=+1.60$ (rose circles) and $-1.90$
(blue squares), yield $S/N=44.1$ and $54.0$, respectively.
Points show model predictions, with error bars from the diagonal
of the simulated covariance; the quoted signal-to-noise values
include correlations between angular bins.
These illustrative configurations show the potential gain from
separately measuring populations with opposite magnification
responses.
}
    \label{fig:bigmag_xcorr1}
\end{figure*}

Building on the framework above, we now consider how background-galaxy selections can be designed to enhance sensitivity to magnification. Because the magnification-induced cross-correlation scales linearly with $C_\mathrm{mag}$, Eq.~(\ref{eq:gg_cross_mag}), broad selections can combine galaxies with positive, negative, and near-zero responses whose contributions partially cancel, reducing the signal-to-noise ratio (Figure~\ref{fig:alpha_distribution}). This motivates grouping background galaxies according to the sign and magnitude of $C_\mathrm{mag}$, so that their contributions add coherently. Since $C_\mathrm{mag}$ can be characterized using sample-selection response techniques such as synthetic source injections, these response-based selections can be constructed and optimized in practice. The principle is the same as in $(\alpha-1)$-weighted estimators \citep{2002A&A...386..784M, 2005ApJ...633..589S}, here applied to responses calibrated for arbitrary selections rather than inferred from the number-count slope, and with the resulting groups kept as separate fields rather than combined into a single weighted sample. 

Figure~\ref{fig:bigmag_xcorr1} illustrates this gain using a simulated clustering cross-correlation between foreground and background samples at $z=0.34$ and $z=1.7$, like those from Fig.~\ref{fig:nzs_simple}, with number densities informed by DES Y6 (more details about this in \S\ref{sec:data_sims}). In the absence of magnification, the cross-correlation of the foreground and background samples is null, as expected, since they do not overlap in redshift. When considering magnification, an unsplit background sample, like the one depicted in Fig.~\ref{fig:alpha_distribution}, produces a magnification signal with modest signal-to-noise because of partial cancellation (left panel). Splitting the same background parent population into subsamples with positive and negative magnification responses substantially increases the signal-to-noise of the measurement (right panel). The construction of these samples from real data is described in \S\ref{sec:data_sims}.

Opposite-response subsamples also provide an internal consistency test of
the origin of the signal. The leading magnification-induced correlation
changes sign with $C_{\rm mag}$, whereas obscuration and dust extinction dim background galaxies \citep{2010MNRAS.405.1025M} and
can reduce the observed background counts without following this sign
reversal \citep{2013MNRAS.429.3230H, 2021MNRAS.503.4964G}. Comparing
subsamples around the same foreground galaxies, while accounting for
differences in their redshift distributions and selection, can therefore
help distinguish magnification from observational contamination. 

On the other hand, samples with $C_{\rm mag}\simeq0$ provide a complementary null test of the leading
magnification contribution, and they would also have applications beyond these
consistency tests. The same response-calibration and selection machinery
used here to enhance magnification can instead be used to construct
samples with a total magnification response close to zero in each
tomographic bin. Applied to the galaxy-position samples in
$3\times2$pt analyses, this would suppress lens-magnification contributions
to galaxy clustering and galaxy--galaxy lensing, reducing their sensitivity
to magnification modelling and response-calibration uncertainties
\citep{Euclidmagnification2021,y3-2x2ptmagnification, Legnani2026}. Furthermore, null-response lens samples would be particularly valuable for
lensing ratio measurements \citep{Jain2003,DES_spt_lensingratio,Unruh_2019}, where the contributions from magnification can reduce the geometrical nature  that makes these measurements so interesting.  

Having established the motivation for response-based selections, we next discuss their relevance to inverse galaxy--galaxy lensing and the impact of observational systematics on these observables.

\subsection{(Inverse) Galaxy-galaxy lensing}
\label{sec:iggl_theory}

Another measurement that can be dominated by weak lensing magnification in particular redshift configurations is galaxy-galaxy lensing (GGL, \citealt{1992ApJ...388..272K}).  The GGL observable is the cross-correlation between the observed galaxy density of a position sample, $\delta_g^\mathrm{obs}$, defined in Eq.~(\ref{eq:delta_g}), and the observed ellipticity of a shape sample, $e^\mathrm{obs}$. In the weak-lensing regime, the latter can be modeled as the sum of the gravitational shear and the intrinsic galaxy ellipticity \citep{Prat_2022},
\begin{align}
\label{eq:ellipticities}
e^{\text{obs}} = \gamma + e^{\text{int}},
\end{align}
where $|\gamma|\ll1$. Correlating $\delta_g^\text{obs}$ with $e^{\text{obs}}$ to get the relevant cross-correlation for GGL, we get four terms representing different physical effects:
\begin{equation}
\label{eq:all_terms}
\begin{split}
 &\left< \delta_g^\text{obs} e^{\text{obs}} \right> = 
 \left< (\delta_g^\text{int} +  C_\mathrm{mag} \kappa) \,  (\gamma + e^{\rm int}) \right> = \\
 &= \underbrace{\left< \delta_g^\text{int} \gamma\right>}_{\text{trad GGL}} 
 + \, \underbrace{C_\mathrm{mag}\left< \kappa \, \gamma \right>}_{\text{lens mag}} + 
 \, \underbrace{\left< \delta_g^\text{int} e^{\rm int}\right>}_{\text{IA}} + 
 \, \underbrace{C_\mathrm{mag}\left< \kappa \, e^{\rm int} \right>.}_{\text{lens mag + IA}}
   \end{split}
\end{equation}
In the traditional GGL configuration, the position (lens) sample lies at lower redshift than the shape (source) sample, and the first term in Eq.~(\ref{eq:all_terms}) dominates. To make magnification the leading contribution, this ordering can be reversed by placing the position sample behind the shape sample. In that case, the first term in Eq.~(\ref{eq:all_terms}) vanishes, and the measurement only depends on weak lensing and intrinsic alignments (IA). This configuration is known as Inverse Galaxy-Galaxy Lensing (IGGL) and is studied in detail by \citet{IGGL}. Later in this work, we combine IGGL with the clustering cross-correlations described in \S\ref{sec:model} to constrain cosmology using weak-lensing magnification.

\subsection{Impact of observational systematics}
\label{sec:systematics}

Having seen how galaxy samples can be selected and optimized according to their magnification coefficients, we now consider the impact of observational systematics on these measurements and selections. In practice, the observed galaxy density field is affected by observational systematics arising from spatial variations of survey depth, seeing, sky brightness, stellar contamination, and other imaging-related effects. These systematics can be modeled schematically as additive and multiplicative contributions to the observed galaxy density contrast \citep[][]{Noah_weights, Berlfein2024_systematics},
\begin{equation}
\label{eq:delta_sys}
\delta_g^\mathrm{obs} = (1+m_g)\,\delta_g^\mathrm{true} + s_g ,
\end{equation}
where $\delta_g^\mathrm{true}$ denotes the systematics-free physical density contrast, including intrinsic clustering and lensing magnification (see Eq.~\ref{eq:delta_g}), while $m_g$ and $s_g$ represent multiplicative and additive systematic fields, respectively. For the qualitative discussion below, we assume that $m_g$ and $s_g$ have zero mean and are statistically uncorrelated with the cosmological density field. This assumption is appropriate for instrumental or Galactic contaminants in an ensemble sense, but excludes effects tied directly to foreground structure, such as blending and obscuration, which must be included explicitly in a complete analysis (see the discussion in \S\ref{sec:discussion_observational_systematics}).

When considering the clustering auto-correlation of a given galaxy sample, observational systematics enter at leading order. Using Eq.~(\ref{eq:delta_sys}), the observed auto-correlation can be written schematically as
\begin{align}
\langle \delta_g^{\rm obs}\delta_g^{\rm obs} \rangle
&=
\left(1+\langle m_gm_g\rangle\right)
\langle \delta_g^{\rm true}\delta_g^{\rm true}\rangle
\nonumber\\
&\quad
+ \langle s_g s_g\rangle
+ \mathcal{O}(m_g s_g).
\end{align}
As a result, multiplicative systematics rescale the intrinsic clustering amplitude, while additive systematics contribute their own spatial correlations. Accurate cosmological inference from the foreground-sample auto-correlation therefore requires these observational systematics to be carefully characterized and mitigated \citep{Noah_weights, Berlfein2024_systematics, DESY6_systematics} 

We now consider the cross-correlation between a foreground sample $i$ and a background sample $j$ in non-overlapping redshift bins, as discussed in \S\ref{sec:model}. Provided that the foreground sample is adequately corrected for observational systematics ($\delta_{g,i}^\mathrm{obs}\approx\delta_{g,i}^\mathrm{true}$), any residual additive or multiplicative contamination in the background sample does not correlate with the intrinsic density of the foreground galaxies. In this case, the expectation value of the observed cross-correlation remains unbiased,
\begin{align}
\langle \delta_{g,i}^{\rm obs} \delta_{g,j}^{\rm obs} \rangle
&=
\left\langle
\delta_{g,i}^{\rm true}
(1+m_{g,j})\delta_{g,j}^{\rm true}
\right\rangle
+
\left\langle
\delta_{g,i}^{\rm true} s_{g,j}
\right\rangle
\nonumber\\
&
\simeq
\left\langle \delta_{g,i}^\mathrm{true} \, \delta_{g,j}^\mathrm{true} \right\rangle.
\end{align}
Therefore, the magnification-induced cross-correlation does not require the background sample to be free of observational systematics in order to obtain an unbiased cross-correlation signal. 

However, although the mean cross-correlation is unaffected, residual systematics in the background sample can increase the variance of the measurement. To illustrate this, we consider the case where the foreground sample is well corrected, $\delta_{g,i}^\mathrm{obs}\simeq \delta_{g,i}^\mathrm{true}$, while the background sample contains residual additive and multiplicative contaminants, as in Eq.~(\ref{eq:delta_sys}). For the calculation, we define
\begin{equation}
X \equiv \delta_{g,i}^\mathrm{obs}\,\delta_{g,j}^\mathrm{obs}
\simeq \delta_{g,i}^\mathrm{true}\Big[(1+m_{g,j})\,\delta_{g,j}^\mathrm{true}+s_{g,j}\Big].
\end{equation}
Under the assumptions stated above (in particular that $m_{g,j}$ and $s_{g,j}$ do not correlate with the cosmological density field), $\langle X\rangle\simeq \langle\delta_{g,i}^\mathrm{true}\delta_{g,j}^\mathrm{true}\rangle$, i.e.\ the mean cross-correlation remains unbiased. The variance, $\mathrm{Var}(X)=\langle X^2\rangle-\langle X\rangle^2$, however, receives additional contributions.

Neglecting cross-terms between additive and multiplicative contaminants for simplicity, and assuming that $m_{g,j}$ is statistically independent of the cosmological fields, we obtain
\begin{align}
\langle X^2 \rangle
&\simeq
\langle (1+m_{g,j})^2 \rangle
\left\langle
(\delta_{g,i}^{\rm true}\delta_{g,j}^{\rm true})^2
\right\rangle
\nonumber\\
&\quad
+
\left\langle
(\delta_{g,i}^{\rm true} s_{g,j})^2
\right\rangle .
\end{align}
and therefore
\begin{align}
\mathrm{Var}(X)
&\simeq
\langle (1+m_{g,j})^2 \rangle\,
\mathrm{Var}\!\left(
\delta_{g,i}^{\rm true}\delta_{g,j}^{\rm true}
\right)
\nonumber\\
&\quad
+
\mathrm{Var}\!\left(
\delta_{g,i}^{\rm true} s_{g,j}
\right)
\nonumber\\
&\quad
+
\mathrm{Var}(m_{g,j})
\left\langle
\delta_{g,i}^{\rm true}\delta_{g,j}^{\rm true}
\right\rangle^2
+ \cdots .
\end{align}

For a mean-subtracted multiplicative field, $\langle m_{g,j}\rangle=0$, the prefactor reduces to $\langle(1+m_{g,j})^2\rangle=1+\langle m_{g,j}^2\rangle$, making explicit that multiplicative residuals rescale the variance of the true cross-correlation measurement. Similarly, the remaining terms contribute to the increase in the measurement variance (and thus the covariance of the estimated cross-correlation), degrading the statistical precision. 

Since the background sample enters the analysis only through cross-correlations, and its auto-correlation is not required for cosmological inference, this framework allows for more aggressive binning strategies at high redshift. In particular, the background sample can be extended to the full depth of the survey without stringent systematics mitigation requirements. At the same time, accurate correction of observational systematics in the foreground sample remains essential, as its auto-correlation is used to constrain the galaxy bias.

The impact of observational systematics on the IGGL observable can be understood similarly. In this configuration the galaxy position sample acts as the background field of the cross-correlation and may contain residual multiplicative and additive contaminants,
\begin{equation}
\delta_g^\mathrm{obs} = (1+m_g)\,\delta_g^\mathrm{true} + s_g,
\end{equation}
while the shape sample is assumed to be well calibrated.
Expanding the estimator,
\begin{equation}
\langle \delta_g^\mathrm{obs} e^\mathrm{obs} \rangle
=
\left\langle
\big[(1+m_g)\delta_g^\mathrm{true} + s_g\big]
(\gamma + e^\mathrm{int})
\right\rangle,
\end{equation}
and assuming that $m_g$ and $s_g$ do not correlate with the cosmological density or shear fields, all mixed terms vanish. The expectation value therefore reduces to the standard IGGL contribution, and the mean signal remains unbiased. However, residual multiplicative and additive contaminants increase the variance of the estimator through terms proportional to $\langle m_g^2\rangle$ and $\langle s_g s_g\rangle$, inflating the covariance without biasing the mean measurement.

\section{Data and simulations}
\label{sec:data_sims}

This section describes the data products and galaxy selections used to motivate the BigMag forecasts.  The empirical and forecast components play distinct roles.  First, DES Year 6 photometry, DES Year 6 \textsc{Balrog} synthetic-source injections, and SOMPZ redshift calibration are used to identify well-separated foreground and background populations and to measure how the number-density response to magnification varies within the high-redshift population.  A simple magnitude partition provides direct evidence that positive and negative magnification contributions can partially cancel in an unsplit sample.  Second, we explore a self-organizing map (SOM) as a more general representation in which both the redshift distribution and magnification response can be associated with regions of colour--magnitude space.  The present \textsc{Balrog} campaign, however, does not provide sufficient statistics for an unbiased optimization of the response of every individual SOM cell.  We therefore distinguish the response separation demonstrated by externally defined magnitude cuts from the larger, noisier apparent separation obtained by ranking SOM cells.  Finally, we construct simulated data vectors with DES-informed redshifts and number densities, but with optimistic target magnification responses.  The cosmological forecasts should consequently be interpreted as conditional forecasts of the information available if future, larger injection campaigns can construct and calibrate samples with this level of response separation.

\subsection{DES Y6 samples and \textsc{Balrog} injections}
\label{sec:data_sims_des_balrog}

We start from the public DES Y6 Gold photometric catalog \citep{Y6Gold} and the corresponding DES Y6 \textsc{Balrog} synthetic-source injection catalogs \citep{Y6Balrog}. To generate realistic injected sources, \textsc{Balrog} uses galaxy models fitted to observations from the DES Y3 deep fields \citep{y3-deepfields}. These synthetic galaxies are injected into the real DES images, after which the same detection, deblending, and measurement pipeline used for the survey data is rerun. This procedure provides an empirical characterization of the survey transfer function: for a galaxy with known injected properties, we can determine whether it is detected, how its properties are measured in the wide-field data, and which sample selection it enters. This information is a key ingredient in both the SOMPZ redshift calibration and the magnification-response calibration, which are formulated in terms of observed galaxy properties.

For the magnification calibration we use two \textsc{Balrog} runs from \citet{Y6Balrog}.  The fiducial run contains the unmodified injected galaxies.  The magnified run contains the same underlying injected galaxies after applying a small positive magnification with convergence $\Delta\kappa=0.01$ to their fluxes and sizes.  We apply the same catalog-level cuts to both measured catalogs: \texttt{FLAGS\_GOLD=0}, \texttt{FLAGS\_FOOTPRINT=1}, \texttt{FLAGS\_FOREGROUND=0}, and \texttt{EXT\_MASH>=2}.  We use the dereddened measured $i$-band magnitude for magnitude selections, and assign each detected object to a $32\times32$ wide-field SOM cell trained on DES wide-field $griz$ photometry.

The sample-selection response is estimated from weighted number counts.  If $S$ denotes a selected set of wide SOM cells and a possible magnitude range, the fiducial and magnified counts are
\begin{equation}
  \widehat N_r(S) =
  \sum_{j\in D_r(S)} \frac{1}{N_{{\rm inj},j}},
  \qquad r\in\{\mathrm{fid},\mathrm{mag}\},
  \label{eq:balrog_weighted_counts}
\end{equation}
where $D_r(S)$ is the set of detected \textsc{Balrog} objects in run $r$ that pass the selection, and $N_{{\rm inj},j}$ is the number of attempted injections of the corresponding underlying deep-field object.  The factor $1/N_{\rm inj}$ ensures that objects injected more often do not receive disproportionate weight.  We then define \citep{y3-2x2ptmagnification, Legnani2026}:
\begin{equation}
  \widehat C_{\rm sample}(S) =
  \frac{\widehat N_{\rm mag}(S)-\widehat N_{\rm fid}(S)}
       {\widehat N_{\rm fid}(S)\,\Delta\kappa}.
  \label{eq:csample_estimator}
\end{equation}
The geometric area term contributes $-2\kappa$ to the observed number-density contrast, so that $C_{\rm mag}=C_{\rm sample}-2$ is the net magnification response. Thus $C_{\rm sample}=2$ is the null-response boundary, and positive and negative $C_{\rm mag}$ produce oppositely signed magnification contributions. 

For estimates of $C_{\rm sample}$ and its covariance, we restrict the fiducial \textsc{Balrog} detections to the same set of wide-field tiles as the magnified run.  This matched-footprint choice removes changes in the count ratio caused by different survey areas rather than by magnifying the same parent population.  In contrast, the redshift calibration uses the full fiducial \textsc{Balrog} footprint, because it is intended to estimate the redshift distribution of the survey sample rather than the response of the matched magnification sub-footprint.

We estimate statistical uncertainties on $C_{\rm sample}$ with a spatial delete-one jackknife.  We construct $K$ \textsc{TreeCorr} patches \citep{treecorr} from the positions of the magnified injections and assign both fiducial and magnified detections to these fixed patches.  Let $\widehat N_{r,k}(S)$ be the weighted count in patch $k$.  The leave-one-out counts and response are
\begin{align}
  \widehat N_{r,-k}(S) &=
  \widehat N_r(S)-\widehat N_{r,k}(S), \\
  \widehat C_{{\rm sample},-k}(S) &=
  \frac{\widehat N_{{\rm mag},-k}(S)-\widehat N_{{\rm fid},-k}(S)}
       {\widehat N_{{\rm fid},-k}(S)\,\Delta\kappa}.
  \label{eq:csample_jackknife_samples}
\end{align}
The corresponding jackknife variance is then
\begin{align}
  \sigma^2_{\rm JK}\!\left(C_{\rm sample}\right)
  &=
  \frac{K-1}{K}
  \sum_{k=1}^{K}
  \left[
    \widehat C_{{\rm sample},-k}
    - \overline C_{\rm sample}^{\rm JK}
  \right]^2,\\
  \qquad
  \overline C_{\rm sample}^{\rm JK}
  &=
  \frac{1}{K}\sum_{k=1}^{K}
  \widehat C_{{\rm sample},-k}.
  \label{eq:csample_jackknife_variance}
\end{align}
This construction preserves the nonlinear ratio in Eq.~\eqref{eq:csample_estimator} for every leave-one-out realization. These jackknife uncertainties characterize the response of a fixed selection.  They do not by themselves include the additional uncertainty or selection bias that arises when the same noisy cell responses are used to define that selection.

\begin{figure}
    \centering
    \includegraphics[width=0.99\linewidth]{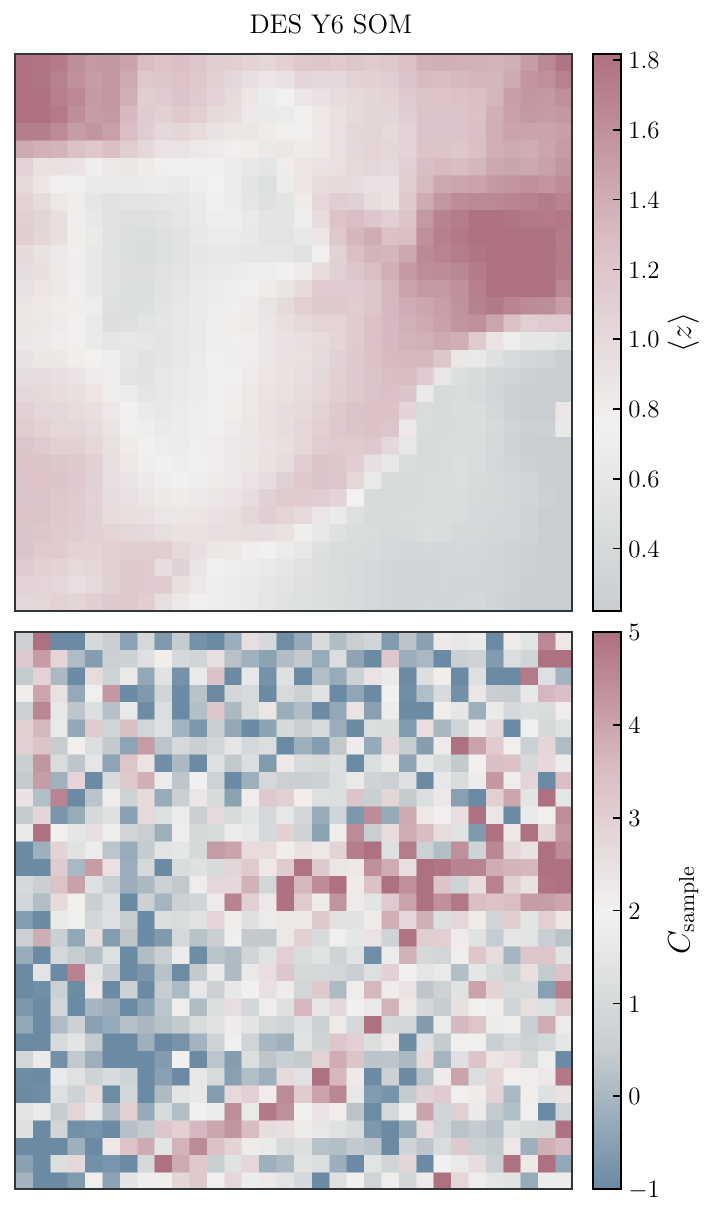}
    \caption{
    Self-organizing map (SOM) of DES Year 6 galaxies used to construct the redshift- and magnification-selected samples. Each pixel corresponds to a SOM cell defined from the observed galaxy fluxes. The top panel shows the mean redshift, $\langle z\rangle$, of the galaxies assigned to each cell, revealing the redshift structure across the color--magnitude manifold. The bottom panel shows the corresponding sample-selection magnification response, $C_{\rm sample}$, estimated with \textsc{Balrog} synthetic injections. The color scale for $C_{\rm sample}$ is centered on $C_{\rm sample}=2$, the value at which the sample-selection response cancels the geometric area term in the observed number density. Cells with different values of $C_{\rm sample}$ can be combined to build samples with positive, negative, or near-null magnification response at fixed redshift. These SOM maps therefore provide the basis for the high-redshift H$^{+}$ and H$^{-}$ selections used in the simulated data vectors.
}
\label{fig:som_y6_z_alpha}
\end{figure}

\subsection{A SOM representation of redshift and magnification response}
\label{sec:data_sims_som}

We use a SOM representation to compress the galaxy color-magnitude space into cells for which redshift and magnification response can be calibrated.  The wide SOM is trained on DES wide-field $griz$ photometry and contains $32\times32$ cells, denoted $\hat c$. This produces a two-dimensional map of the galaxy population in which each cell carries the quantities needed for the magnification analysis:
\begin{equation}
\label{eq:som_cell_properties}
\hat c \quad \longrightarrow \quad
\left\{ n_{\hat c}(z), \; \bar z_{\hat c}, \; n_{\rm gal}^{\hat c}, \; C_{\rm mag}^{\hat c} \right\}.
\end{equation}
Here $\bar z_{\hat c}$ is the estimated mean redshift for the cell and $n_{\rm gal}^{\hat c}$ is the angular number density. The magnification coefficient $C_{\rm mag}^{\hat c}$ is estimated from \textsc{Balrog} counts as explained in \S~\ref{sec:data_sims_des_balrog}.

For the redshift calibration we also use a deep SOM, denoted $c$, trained on the DES deep-field optical and near-infrared photometry.  The deep-field catalog is restricted to objects passing the deep field mask and flag cuts, and we keep objects classified as galaxies by the deep-field nearest-neighbor classifier.  Deep objects are matched to redshift information from the DES redshift calibration sample when available, while the \textsc{Balrog} transfer function links their deep-field properties to their measured wide-field SOM cells. The SOMPZ calibration follows the DES Y3 and DES Y6 methodology \citep{Buchs2019, Myles:2021:SOMPZ, highz, Campos2024, Giannini:2024:DES, DESY6lenspz,DESY6sourcepz}.

For a sample defined in wide cells, the redshift distribution of a target ensemble is written as a mixture over deep SOM cells,
\begin{equation}
  p(z\mid S) =
  \sum_{\hat c\in S} p(\hat c\mid S)
  \sum_c p(z\mid c,\hat c,S)\,p(c\mid \hat c,S),
  \label{eq:sompz_wide_mixture}
\end{equation}
where $p(c\mid\hat c,S)$ is measured from \textsc{Balrog}, and $p(z\mid c,\hat c,S)$ is calibrated from the redshift sample in the deep fields. In practice, we use the DES SOMPZ reweighting form, in which the \textsc{Balrog} distribution of deep cells is reweighted to match the deep-field population:
\begin{equation}
  p_S(z\mid \hat c)
  \propto
  \sum_c
  p(z\mid c,\hat c,S)
  p_{\rm B}(c\mid \hat c,S)
  \frac{p_{\rm D}(c\mid S)}{p_{\rm B}(c\mid S)}.
  \label{eq:sompz_balrog_reweighting}
\end{equation}
Here the subscripts ${\rm B}$ and ${\rm D}$ denote probabilities measured from
\textsc{Balrog} and from the deep-field sample, respectively.  The final
redshift distribution for a sample is
\begin{equation}
  n_S(z) =
  \sum_{\hat c\in S}
  p_{\rm W}(\hat c\mid S)\,
  p_S(z\mid \hat c),
  \qquad
  \int n_S(z)\,dz = 1,
  \label{eq:sompz_sample_nz}
\end{equation}
where $p_{\rm W}(\hat c\mid S)$ is the weighted abundance of the selected wide-SOM cell in the fiducial \textsc{Balrog} catalog. 

The SOM maps in Fig.~\ref{fig:som_y6_z_alpha} summarize the mean redshift and sample-selection response to magnification across the $32\times32$ wide-cell grid. As seen there, the redshift structure is sufficiently smooth and well sampled for defining high- and low-redshift selections. The magnification response map shows that individual cell estimates are noisy, although an aggregate response can be measured with useful precision after many cells are combined. However, selecting extreme cells and evaluating their aggregate response with the same injections preferentially selects upward and downward noise fluctuations, and their true response variation should be estimated out of sample.

\subsection{High-redshift samples and direct evidence for response cancellation} \label{sec:data_sims_sample_selection}

\begin{figure}
    \centering
    \includegraphics[width=0.99\linewidth]{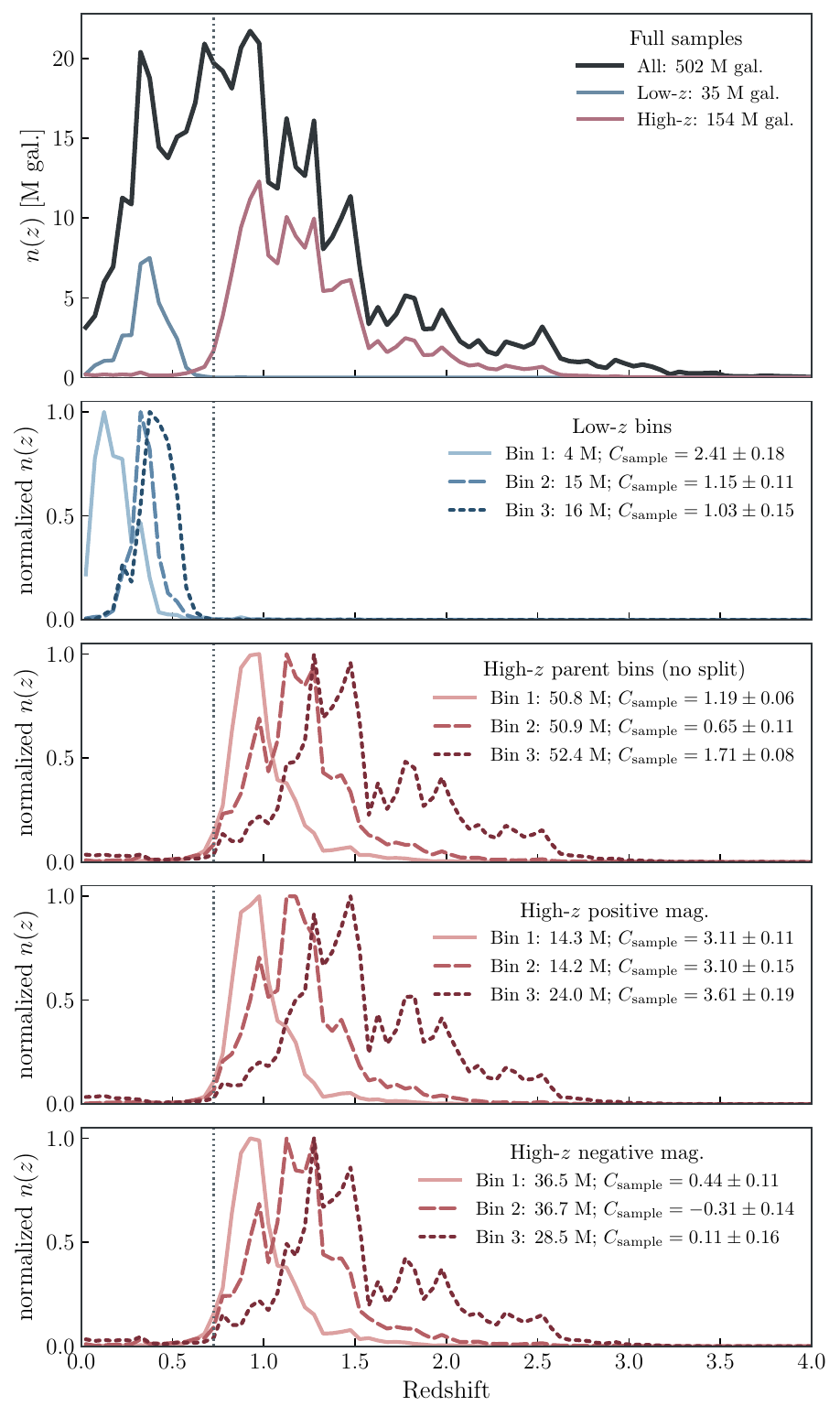}
    \caption{
Redshift distributions and magnification responses of the SOM-selected DES Y6 samples.  The top panel shows the SOMPZ redshift distributions for the full $i<27$ parent sample, the union of the low-redshift lens bins, and the union of the high-redshift density bins, with the vertical dotted line marking the boundary used to isolate the high-redshift selections.  The second panel shows the three low-redshift lens bins selected from the brighter $i<23$ sample.  The third panel shows the three high-redshift bins before splitting by magnification response.  The bottom two panels show the same high-redshift bins after splitting wide SOM cells at $C_{\rm sample}=2$, giving branches with positive and negative net magnification response.  Curves in the lower panels are normalized to unit peak for visual comparison.  Legend entries report the number density of each sample and the measured $C_{\rm sample}\pm\sigma_{\rm JK}(C_{\rm sample})$, where the uncertainty is estimated from the spatial delete-one jackknife of the matched \textsc{Balrog} counts. The response values were evaluated using the same injection catalog used to assign cells to the branches, and so they are apparent in-sample estimates and should not be interpreted as unbiased estimates of the response separation achievable on an independent realization.
}
\label{fig:y6_samples_nz_csample}
\end{figure}

We define each galaxy sample as a union of SOM cells. We separate the low-redshift lenses from the high-redshift position samples by requiring the redshift probability leaking across a boundary near $z=0.7$ to be small. We adopt this boundary after visually inspecting the DES colour information and finding that the available filters provide particularly effective redshift discrimination around this value. Minimizing the overlap between the two redshift distributions is essential for suppressing the intrinsic galaxy-clustering signal from physically associated galaxies. We then subdivide the samples according to whether $C_{\rm sample}^{\hat c}$ lies above or below 2.

The lens samples are built from the brighter magnitude threshold $i<23$. We first select low-redshift cells satisfying $P(z<0.60\mid\hat c)>0.98$.  These cells are then split into three lens bins using cumulative redshift-probability cuts: cells with $P(z<0.35\mid\hat c)\ge 0.80$, cells failing that condition but satisfying $P(z<0.45\mid\hat c)\ge 0.80$, and the remaining selected cells. This produces three low-redshift lens bins with increasing mean redshift.

The high-redshift samples are built from a deeper selection without an additional magnitude threshold.  We select cells for which $P(z<0.70\mid\hat c)<0.05$, sort them by their SOMPZ mean redshift, and divide the ordered cells into three bins with approximately equal cumulative wide-cell abundance.  These three parent samples form the basis for both the empirical response tests and the forecast samples below.

Starting from the top, Figure~\ref{fig:y6_samples_nz_csample} shows the corresponding redshift distributions, number counts, and measured values and corresponding errors of $C_{\rm sample}$ of the overall galaxy population, along with the low redshift and high redshift position samples we just described. The bottom two panels show versions of the high-redshift samples after splitting them on magnification response, which is described in \S\ref{sec:data_sims_som_response_validation}.

\subsubsection{Magnitude partitions}
\label{sec:data_sims_magnitude_partitions}

\begin{figure*}
  \centering
  \includegraphics[width=0.99\linewidth,height=0.68\textheight,keepaspectratio]
  {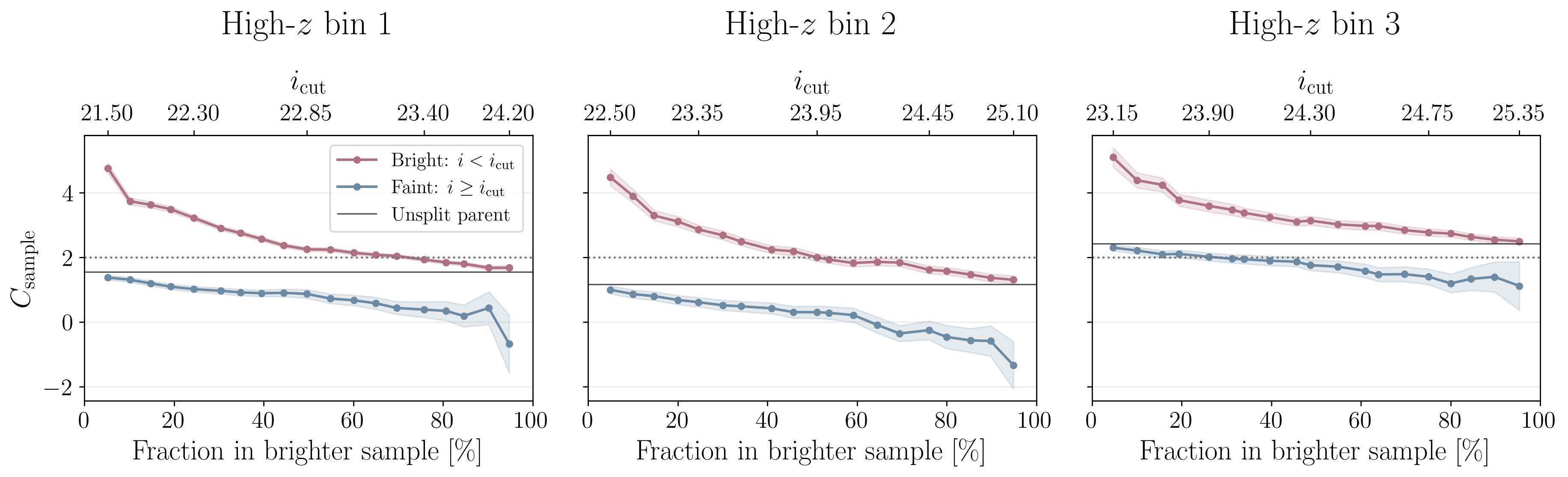}
  \caption{Measured sample-selection response for complementary magnitude partitions of the three DES Y6 high-redshift parent samples.  The horizontal axis gives the fraction of the fiducial parent assigned to the brighter branch; the upper axis gives the corresponding $i$-band threshold.  Rose and blue curves show the responses of the bright and faint branches, respectively, and the dotted line marks $C_{\rm sample}=2$ or $C_{\rm mag}=0$.  Because the thresholds are defined by magnitude and sample fraction rather than by noisy per-cell response estimates, the figure provides direct evidence for response cancellation.  At an illustrative 25/75 split, the bright/faint $C_{\rm mag}$ values are approximately $(+1.21,-0.99)$, $(+0.86,-1.39)$, and $(+1.59,+0.01)$ for high-redshift bins 1--3.  The first two bins therefore contain substantial positive and negative contributions, while the third is predominantly positive.  The shaded bands show the spatial jackknife uncertainty on each fixed magnitude selection.}
  \label{fig:y6_magnitude_fraction_splits}
\end{figure*}

A direct and interpretable way to test for response cancellation is to partition each high-redshift parent sample at an observed $i$-band magnitude threshold. For each chosen threshold $i_{\rm cut}$, the bright and faint selections are complementary:
\begin{align}
  H_i^{\rm bright}=\{g\in H_i:i<i_{\rm cut}\},\\
  H_i^{\rm faint}=\{g\in H_i:i\geq i_{\rm cut}\}.
  \label{eq:magnitude_partition}
\end{align}
We choose thresholds corresponding to a predefined grid of fractions in the brighter branch and then measure the response of each complete branch using Eq.~\eqref{eq:csample_estimator}.  Unlike the SOM-response split considered below, the membership of these branches is not determined by a noisy estimate of $C_{\rm sample}$ for each cell.

The physical origin of the separation is simple.  To first order, magnification changes an observed magnitude by
\begin{equation}
  \delta m\simeq-\frac{5}{\ln 10}\,\kappa.
  \label{eq:magnitude_shift}
\end{equation}
Objects therefore move across an internal magnitude boundary: the brighter branch gains objects from the fainter branch, while the fainter branch loses them.  When the two branches are recombined, this internal transfer cancels. The remaining response of the unsplit parent is set by its external detection and selection boundaries, as well as size, multiband signal-to-noise, photometric-redshift assignment, blending, and other measurement effects.  The response of a simple flux-limited sample is governed by the local slope of the cumulative number counts; in these more complicated selections, the matched \textsc{Balrog} measurement propagates all of these effects jointly.  Evolution of the luminosity function and selection boundaries with redshift can therefore produce two canceling branches in some tomographic bins but a mainly positive response in another.

The magnitude partitions shown in Figure~\ref{fig:y6_magnitude_fraction_splits} establish that significant response separation is already available in a one-dimensional selection.  They should not be regarded as the final optimized samples: the best selection may also use size, multiband fluxes, signal-to-noise, morphology, or other measured properties. They nevertheless demonstrate the physical cancellation that motivates BigMag without relying on the noisy ordering of individual SOM-cell responses. This is the photometric-galaxy analogue of the magnitude-dependent sign change observed in quasar and Lyman-break samples \citep{2005ApJ...633..589S, 2009A&A...507..683H}, although here the response includes effects from the full selection. The 25/75 point in the caption is quoted as an illustrative common comparison across bins, not as a claim that this threshold maximizes cosmological information.

\subsubsection{SOM-response partitions}
\label{sec:data_sims_som_response_validation}

The SOM provides a higher-dimensional route to response selection.  Within each high-redshift parent bin, an exploratory split can be formed from the cell point estimates,
\begin{align}
  H_i^{\rm +} &= \{\hat c\in H_i:
  \widehat C_{\rm sample}(\hat c)>2\},\\
  H_i^{\rm -} &= \{\hat c\in H_i:
  \widehat C_{\rm sample}(\hat c)<2\}.
  \label{eq:highz_csample_split_candidates}
\end{align}
When the same injection catalog is used to define and evaluate the branches, the $H_i^+$ and $H_i^-$ samples have positive and negative apparent net magnification responses, respectively, in the convention of Sec.~\ref{sec:gal_density_mag}.  Their redshift distributions, number counts, and all-injection estimates of $C_{\rm sample}$ and their jackknife uncertainties are shown in Fig.~\ref{fig:y6_samples_nz_csample}. 

The split at $C_{\rm sample}=2$ introduces a direct tradeoff between magnification response and number density.  Cells with the most extreme values of $C_{\rm sample}$ provide the largest magnification signal per galaxy, but they occupy only a small fraction of the SOM and therefore yield low-density samples.  Conversely, including cells progressively closer to $ C_{\rm sample}=2$ increases the number of galaxies, but reduces the net magnification response of the sample. The null-response partition used here is therefore a simple, physically motivated choice rather than a claim of optimality.  An optimized selection could instead combine the SOM coordinates with magnitude, size, signal-to-noise, or other observed properties, and should maximize the expected cosmological information while propagating response and redshift-calibration uncertainty.

The tests in Appendix~\ref{app:csample-flagship} and Appendix~\ref{app:csample-heldout-validation} help distinguish between the presence of response cancellation and our ability to reliably identify the optimal response branches. In a DES-like Flagship sample, we recover the same qualitative trends with magnitude seen in the data: the first two high-redshift bins contain substantial positive and negative contributions, while the third is dominated by a positive response. When the number of galaxies per SOM cell is large, partitioning the sample by SOM response produces a strong and stable separation, with good agreement between in-sample and held-out measurements. However, once the Flagship sample is degraded to Balrog-like statistics, the contrast of the held-out responses is significantly reduced, closely resembling the behavior seen in DES. This shows that the SOM ordering does contain genuine information about the response, and that current DES injection statistics are not sufficient to fully optimize the partition of each tomographic bin by magnification response. Optimally separating each tomographic bin for magnification measurements requires a larger injection campaign with disjoint optimization and validation subsets. Appendix~\ref{app:csample-injection-scaling} shows that the statistical uncertainty of the response follows the expected inverse-square-root scaling with the number of injections, supporting this strategy and quantifying the increase in injection statistics required for a precise calibration.

\subsection{Simulated data vectors}
\label{sec:data_sims_inputs}

\begin{table*}
\centering
\refstepcounter{table}
\label{tab:data_sims_samples}

{\normalfont Table~\thetable. Fiducial galaxy samples used to construct the simulated data vectors. The table
summarizes the low-redshift lens samples (L1--L3), the high-redshift lens
samples split by magnification response (H$^{+}$ and H$^{-}$), and the four
source samples used for galaxy--galaxy lensing and cosmic shear. The low-redshift lens magnification responses are set to
$C_{\rm mag}=0$ in the fiducial simulations, while the high-redshift values are
informed by the DES Y6+\textsc{Balrog} selections. The H$^{+}$ samples therefore have
$C^{\rm mag}>0$, while the H$^{-}$ samples have $C^{\rm mag}<0$. All redshift
distributions are modeled as Gaussian approximations with data-informed means
$\langle z\rangle$ and widths $\sigma_z$. Source samples use the effective
number densities and shape-noise values adopted in the covariance simulations.
\par}

\vspace{0.8em}

\renewcommand{\arraystretch}{1.25}
\setlength{\tabcolsep}{4pt}
\begin{tabular*}{\textwidth}{
@{\extracolsep{\fill}}
l
ccc|
ccc|
ccc|
cccc
}
\hline
&
\multicolumn{3}{c|}{Low-$z$ lenses} &
\multicolumn{3}{c|}{High-$z$ lenses ($C_{\rm mag}>0$)} &
\multicolumn{3}{c|}{High-$z$ lenses ($C_{\rm mag}<0$)} &
\multicolumn{4}{c}{Sources}
\\
\cline{2-14}

Property
& $L_1$ & $L_2$ & $L_3$
& $H_1^{+}$ & $H_2^{+}$ & $H_3^{+}$
& $H_1^{-}$ & $H_2^{-}$ & $H_3^{-}$
& $S_1$ & $S_2$ & $S_3$ & $S_4$
\\
\hline

$\langle z\rangle$
& 0.20 & 0.34 & 0.44
& 0.95 & 1.30 & 1.70
& 0.95 & 1.30 & 1.70
& 0.414 & 0.538 & 0.846 & 1.157
\\[2pt]

$\sigma_z$
& 0.06 & 0.04 & 0.08
& 0.10 & 0.13 & 0.18
& 0.10 & 0.13 & 0.18
& 0.15 & 0.15 & 0.15 & 0.15
\\[2pt]

$n_{\rm gal}\,[{\rm arcmin}^{-2}]$
& 0.224 & 0.842 & 0.842
& 0.673 & 0.617 & 1.403
& 2.132 & 2.188 & 1.515
& 2.05 & 2.10 & 2.14 & 2.32
\\[2pt]

$C_{\rm mag}$
& 0.00 & 0.00 & 0.00
& 1.10 & 1.10 & 1.60
& $-1.50$ & $-2.25$ & $-1.90$
& -- & -- & -- & --
\\[2pt]

$\sigma_e$
& -- & -- & --
& -- & -- & --
& -- & -- & --
& 0.265 & 0.287 & 0.282 & 0.347
\\
\hline
\end{tabular*}
\end{table*}

\begin{figure*}
\centering
\begin{tikzpicture}
\begin{axis}[
    width=0.9\linewidth,
    height=0.32\linewidth,
    xmin=0, xmax=2.3,
    ymin=-1.08, ymax=1.08,
    axis lines=middle,
    axis x line*=middle,
    axis y line=none,
    xlabel={\normalsize Redshift $z$},
    ylabel={},
    xlabel style={at={(axis description cs:0.98,0.48)}, anchor=north east},
    xtick={0.2,0.5,0.8,1.1,1.4,1.7},
    ytick=\empty,
    tick align=outside,
    axis line style={line width=0.8pt},
    tick style={line width=0.8pt},
    ticklabel style={font=\footnotesize},
    every axis plot/.append style={line width=1.15pt, samples=220},
    legend style={
        draw=none,
        fill=none,
        font=\footnotesize,
        at={(0.985,0.985)},
        anchor=north east,
        legend columns=5
    },
    legend image code/.code={
        \draw[#1, line width=1.15pt] (0cm,0cm) -- (0.48cm,0cm);
    },
    clip mode=individual,
    clip=false,
]

% Custom vertical axis with arrows in both directions
\draw[
    {Latex[length=2.4mm]}-{Latex[length=2.4mm]},
    line width=0.8pt
]
(axis cs:0,-1.05) -- (axis cs:0,1.05);

% Labels next to the two y-axis arrows
\node[anchor=east, font=\normalsize] at (axis cs:-0.055,0.92)
    {$n^{\mathrm{lens}}(z)$};

\node[anchor=east, font=\normalsize] at (axis cs:-0.055,-0.92)
    {$n^{\mathrm{source}}(z)$};

% Shared redshift axis
\addplot[name path=zero, domain=0:2.3, draw=black, line width=0.8pt, forget plot] {0};

% Lens bins: peak-normalized Gaussians above the redshift axis.
% First 3 lenses: color1, alpha=0.5

\addplot[name path=Lone, domain=0:0.45, lensLow] 
    {exp(-((x-0.20)^2)/(2*(0.055^2)))};
\addplot[draw=none, fill=color1, fill opacity=0.2, forget plot] 
    fill between[of=Lone and zero];

\addplot[name path=Ltwo, domain=0.18:0.58, lensLow] 
    {exp(-((x-0.34)^2)/(2*(0.040^2)))};
\addplot[draw=none, fill=color1, fill opacity=0.2, forget plot] 
    fill between[of=Ltwo and zero];

\addplot[name path=Lthree, domain=0.18:0.72, lensLow] 
    {exp(-((x-0.44)^2)/(2*(0.080^2)))};
\addplot[draw=none, fill=color1, fill opacity=0.2, forget plot] 
    fill between[of=Lthree and zero];

% Remaining lenses: color1, alpha=0.9

\addplot[name path=Lfour, domain=0.56:1.24, lensHigh] 
    {exp(-((x-0.95)^2)/(2*(0.090^2)))};
\addplot[draw=none, fill=color1, fill opacity=0.4, forget plot] 
    fill between[of=Lfour and zero];

\addplot[name path=Lfive, domain=0.56:1.24, lensHigh] 
    {exp(-((x-0.95)^2)/(2*(0.090^2)))};
\addplot[draw=none, fill=color1, fill opacity=0.4, forget plot] 
    fill between[of=Lfive and zero];

\addplot[name path=Lsix, domain=0.92:1.62, lensHigh] 
    {exp(-((x-1.30)^2)/(2*(0.125^2)))};
\addplot[draw=none, fill=color1, fill opacity=0.4, forget plot] 
    fill between[of=Lsix and zero];

\addplot[name path=Lseven, domain=0.92:1.62, lensHigh] 
    {exp(-((x-1.30)^2)/(2*(0.125^2)))};
\addplot[draw=none, fill=color1, fill opacity=0.4, forget plot] 
    fill between[of=Lseven and zero];

\addplot[name path=Leight, domain=1.22:2.20, lensHigh] 
    {exp(-((x-1.70)^2)/(2*(0.180^2)))};
\addplot[draw=none, fill=color1, fill opacity=0.4, forget plot] 
    fill between[of=Leight and zero];

\addplot[name path=Lnine, domain=1.22:2.20, lensHigh] 
    {exp(-((x-1.70)^2)/(2*(0.180^2)))};
\addplot[draw=none, fill=color1, fill opacity=0.4, forget plot] 
    fill between[of=Lnine and zero];

% Source bins: color2, alpha=0.8

\addplot[name path=Sone, domain=0:0.88, sourceStyle] 
    {-exp(-((x-0.415)^2)/(2*(0.140^2)))};
\addplot[draw=none, fill=color2, fill opacity=0.4, forget plot] 
    fill between[of=Sone and zero];

\addplot[name path=Stwo, domain=0.08:1.02, sourceStyle] 
    {-exp(-((x-0.538)^2)/(2*(0.140^2)))};
\addplot[draw=none, fill=color2, fill opacity=0.4, forget plot] 
    fill between[of=Stwo and zero];

\addplot[name path=Sthree, domain=0.40:1.30, sourceStyle] 
    {-exp(-((x-0.846)^2)/(2*(0.135^2)))};
\addplot[draw=none, fill=color2, fill opacity=0.4, forget plot] 
    fill between[of=Sthree and zero];

\addplot[name path=Sfour, domain=0.76:1.60, sourceStyle] 
    {-exp(-((x-1.157)^2)/(2*(0.135^2)))};
\addplot[draw=none, fill=color2, fill opacity=0.4, forget plot] 
    fill between[of=Sfour and zero];

%\node[anchor=east, font=\normalsize] at (axis cs:0.17,0.72) {lenses};
%\node[anchor=east, font=\normalsize] at (axis cs:0.17,-0.72) {sources};

%----------------------------------
% Bin labels
%----------------------------------

% Low-z lens bins
\node[font=\footnotesize, text=color1] at (axis cs:0.20,1.12) {$L_1$};
\node[font=\footnotesize, text=color1] at (axis cs:0.34,1.12) {$L_2$};
\node[font=\footnotesize, text=color1] at (axis cs:0.44,1.12) {$L_3$};

% High-z lens bins
\node[font=\footnotesize, text=color1] at (axis cs:0.95,1.12) {$H_1$};
\node[font=\footnotesize, text=color1] at (axis cs:1.30,1.12) {$H_2$};
\node[font=\footnotesize, text=color1] at (axis cs:1.70,1.12) {$H_3$};

% Source bins
\node[font=\footnotesize, text=color2] at (axis cs:0.415,-1.12) {$S_1$};
\node[font=\footnotesize, text=color2] at (axis cs:0.538,-1.12) {$S_2$};
\node[font=\footnotesize, text=color2] at (axis cs:0.846,-1.12) {$S_3$};
\node[font=\footnotesize, text=color2] at (axis cs:1.157,-1.12) {$S_4$};

\end{axis}
\end{tikzpicture}
\caption{
Gaussian redshift distributions used for the simulated data vectors. The low-redshift lens samples L1--L3 and high-redshift lens samples H1--H3 are shown above the horizontal axis, while the four source samples S1--S4 are mirrored below the axis for visual clarity. The high-redshift lens distributions are used for both the positive- and negative-magnification selections listed in Table~\ref{tab:data_sims_samples}; each H$^{+}$/H$^{-}$ pair shares the same redshift distribution but differs in number density and magnification coefficient $C_{\rm mag}$.
}
\label{fig:nzs_all}
\end{figure*}
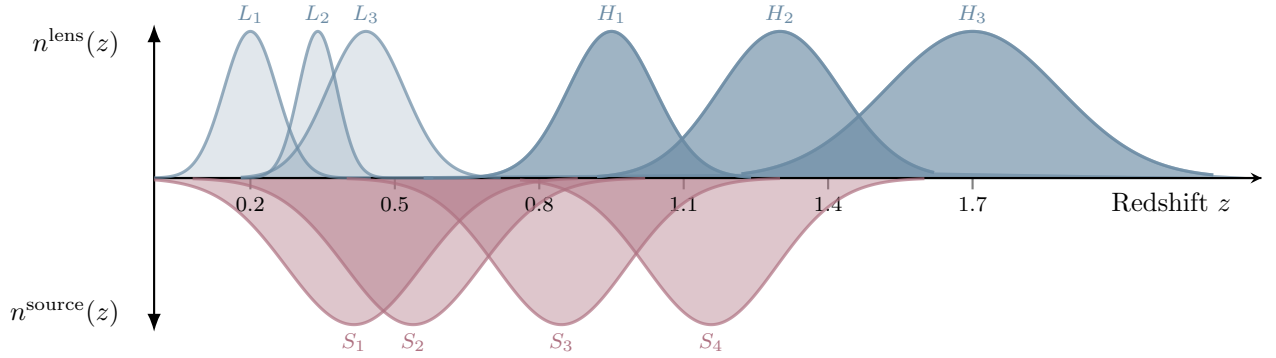

The empirical results above establish significant cancellation and demonstrate that galaxy-property selections can separate the response. Therefore, we use the DES Y6 SOM selections described in the previous sections and shown in Figure~\ref{fig:y6_samples_nz_csample} to provide the empirical inputs used to generate the simulated observables considered in this work. These observables include galaxy clustering, galaxy--galaxy lensing, and cosmic shear, measured for the following classes of galaxy samples:
\begin{enumerate}
    \item Low-$z$ lens samples: galaxy-position samples at redshift $z<0.7$,
    with moderate number densities. We assume that these samples can be
    corrected for imaging systematics, since their clustering autocorrelations
    are included in the analysis.

    \item High-$z$ lens samples with positive/negative magnification response:
    galaxy-position samples at redshift $z>0.7$, with high number densities.
    These samples are used only through cross-correlations with lower-redshift
    samples, so residual imaging systematics primarily affect the covariance
    rather than biasing the mean signal, under the assumptions discussed in
    Sec.~\ref{sec:systematics}. They are selected to have either positive or
    negative magnification response, as described earlier in this section.

    \item Source galaxies: galaxy-shape samples, mimicking the four source bins
    from the DES Y6 fiducial weak-lensing analysis.
\end{enumerate}

For these samples, we take the mean redshifts from the DES Y6+\textsc{Balrog} sample selections described above and use approximate widths to construct Gaussian redshift distributions. The galaxy number densities are derived from the data. The magnification coefficients are motivated by the DES Y6+\textsc{Balrog} measurements and are adopted as
optimistic target responses, rather than independently validated
calibrations of the final data samples. We denote these coefficients by $C^{\rm mag}_i$, following the convention of Sec.~\ref{sec:theory}, where $C^{\rm mag}_i=0$ corresponds to zero net magnification response. The sign of $C^{\rm mag}_i$ therefore determines the sign of the magnification contribution: samples with $C^{\rm mag}_i>0$ have positive magnification response, while samples with $C^{\rm mag}_i<0$ have negative magnification response.

The details of the simulated galaxy samples are summarized in Table~\ref{tab:data_sims_samples}, and the redshift distributions corresponding to these fiducial samples are shown in Fig.~\ref{fig:nzs_all}, with the lens samples displayed above the axis and the source samples mirrored below it for visual clarity. The first three bins are low-redshift foreground samples. The remaining six lens bins are three high-redshift parent samples split by their magnification response. Bins H1$^{+}$ and H1$^{-}$ share the same Gaussian redshift distribution centered at $z=0.95$, bins H2$^{+}$ and H2$^{-}$ share $z=1.30$, and bins H3$^{+}$ and H3$^{-}$ share $z=1.70$. Within each pair, the two subsamples have different number densities and different magnification coefficients, inherited from the SOM+\textsc{Balrog} selection. We also include four source bins matching the mean redshifts, effective number densities, and shape noise of the DES Y6 fiducial weak-lensing analysis \citep{Y6Mdet}.

\begin{figure*}
    \centering
    \includegraphics[width=0.99\linewidth]{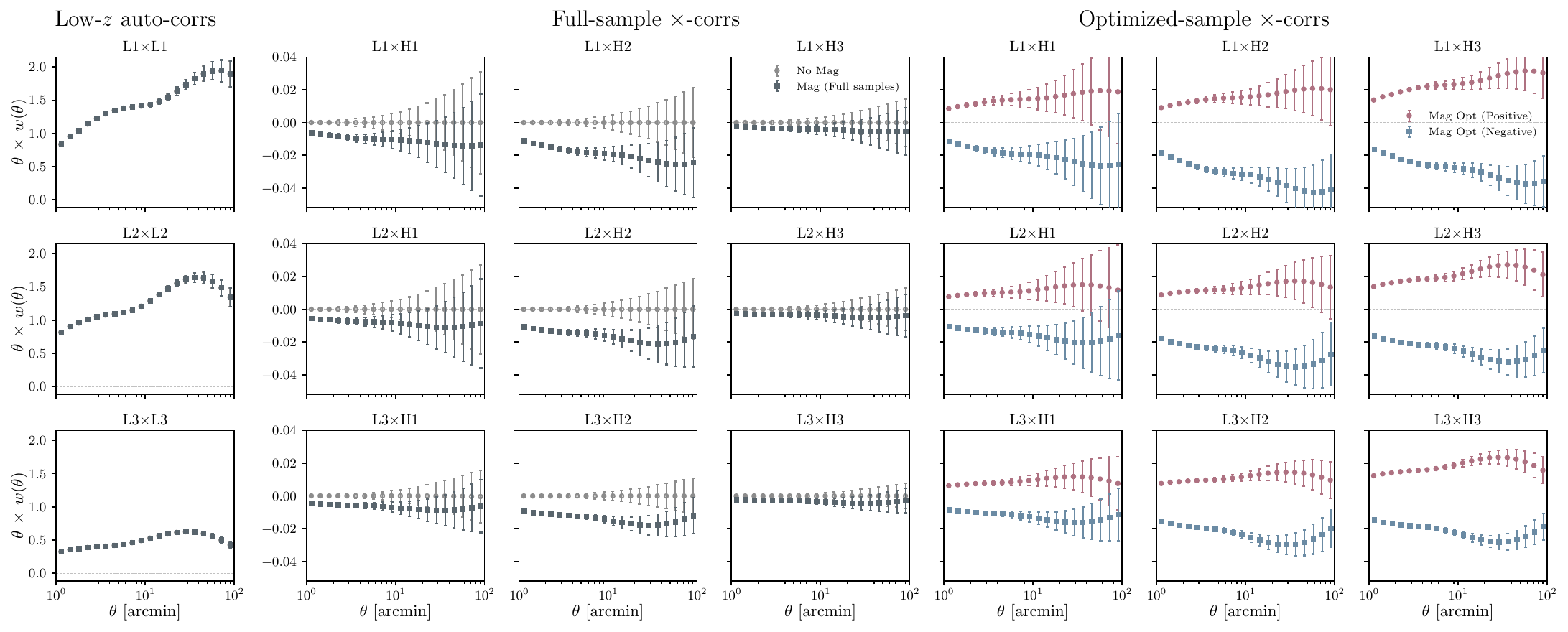}
\caption{
    Simulated galaxy-clustering measurements and diagnostic comparison curves. The fiducial data vector used in the likelihood consists of the low-redshift autocorrelations shown in the left column and the optimized low-$z$--high-$z$ cross-correlations shown in the right panels (Optimized-sample $\times$-corrs.), where the high-redshift samples are split by magnification response. The middle panels (Full-sample $\times$-corrs.) show the corresponding cross-correlations for the unsplit high-redshift parent samples, while the open grey points show the same parent-sample correlations with the magnification response artificially set to zero. These middle-panel curves are included only as diagnostics and are not additional observables in the fiducial likelihood. Error bars show the square root of the diagonal of the simulated covariance, and all points are shown before applying analysis scale cuts.
}
    \label{fig:wtheta_all}
\end{figure*}
\begin{figure*}
 \centering
    \includegraphics[width=0.99\linewidth]{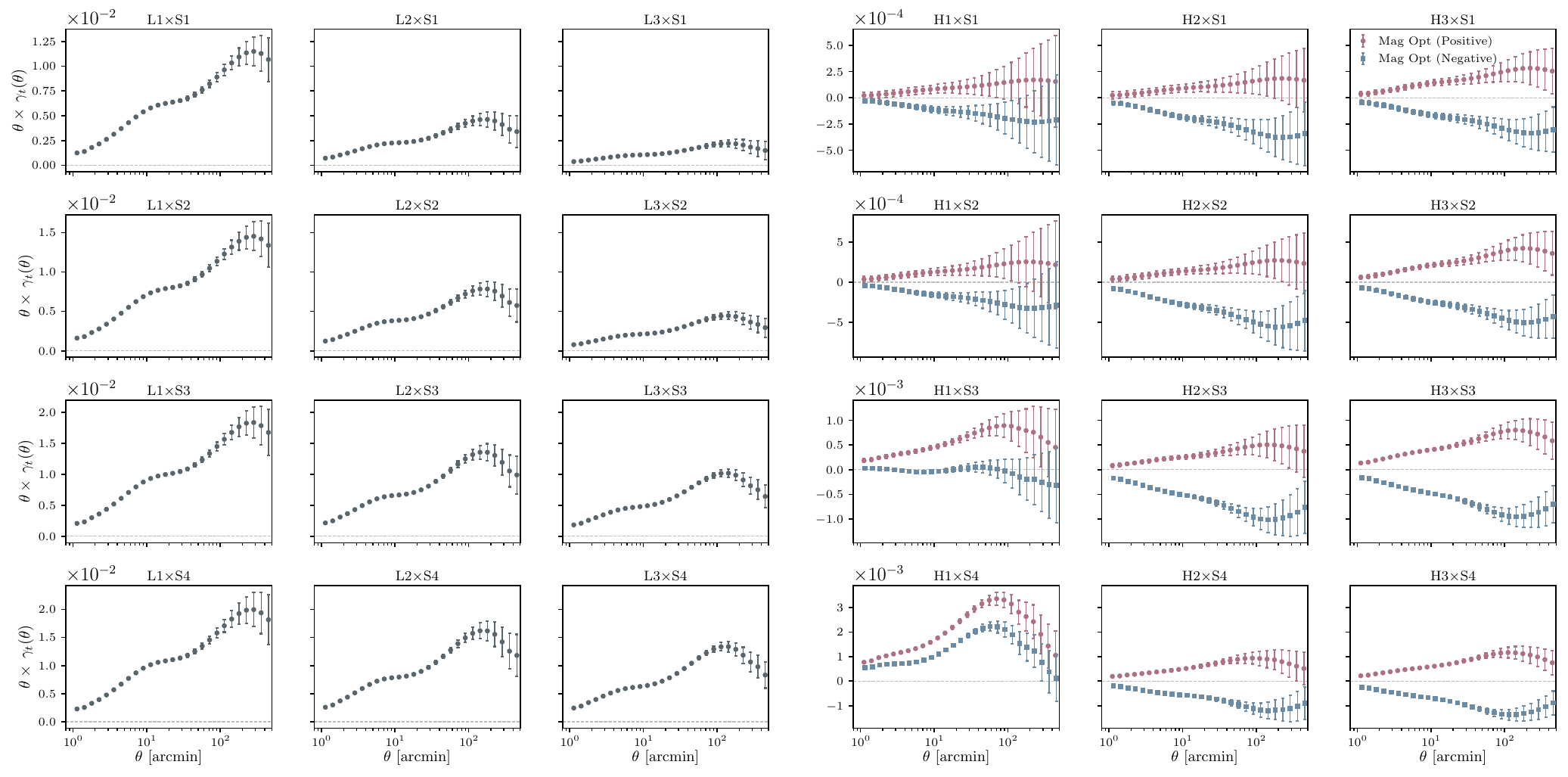}
    \caption{
    Simulated galaxy--galaxy lensing measurements for the fiducial galaxy samples. Each panel shows $\theta\,\gamma_t(\theta)$ for one position--source combination. The left block contains the low-redshift lens samples L1--L3 cross-correlated with the four source bins S1--S4; these are traditional galaxy--galaxy lensing measurements because the position samples lie in front of the source samples. The right block shows the high-redshift position samples H1--H3 cross-correlated with the same source bins. Most of these combinations are inverse galaxy--galaxy lensing (IGGL) measurements, with the position sample behind the source sample, so the signal is driven by the magnification response of the high-redshift sample and changes sign between the positive- and negative-magnification selections. The H1$\times$S4 case is an exception: H1 lies below S4 in redshift, so this panel behaves as a traditional GGL measurement. Error bars show the diagonal uncertainty from the simulated covariance.
}
\label{fig:sims_ggl_iggl}
\end{figure*}

We generate the simulated two-point measurements with \textsc{CosmoSIS} \citep{Cosmosis}, using the redshift distributions, number densities, shape noise, galaxy biases, and magnification coefficients listed in Table~\ref{tab:data_sims_samples}. The fiducial theory model uses a flat $\Lambda$CDM cosmology with values from the DES Y3 3x2pt configuration \citep{y3-generalmethods}, as distributed with the \textsc{CosmoSIS} standard library. We use linear galaxy bias $b_i=1.5$ for all position samples, the magnification responses $C_{\rm mag}$ given in the table, and an NLA intrinsic-alignment model with fiducial amplitude $A_1=0$. Non-linear matter clustering and baryonic effects are included using the model from \citet{Mead_2021} with $\log_{10}(T_{\rm AGN}/{\rm K})=7.8$. The angular correlation functions are produced in 27 logarithmic angular bins spanning $1$--$500$ arcmin for galaxy clustering, galaxy--galaxy lensing, and cosmic shear, corresponding to the observables $w(\theta)$, $\gamma_t(\theta)$, $\xi_+(\theta)$, and $\xi_-(\theta)$.

The simulated covariance is computed for the full joint data vector, including all auto- and cross-covariances between $w(\theta)$, $\gamma_t(\theta)$, $\xi_+$, and $\xi_-$. We assume the best-case scenario where all systematics have been corrected in all samples. We assume $f_{\rm sky}=0.12$, from DES, and include the relevant shot-noise and shape-noise contributions using the densities and ellipticity dispersions in Table~\ref{tab:data_sims_samples}. For reference, we quote the raw signal-to-noise ratios of the simulated data-vector blocks over the full angular range used to generate the measurements, before applying any analysis scale cuts. Using all 27 angular bins between $1$ and $500$ arcmin, we find $S/N\simeq1254$ for the full galaxy-clustering block, $S/N\simeq484$ for the full galaxy--galaxy lensing block, and $S/N\simeq165$ for the cosmic-shear block.

The clustering measurements are shown in Fig.~\ref{fig:wtheta_all}, illustrating both the low-redshift autocorrelations and the magnification-driven cross-correlations with the high-redshift samples. Figure~\ref{fig:wtheta_all} also shows two diagnostic comparison cases that are not included as separate observables in the fiducial likelihood: the cross-correlations using the unsplit high-redshift parent samples, and the same cross-correlations with the magnification response artificially set to zero. These curves (shown in the middle panels titled 'Full-sample $\times$-corrs.') are shown only to illustrate the origin of the signal and the gain from the magnification split. Restricting to the low-$z$--high-$z$ cross-correlations, the optimized magnification split increases the raw signal-to-noise from $S/N\simeq64$ for the unsplit parent samples to $S/N\simeq194$ for the fiducial split samples (with the corresponding no-magnification case having negligible signal-to-noise).

The simulated galaxy--galaxy lensing measurements are shown in Fig.~\ref{fig:sims_ggl_iggl}. The measurements naturally split into two redshift regimes. When the mean redshift of the position sample is lower than that of the source sample, the signal corresponds to traditional galaxy--galaxy lensing: the dominant contribution is the correlation between foreground galaxy density and background shear. This is the case for all low-$z$ lens--source combinations, L$\times$S, shown in the left part of the figure. When the position sample is instead at higher redshift than the source sample, the standard density--shear contribution is suppressed and the measurement becomes an inverse galaxy--galaxy lensing (IGGL) signal, as defined in Sec.~\ref{sec:theory} and \citet{IGGL}. In this regime, the signal is sourced by the magnification contribution to the high-redshift position sample, and therefore changes sign between the positive- and negative-magnification selections. This is the dominant configuration for most H$\times$S combinations shown in the right part of the figure.

There are, however, mixed cases among the high-redshift lenses. In particular, H1 lies below the highest-redshift source bin S4, so the H1$\times$S4 combination behaves as a traditional GGL measurement rather than as IGGL. This explains why this panel has the same qualitative sign for the positive- and negative-magnification selections, unlike the genuine IGGL configurations where the two selections trace opposite-sign signals. Separating the data vector into these two regimes, we find raw signal-to-noise ratios of $S/N\simeq470$ for the traditional GGL subset and $S/N\simeq149$ for the IGGL subset; the latter is comparable to the signal-to-noise of the simulated cosmic shear measurements, which are not shown because we will only use them for comparison/combination.

\section{Cosmological constraints from optimized magnification probes}
\label{sec:results}

We now investigate the cosmological information carried by the optimized magnification observables introduced in the previous sections. We first consider the clustering-only BigMag data vector, isolating the role of the magnification-selected samples in breaking the degeneracy between galaxy bias and the amplitude of matter fluctuations, \S\ref{sec:results_bigmag}. We then study how its constraining power depends on the calibration of the magnification coefficients and compare it directly with traditional $2\times2$pt, \S\ref{sec:results_bigmag_calibration}, which provides a natural shear-based counterpart to BigMag. We subsequently add IGGL, study the complementarity between the two magnification probes, and compare their joint constraints with those from a traditional $3\times2$pt analysis, \S\ref{sec:results_bigmag_iggl}.

For each combination of observables, we compute theoretical predictions using \textsc{CosmoSIS} \citep{Zuntz2015} and evaluate a Gaussian likelihood,
\begin{equation}
-2\ln\mathcal{L}
=
\left(\mathbf{d}-\mathbf{t}(\boldsymbol{\vartheta})\right)^{\rm T}
\mathbf{C}^{-1}
\left(\mathbf{d}-\mathbf{t}(\boldsymbol{\vartheta})\right),
\end{equation}
where $\mathbf{d}$ is the simulated data vector described in Sec.~\ref{sec:data_sims_inputs}, $\mathbf{t}(\boldsymbol{\vartheta})$ is its theoretical prediction, and $\mathbf{C}$ is the fixed simulated covariance. We explore the posterior using the \textsc{Nautilus} nested sampler \citep{Nautilus_Lange23}, adopting 1500 live points and a target effective posterior sample size of 5000 for each chain.

We sample the flat $\Lambda$CDM parameter space used to construct the simulations, varying $\Omega_{\rm m}$, $h$, $\Omega_{\rm b}$, $n_{\rm s}$, $A_{\rm s}$, and the sum of the neutrino masses, and derive $\sigma_8$ from the resulting matter power spectrum. The relevant nuisance parameters for each data-vector combination are varied simultaneously. These include the galaxy biases, redshift-distribution shifts, and magnification coefficients of the position samples, as well as shear-calibration, intrinsic-alignment, and baryonic-feedback parameters when shear measurements are included. The fiducial values and priors adopted for all sampled parameters are summarized in Table~\ref{tab:analysis_parameters}.

% Requires \usepackage{booktabs}

\begin{table}
\centering
\caption{
Fiducial values and baseline priors for the cosmological
and nuisance parameters. Uniform priors are denoted by
$\mathcal{U}[a,b]$, and Gaussian priors by
$\mathcal{N}(\mu,\sigma)$, with mean $\mu$ and standard
deviation $\sigma$. The Gaussian priors on redshift shifts
and shear-calibration parameters are truncated to
$[-0.1,0.1]$. The magnification-response rows give the
fiducial $\sigma_C=0.2$ priors; alternative calibration
scenarios and the parameters varied in each analysis
are specified below and in Table~\ref{tab:chain_summary}.
}
\label{tab:analysis_parameters}
\footnotesize
\renewcommand{\arraystretch}{1.08}
\begin{tabular}{@{}lll@{}}
\toprule
Parameter & Fiducial value & Prior \\
\midrule

\multicolumn{3}{l}{\textbf{Cosmology}} \\[1pt]
$\Omega_{\rm m}$
    & $0.30$
    & $\mathcal{U}[0.10,0.90]$ \\
$h$
    & $0.69$
    & $\mathcal{U}[0.55,0.91]$ \\
$\Omega_{\rm b}$
    & $0.048$
    & $\mathcal{U}[0.03,0.07]$ \\
$n_{\rm s}$
    & $0.97$
    & $\mathcal{U}[0.87,1.07]$ \\
$10^{9}A_{\rm s}$
    & $2.19$
    & $\mathcal{U}[0.50,5.00]$ \\
$\sum m_\nu\ [{\rm eV}]$
    & $0.0773$
    & $\mathcal{U}[0.06,0.60]$ \\
$\sigma_8$
    & $0.82355$
    & Derived \\

\addlinespace[2pt]
\multicolumn{3}{l}{\textbf{Position samples}} \\[1pt]
$b_i\;(i=1,\ldots,9)$
    & $1.5$
    & $\mathcal{U}[0.8,4.0]$ \\
$b_{2,i}\;(i=1,\ldots,9)$
    & $0$
    & $\mathcal{U}[-5,5]$ \\
$\Delta z_i^{\rm pos}\;(i=1,\ldots,9)$
    & $0$
    & $\mathcal{N}(0,0.01)$ \\
$C^{\rm mag}_{L_i}\;(i=1,2,3)$
    & $0$
    & $\mathcal{N}(0,0.2)$ \\
$C^{\rm mag}_{H_1^+}$
    & $1.10$
    & $\mathcal{N}(1.10,0.2)$ \\
$C^{\rm mag}_{H_1^-}$
    & $-1.50$
    & $\mathcal{N}(-1.50,0.2)$ \\
$C^{\rm mag}_{H_2^+}$
    & $1.10$
    & $\mathcal{N}(1.10,0.2)$ \\
$C^{\rm mag}_{H_2^-}$
    & $-2.25$
    & $\mathcal{N}(-2.25,0.2)$ \\
$C^{\rm mag}_{H_3^+}$
    & $1.60$
    & $\mathcal{N}(1.60,0.2)$ \\
$C^{\rm mag}_{H_3^-}$
    & $-1.90$
    & $\mathcal{N}(-1.90,0.2)$ \\

\addlinespace[2pt]
\multicolumn{3}{l}{\textbf{Source samples}} \\[1pt]
$\Delta z_s^{\rm src}\;(s=1,\ldots,4)$
    & $0$
    & $\mathcal{N}(0,0.01)$ \\
$m_s\;(s=1,\ldots,4)$
    & $0$
    & $\mathcal{N}(0,0.01)$ \\

\addlinespace[2pt]
\multicolumn{3}{l}{\textbf{Intrinsic alignments}} \\[1pt]
$A_1$
    & $0$
    & $\mathcal{U}[-5,5]$ \\
$\alpha_1^{\rm IA}$
    & $0$
    & Fixed \\
$z_{\rm piv}$
    & $0.3$
    & Fixed \\

\addlinespace[2pt]
\multicolumn{3}{l}{\textbf{Baryonic feedback}} \\[1pt]
$\log_{10}(T_{\rm AGN}/{\rm K})$
    & $7.8$
    & $\mathcal{U}[7.0,10.0]$ \\

\bottomrule
\end{tabular}
\vspace{0.5em}

\begin{minipage}{\linewidth}
\footnotesize
The second-order coefficients $b_{2,i}$ are varied only
in the nonlinear-bias tests; their linear coefficients
$b_{1,i}$ have the same fiducial values and priors as
$b_i$. Position-sample nuisance parameters are varied
for the samples included in each likelihood:
all nine position samples for BigMag and the joint
analyses, the six high-redshift samples for IGGL alone,
and the three foreground samples for traditional
$2\times2$pt and $3\times2$pt.

\smallskip
The magnification-calibration tests additionally use
$\sigma_C=0.1$ or $0.02$, uniform priors
$C^{\rm mag}\in[-12,8]$, or responses fixed to their
input values, as indicated in Table~\ref{tab:chain_summary}.
The Gaussian response priors have the same hard bounds.

\smallskip
Source-redshift, shear-calibration, and intrinsic-alignment
parameters are included in analyses containing shape
measurements. These analyses also vary baryonic feedback
with the prior shown above; BigMag-only analyses fix
$\log_{10}(T_{\rm AGN}/{\rm K})=7.8$.
Independent point-mass amplitudes are analytically
marginalized for each retained position--shape pair
and are not included among the sampled parameters.
\end{minipage}
\end{table}

Unless otherwise stated, we model the position-sample clustering using linear galaxy bias and restrict clustering measurements to transverse scales larger than $4\,h^{-1}{\rm Mpc}$. This is an aggressive choice, but it is applied consistently to the clustering pieces of the BigMag, traditional $2\times2$pt, and traditional $3\times2$pt analyses, so the relative comparisons are meaningful. As a robustness test, we repeated the BigMag clustering analysis with a nonlinear galaxy-bias model including an additional second-order bias parameter, and as discussed in \S\ref{sec:results_bigmag}, found nearly the same $S_8$ constraint. For the lensing observables we use $4\,h^{-1}{\rm Mpc}$ cuts for traditional galaxy--galaxy lensing, and $2\,h^{-1}{\rm Mpc}$ cuts for IGGL and cosmic shear, since these measurements are not directly sensitive to galaxy-bias modeling.

\subsection{Clustering-only constraints: breaking the galaxy-bias degeneracy}
\label{sec:results_bigmag}

We begin with the clustering-only data vector, consisting of the autocorrelations of the three low-redshift lens samples and their cross-correlations with the high-redshift samples. As discussed in Sec.~\ref{sec:model}, the amplitudes of these two classes of measurements can be written schematically, around the fiducial cosmology, as
\begin{align}
w^{L_iL_i}(\theta)
&\propto
b_i^2\,\sigma_8^2\,
\mathcal{K}^{\delta\delta}_i
\left(\theta;\Omega_{\rm m},\ldots\right),
\\
w^{L_iH_j}(\theta)
&\propto
b_i\,C^{\rm mag}_j\,\sigma_8^2\,
\mathcal{K}^{\delta\kappa}_{ij}
\left(\theta;\Omega_{\rm m},\ldots\right).
\end{align}
Here $\mathcal{K}^{\delta\delta}$ and $\mathcal{K}^{\delta\kappa}$ represent the corresponding projection kernels and their residual dependence on cosmology after factoring out the dominant $\sigma_8^2$ scaling. In particular, they contain the dependence on $\Omega_{\rm m}$ through the shape and evolution of the matter power spectrum, the distance--redshift relation, and, for $\mathcal{K}^{\delta\kappa}$, the lensing-efficiency kernel. The two observables therefore have different dependences on both the nuisance parameters and cosmology.

The low-redshift autocorrelations alone constrain approximately the combination $b_i\sigma_8$, modulated by their dependence on $\Omega_{\rm m}$ and the remaining cosmological parameters. This produces the familiar degeneracy between galaxy bias and the amplitude of matter fluctuations. The magnification-dominated cross-correlations provide a second set of measurements, with a different dependence on $b_i$ and on the cosmological kernels. If $C^{\rm mag}_j$ is calibrated, jointly fitting the autocorrelations and cross-correlations can therefore separate the foreground galaxy biases from the cosmological lensing amplitude.

We emphasize that the analysis does not construct or fit a ratio of these observables. The schematic scalings above only illustrate why the two sets of measurements contain complementary parameter dependences. In the inference, all autocorrelations and cross-correlations are fitted jointly, with their full covariance and their complete scale- and redshift-dependent theoretical predictions. The result is consequently a constraint on a combination of $\sigma_8$ and $\Omega_{\rm m}$, rather than on $\sigma_8$ alone.

The BigMag and traditional $2\times2$pt posteriors follow
somewhat different degeneracy directions in the
$\Omega_{\rm m}$--$\sigma_8$ plane from that commonly found
in cosmic-shear analyses. To summarize their constraints, we define
\begin{equation}
S'_8
\equiv
\sigma_8
\left(\frac{\Omega_{\rm m}}{0.3}\right)^{0.7},
\label{eq:s8prime}
\end{equation}
where we adopt $\eta=0.7$ as a representative exponent for BigMag and
its comparison with traditional $2\times2$pt in Sections~\ref{sec:results_bigmag}--\ref{sec:results_bigmag_calibration}.
This is a common reference choice, rather than an exponent optimized
for each posterior: the value that minimizes the marginalized
uncertainty depends on the observables and magnification priors.
Keeping the definition fixed ensures that all quoted improvements
within these comparisons refer to the same parameter. The preferred
exponent for each chain is reported separately in
Table~\ref{tab:chain_summary}.

The prime distinguishes this quantity from the conventional
$S_8=\sigma_8(\Omega_{\rm m}/0.3)^{1/2}$, which we use when introducing
IGGL and comparing with traditional $3\times2$pt in
Sections~\ref{sec:results_bigmag_iggl}--\ref{sec:results_bigmag_3x2pt}. For the fiducial cosmology used in the simulations,
$\Omega_{\rm m}=0.3$ and $\sigma_8=S'_8=S_8=0.82355$.

\begin{figure*}
    \centering
    \includegraphics[width=0.95\linewidth]{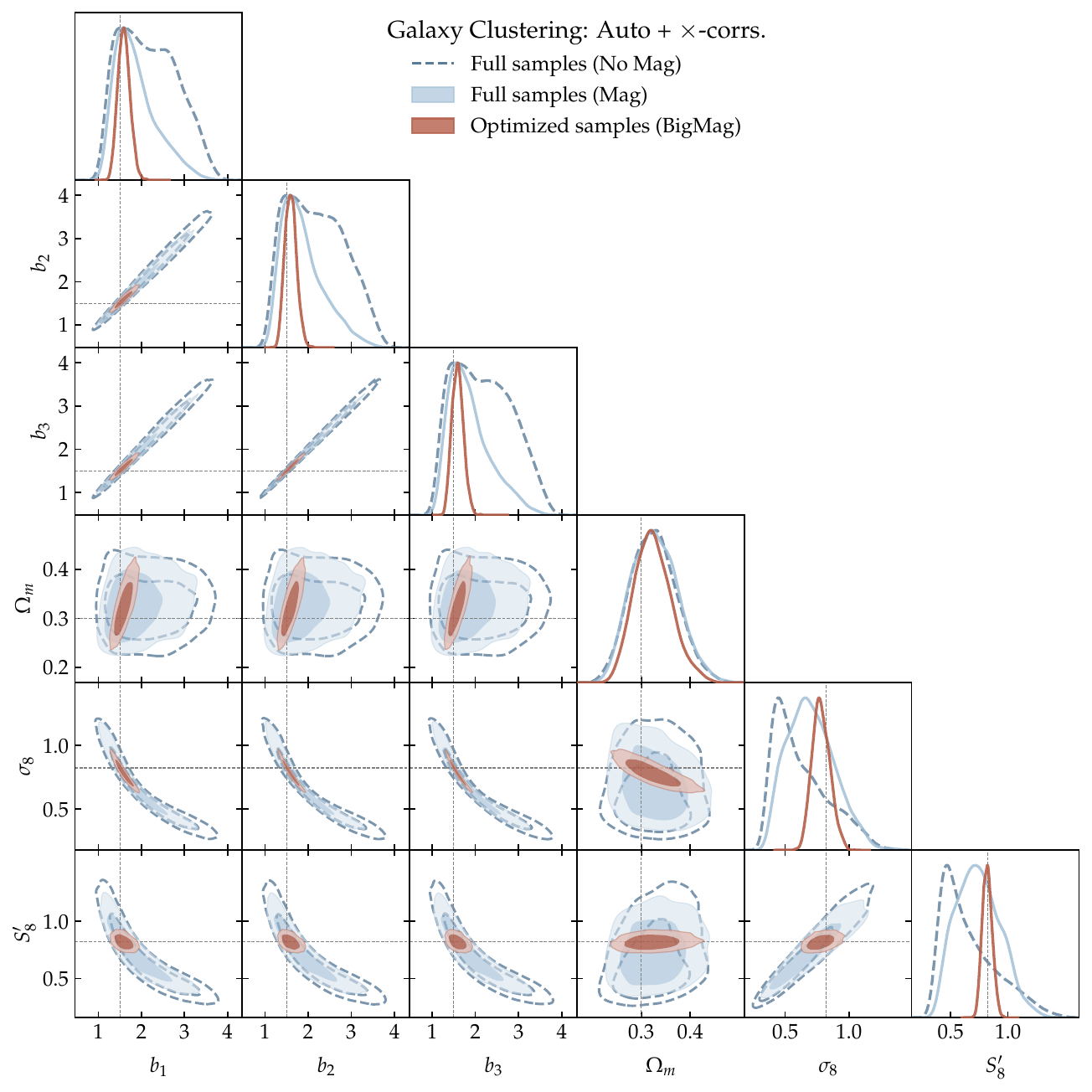}
    \caption{
    Marginalized posterior constraints from the clustering-only data vector, consisting of the low-redshift lens autocorrelations and their cross-correlations with the high-redshift samples. Contours show the $68\%$ and $95\%$ credible regions for the three low-redshift galaxy biases, $\Omega_{\rm m}$, $\sigma_8$, and the derived parameter $S'_8=\sigma_8(\Omega_{\rm m}/0.3)^{0.7}$. Dashed contours show the case in which the unsplit high-redshift samples have no magnification response; light-blue contours include magnification for the same unsplit samples; and red contours use the positive- and negative-response samples defining the optimized BigMag data vector. Without magnification, the $b_i$--$\sigma_8$ degeneracy remains unbroken. Magnification in the unsplit samples provides only limited additional information because of cancellation between galaxies with opposite responses. The optimized split removes this cancellation, allowing the magnification-dominated cross-correlations to break the galaxy-bias degeneracy and constrain a combination of $\Omega_{\rm m}$ and $\sigma_8$. Dashed grey lines indicate the input parameter values.
    }
    \label{fig:bigmag_clustering_constraints}
\end{figure*}

Figure~\ref{fig:bigmag_clustering_constraints} demonstrates how the optimized magnification split makes this cosmological information accessible. When magnification is removed from the high-redshift samples, their cross-correlations with the low-redshift lenses vanish, up to the negligible redshift overlap between the samples. The posterior then retains a nearly perfect $b_i$--$\sigma_8$ degeneracy. Including magnification for the unsplit high-redshift parent samples adds some information, but galaxies with positive and negative magnification responses partially cancel within each parent sample. The remaining cross-correlation signal provides only limited
leverage on the galaxy-bias degeneracy, leaving a broad
constraint on $S'_8$.

The result changes qualitatively for the optimized BigMag samples. Splitting each high-redshift sample into positive- and negative-response populations prevents this cancellation and produces high signal-to-noise cross-correlations with opposite signs. In combination with the low-redshift autocorrelations, these measurements tightly constrain the three foreground galaxy biases and substantially reduce their degeneracy with the cosmological amplitude. Thus, it is not merely the presence of magnification that enables clustering-only cosmological inference: selecting samples according to their magnification response is essential. This is one of the central results of this work.

With the fiducial magnification-coefficient priors,
$\sigma\!\left(C^{\rm mag}_i\right)=0.2$
(see also Table~\ref{tab:analysis_parameters}), the optimized
clustering data vector yields
\begin{equation}
S'_8 = 0.818^{+0.045}_{-0.042}.
\end{equation}
This constraint is consistent with the input value
$S'_8=0.82355$.

As a robustness test, we repeated the BigMag clustering analysis
using a nonlinear galaxy-bias model with an additional second-order
bias parameter for each galaxy sample, retaining the same scale
cuts and magnification priors. Using the same fixed definition of
$S'_8$ with $\eta=0.7$, we obtain
\begin{equation}
S'_8 = 0.800^{+0.046}_{-0.042}.
\end{equation}
The half-width of the central $68\%$ credible interval increases
from $0.0434$ for the linear-bias model to $0.0441$ for the
nonlinear model, a degradation of approximately $2\%$.
Although the posterior median shifts to a lower value, the input
cosmology remains within the central $68\%$ credible interval.
The precision of the BigMag amplitude constraint is therefore
largely preserved when allowing for second-order galaxy bias
on the scales considered here.

We also tested the inclusion of cross-correlations between
distinct low-redshift bins, $L_1\times L_2$, $L_1\times L_3$,
and $L_2\times L_3$, using the same nonlinear-bias model,
scale cuts, and magnification priors. These measurements
provide little additional information on $S'_8$: the
central $68\%$ credible-interval half-width decreases from
$0.0441$ to $0.0429$, an improvement of approximately $3\%$,
while the median remains $S'_8\simeq0.800$.
We retain these cross-correlations in the joint analyses
introduced below, but this test indicates that their
inclusion has only a modest effect on the standalone
BigMag amplitude precision.

\subsection{Magnification calibration and comparison with traditional
$2\times2$pt}
\label{sec:results_bigmag_calibration}

We next examine how uncertainty in the magnification coefficients affects the BigMag constraints and compare its limiting performance with that of a traditional $2\times2$pt analysis. The two analyses have analogous structures: both combine foreground galaxy clustering with a lensing-induced cross-correlation that helps separate galaxy bias from the amplitude of matter fluctuations. Traditional $2\times2$pt uses the cross-correlation between foreground positions and background galaxy shapes, whereas BigMag uses magnification-induced correlations between foreground and background galaxy positions. 

In both cases, the amplitude of the lensing observable depends on a response calibration. Galaxy--galaxy lensing depends on the multiplicative shear calibration factor $1+m$, while the BigMag cross-correlations depend linearly on $C^{\rm mag}$. There is, however, an important physical difference between these parameters. The shear response remains close to unity and does not change the sign of the galaxy--galaxy lensing signal, whereas $C^{\rm mag}$ can be positive, negative, or close to zero. This sign dependence enables the sample optimization considered here: separating galaxies with positive and negative magnification responses prevents signal cancellation and produces cross-correlations with opposite signs.

A further distinction concerns sensitivity to small scales. Tangential shear is non-local, so uncertainties below the minimum modeled scale propagate into the retained galaxy--galaxy lensing measurements. We mitigate this contribution by analytically marginalizing over an independent point-mass template for each retained position--source tomographic-bin pair \citep{MacCrann_2019,Prat:2022:GGL_Y3,Prat_2023}. By contrast, BigMag traces convergence through magnification and has no analogous enclosed-mass contribution, although it remains subject to the usual nonlinear and baryonic modeling uncertainties on the retained scales.

\begin{figure}[t]
    \centering
    \includegraphics[width=0.99\linewidth]
    {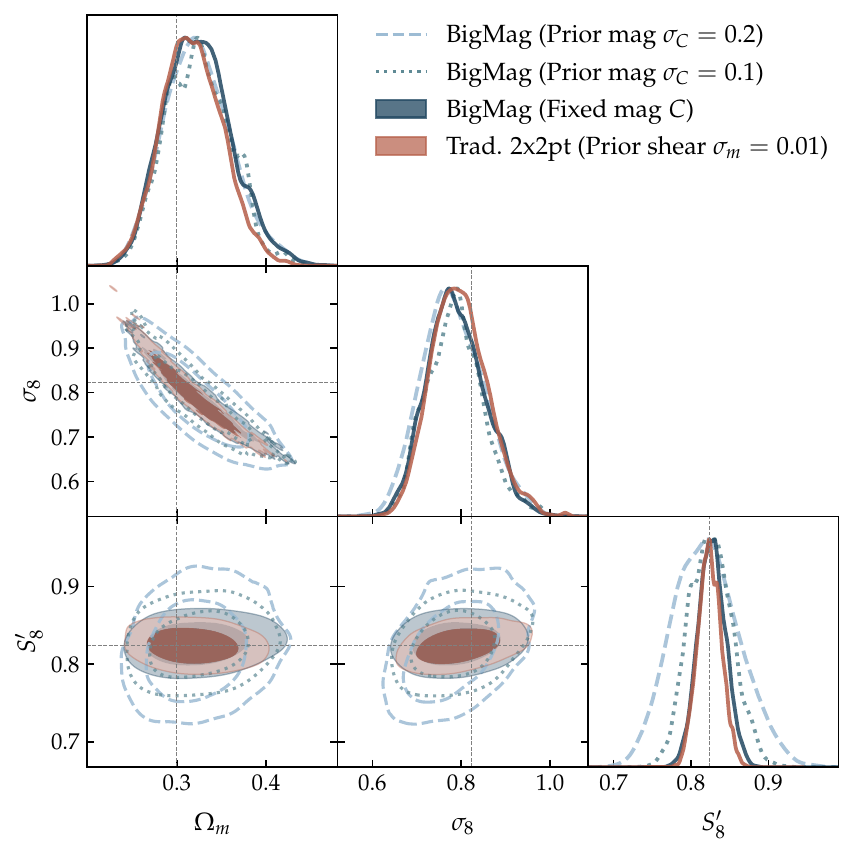}
    \caption{
    Dependence of the BigMag cosmological constraints on the calibration of the magnification coefficients. Contours show the $68\%$ and $95\%$ credible regions in $\Omega_{\rm m}$, $\sigma_8$, and $S'_8=\sigma_8(\Omega_{\rm m}/0.3)^{0.7}$. The dashed and dotted contours show BigMag with independent Gaussian magnification priors of $\sigma(C^{\rm mag})=0.2$ and $0.1$, respectively; the latter is obtained by importance-reweighting the fiducial chain. The dark-blue contours show the limiting case in which the magnification coefficients are fixed to their input values. For comparison, the red contours show a traditional $2\times2$pt analysis with Gaussian multiplicative shear-calibration priors of $\sigma_m=0.01$. Traditional $2\times2$pt combines lens clustering with shear-based galaxy--galaxy lensing, whereas BigMag combines lens clustering with magnification-induced position cross-correlations. All cases use the same fixed exponent $\eta=0.7$, allowing
a direct comparison of their marginalized $S'_8$ constraints.
Tighter magnification calibration substantially improves
the precision: with the magnification coefficients fixed,
the central $68\%$ interval is approximately $17\%$ wider
than for the reference traditional $2\times2$pt analysis,
which retains finite shear-calibration uncertainty.
Dashed grey lines indicate the input cosmology.
    }
    \label{fig:bigmag_cmag_priors}
\end{figure}

Having established these similarities and differences, we now quantify the role of magnification calibration. Our fiducial analysis adopts independent Gaussian priors with $\sigma(C^{\rm mag})=0.2$, motivated by the calibration uncertainties obtained from the DES Y6+\textsc{Balrog} measurements and sample selections described in Sec.~\ref{sec:data_sims}.
Because $C^{\rm mag}$ controls the amplitude of each magnification cross-correlation, its uncertainty propagates directly into the inferred cosmological lensing amplitude. We therefore use Fig.~\ref{fig:bigmag_cmag_priors} to examine how the constraints improve as the magnification calibration is made more precise.

Tightening the magnification-coefficient priors from
$\sigma(C^{\rm mag})=0.2$ to $0.1$ reduces the half-width
of the central $68\%$ credible interval on $S'_8$ from
$0.0434$ to $0.0278$, an improvement of approximately
$36\%$, yielding
\begin{equation}
S'_8 = 0.825^{+0.028}_{-0.028}.
\end{equation}
This result is obtained by importance-reweighting the fiducial
BigMag chain. In the limiting case in which all magnification
coefficients are fixed to their input values, the constraint
becomes
\begin{equation}
S'_8 = 0.827^{+0.018}_{-0.018},
\end{equation}
with a credible-interval half-width of $0.0179$, a factor
of approximately $2.4$ smaller than in the fiducial analysis.
For comparison, the traditional $2\times2$pt analysis gives
\begin{equation}
S'_8 = 0.824^{+0.015}_{-0.015},
\end{equation}
with a half-width of $0.0153$. Thus, the fixed-response
BigMag uncertainty is approximately $17\%$ larger than
that of the traditional $2\times2$pt analysis.
All of these comparisons use the same fixed definition of
$S'_8$ with $\eta=0.7$, rather than separately optimized
amplitude combinations.

The reference $2\times2$pt analysis adopts independent
Gaussian shear-calibration priors with $\sigma_m=0.01$.
The numerical widths of the $m$ and $C^{\rm mag}$ priors
should not be compared directly because these parameters
have different normalizations. The parameter $m$ describes
a fractional multiplicative correction to the shear signal,
whereas $C^{\rm mag}$ is the full magnification response.
For the high-redshift split samples used here,
$|C^{\rm mag}|$ ranges from $1.1$ to $2.25$, so the
fiducial prior width $\sigma(C^{\rm mag})=0.2$ corresponds
to fractional response uncertainties of approximately
$9$--$18\%$. This interpretation does not apply to the
low-redshift samples, whose fiducial magnification
responses are zero. Moreover, the fixed-$C^{\rm mag}$
analysis assumes perfect magnification calibration,
whereas the traditional $2\times2$pt analysis retains
finite shear-calibration uncertainty. The comparison
therefore demonstrates the limiting constraining power
of BigMag under the adopted analysis choices, rather
than performance under matched calibration uncertainties.

Like the shear response, magnification coefficients can be calibrated empirically. Here, this calibration is provided by synthetic-source injections and the SOM-based methodology described in Sec.~\ref{sec:data_sims}. Increasing the number of injections and the calibration area, while improving the characterization of the relevant selection responses, could enable tighter magnification priors. We discuss this further in \S\ref{sec:discussion_response_calibration}. These results demonstrate that, with sufficiently precise magnification calibration, magnification-selected clustering can approach the constraining power of traditional shear-based $2\times2$pt analyses.

\subsection{Adding IGGL: self-calibration of magnification and cosmology}
\label{sec:results_bigmag_iggl}

\begin{figure*}
    \centering
    \includegraphics[width=0.99\linewidth]
    {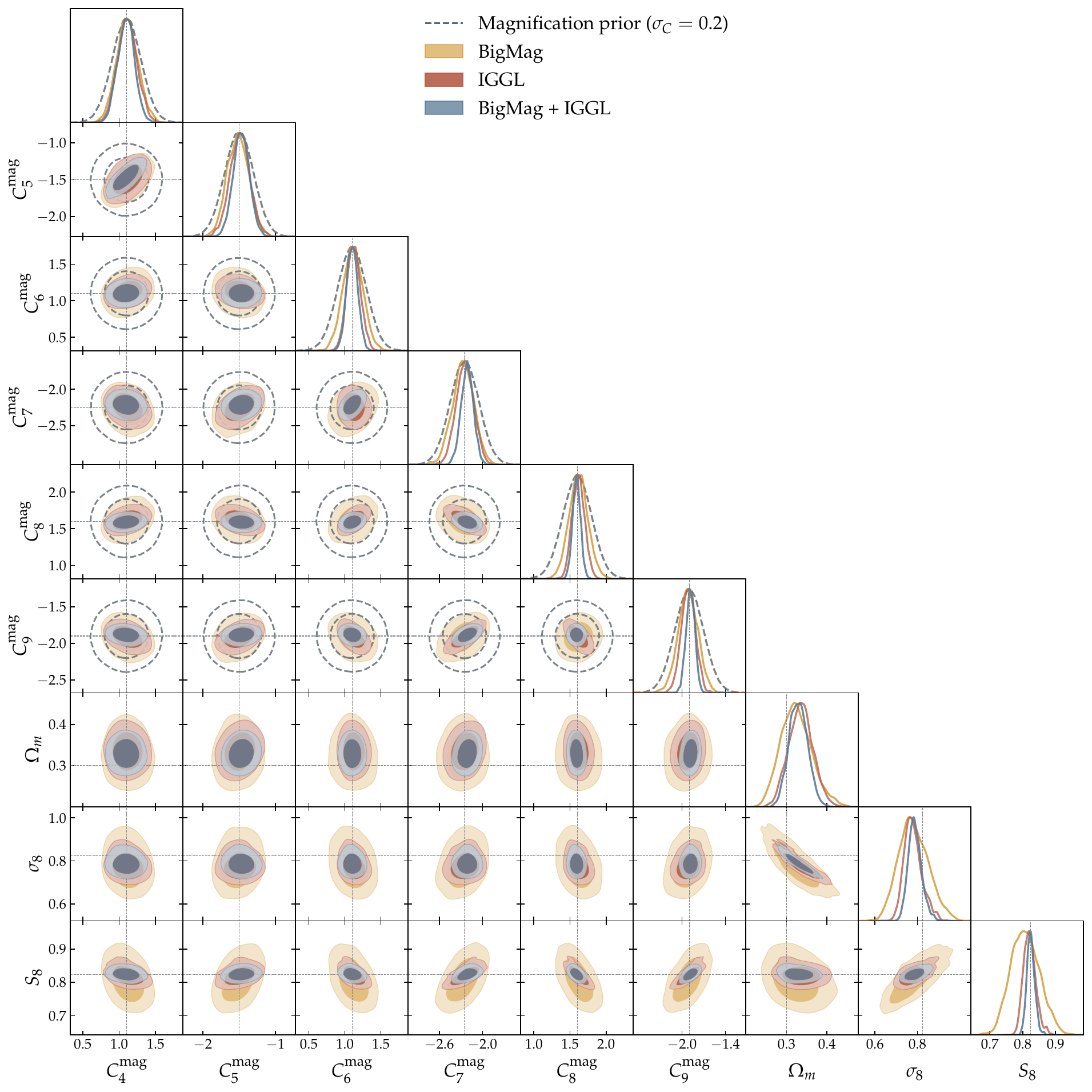}
    \caption{
Self-calibration of the high-redshift magnification
coefficients and resulting cosmological constraints
from BigMag and IGGL. Off-diagonal panels show the
$68\%$ and $95\%$ credible regions; diagonal panels
show the marginalized one-dimensional distributions.
The coefficients $C^{\rm mag}_4$--$C^{\rm mag}_9$
correspond, respectively, to H1$^{+}$, H1$^{-}$,
H2$^{+}$, H2$^{-}$, H3$^{+}$, and H3$^{-}$.
All three analyses use the conventional amplitude
$S_8=\sigma_8(\Omega_{\rm m}/0.3)^{1/2}$.
Dashed grey curves and contours show the independent
Gaussian magnification priors with
$\sigma(C^{\rm mag})=0.2$, while gold, red, and blue
show the BigMag, IGGL, and joint posteriors, respectively.
Both probes partially self-calibrate the magnification
coefficients relative to the common priors. Their
combination further tightens the response constraints
and reduces the uncertainty on $S_8$.
IGGL retains position--shape pairs satisfying
$\bar z_{\rm position}>\bar z_{\rm shape}$ and includes
independent point-mass marginalization for each retained
pair. Dotted grey lines indicate the input parameter values.
}
    \label{fig:bigmag_iggl_cmag}
\end{figure*}

We next investigate the information gained by adding inverse galaxy--galaxy lensing to the BigMag clustering data vector. In the IGGL configuration, \S\ref{sec:iggl_theory}, the galaxy-position sample lies behind the galaxy-shape sample, $z_{\rm position}>z_{\rm shape}$, suppressing the conventional correlation between the intrinsic galaxy density and the gravitational shear. The leading signal of this position--shear correlation instead arises from the magnification of the high-redshift position sample. Schematically, we write
\begin{equation}
\gamma_{\rm t}^{H_jS_s}(\theta)
\simeq
C^{\rm mag}_j\,\sigma_8^2\,
\mathcal{K}^{\kappa\gamma}_{js}
\left(\theta;\Omega_{\rm m},\ldots\right)
+
\gamma_{{\rm t},\,js}^{\rm IA}(\theta),
\label{eq:iggl_schematic}
\end{equation}
where $\mathcal{K}^{\kappa\gamma}_{js}$ contains the redshift-dependent lensing kernels and their dependence on geometry, growth, and the matter power spectrum. The second term represents the contribution involving intrinsic source ellipticities, $\left< (\delta_g^\text{int} +  C_\mathrm{mag} \kappa) \, e^{\rm int} \right>$.

We construct the IGGL data vector using only position--source combinations for which the mean redshift of the high-redshift position sample exceeds that of the source sample. Consequently, we retain H1$^{\pm}\times$S1--S3 and H2$^{\pm}$, H3$^{\pm}\times$S1--S4. We exclude H1$^{\pm}\times$S4 because S4 lies behind H1 in redshift, making this combination predominantly traditional galaxy--galaxy lensing rather than IGGL. We also exclude all low-redshift L$\times$S measurements from the IGGL-only data vector. For these measurements, we adopt a minimum transverse scale of $2\,h^{-1}{\rm Mpc}$ and, very conservatively, also marginalize over an independent point-mass contribution for each retained position--source tomographic-bin pair. This protects against the subdominant traditional galaxy--galaxy lensing contribution generated by residual redshift overlap, whereas the dominant magnification-induced signal is not subject to the same lens-associated enclosed-mass contribution.

BigMag and IGGL constrain the same high-redshift magnification coefficients through different physical correlations. To leading order, BigMag probes $b_iC^{\rm mag}_j\langle\delta_{L_i}\kappa_{H_j}\rangle$ through several low-redshift lens bins, whose biases are constrained by their clustering autocorrelations. IGGL instead probes $C^{\rm mag}_j\langle\kappa_{H_j}\gamma_{S_s}\rangle$, together with the corresponding magnification--intrinsic-alignment contribution, through several foreground source bins. IGGL is therefore independent of the low-redshift galaxy biases but has additional dependences on shear calibration and intrinsic alignments. The multiple redshift combinations and different cosmological dependences of the two probes allow each to partially self-calibrate $C^{\rm mag}_j$ and, when analyzed jointly, further break degeneracies between the magnification coefficients and the cosmological lensing amplitude. The joint likelihood includes the covariance between the BigMag and IGGL measurements. The joint clustering block also retains the cross-correlations
between distinct low-redshift bins. As shown in
Sec.~\ref{sec:results_bigmag}, these provide only a modest
improvement in the standalone nonlinear-bias BigMag test.

Figure~\ref{fig:bigmag_iggl_cmag} illustrates this complementarity. The dashed grey curves and contours show the independent
Gaussian $\sigma(C^{\rm mag})=0.2$ priors adopted in all
three analyses. BigMag alone reduces the half-widths of the
central $68\%$ credible intervals on the six high-redshift
coefficients to approximately $0.13$--$0.15$, while IGGL
alone gives $0.09$--$0.14$. Their joint analysis tightens
these intervals further, to approximately $0.055$--$0.12$,
with the strongest constraints obtained for the
highest-redshift samples. All six coefficients remain
consistent with their input values. The narrowing relative
to the common priors demonstrates partial self-calibration
within the adopted cosmological and nuisance-parameter model.

The improvement in the magnification calibration propagates directly into the cosmological constraints. With the inclusion of IGGL, we adopt the conventional amplitude
parameter $S_8=\sigma_8(\Omega_{\rm m}/0.3)^{1/2}$, which more closely
follows the degeneracy direction of the shear-based measurements.
We retain this common definition throughout the IGGL and traditional
$3\times2$pt comparisons in Sections~\ref{sec:results_bigmag_iggl}--\ref{sec:results_bigmag_3x2pt}, without optimizing
the exponent separately for each analysis.

Using the same fiducial $\sigma(C^{\rm mag})=0.2$ priors, we find
\begin{align}
S_8 &= 0.807^{+0.046}_{-0.042},
&& \text{BigMag},\\
S_8 &= 0.820^{+0.019}_{-0.018},
&& \text{IGGL},\\
S_8 &= 0.824^{+0.013}_{-0.011},
&& \text{BigMag+IGGL}.
\end{align}
The joint constraint is consistent with the input value
$S_8=0.82355$. The central $68\%$ credible-interval
half-widths are $0.0442$, $0.0185$, and $0.0117$ for
BigMag, IGGL, and their joint analysis, respectively.
The joint analysis therefore improves the marginalized
$S_8$ precision by factors of approximately $3.8$ and
$1.6$ relative to BigMag and IGGL alone.

The improvement reflects both the additional lensing
information supplied by IGGL and the complementary
constraints on the magnification coefficients shared
by the two probes. Their different parameter dependences
reduce degeneracies between $C^{\rm mag}$ and the
cosmological lensing amplitude, yielding tighter joint
constraints on both. These results include simultaneous
variation of shear calibration, source-redshift
uncertainties, baryonic feedback, and the
intrinsic-alignment amplitude $A_1$ in the IGGL and
joint analyses, with priors summarized in
Table~\ref{tab:analysis_parameters}.
The different systematic dependencies of the two
observables also motivate their use as an internal
consistency test in applications to observational data.

Finally, Table~\ref{tab:chain_summary} summarizes the analysis
configurations and reports the exponent $\eta_*$
that minimizes the central $68\%$ interval width of
$S_{8,\eta_*}$ for each chain. The preferred exponent
ranges from approximately $0.61$ to $0.67$ for the
BigMag calibration cases, compared with $0.50$ for
IGGL and $0.53$ for the fiducial BigMag+IGGL analysis, showcasing the changing posterior geometry. 
The precision comparisons above instead use the common
parameter $S_8$. IGGL and traditional $2\times2$pt
have similar raw signal-to-noise, but this quantity
alone does not determine cosmological precision:
it does not account for point-mass marginalization,
calibration uncertainties, or degeneracies with
other nuisance parameters.

\subsection{Comparison and combination with traditional $3\times2$pt}
\label{sec:results_bigmag_3x2pt}

Having established the complementarity of BigMag and IGGL, we now compare their joint cosmological constraints with those from traditional $3\times2$pt and investigate the information gained by combining these observables. Traditional $3\times2$pt combines foreground galaxy clustering, galaxy--galaxy lensing (the 2$\times$2pt case from Sec.~\ref{sec:results_bigmag_calibration}), and cosmic shear, and it represents the main weak lensing analysis for imaging surveys like DES \citep{DESY6_3x2pt}. The combination of $3\times2$pt with BigMag and IGGL contains the union of the traditional and magnification-based data vectors, with shared measurements counted once and their cross-covariance included.

We retain the cosmological and nuisance-parameter treatment described above, including marginalization over galaxy bias, redshift uncertainties, shear calibration, intrinsic alignments, baryonic feedback, and point-mass contributions. We consider independent Gaussian magnification-response priors with widths $\sigma_C\equiv\sigma(C^{\rm mag})=0.2$ and $0.02$, together with a broad uniform-prior case (flat $C$). These choices apply to all nine position-sample responses in the magnification analyses. The traditional $3\times2$pt reference retains the fiducial
$\sigma_C=0.2$ priors for its three foreground bins.
Throughout this subsection, we retain the conventional
$S_8$ definition adopted in
Sec.~\ref{sec:results_bigmag_iggl} and denote the half-width
of its central $68\%$ credible interval by
$\sigma_{68}(S_8)$. All precision comparisons use this
same parameter and uncertainty statistic.

\begin{figure}
    \centering
    \includegraphics[width=0.99\linewidth]
    {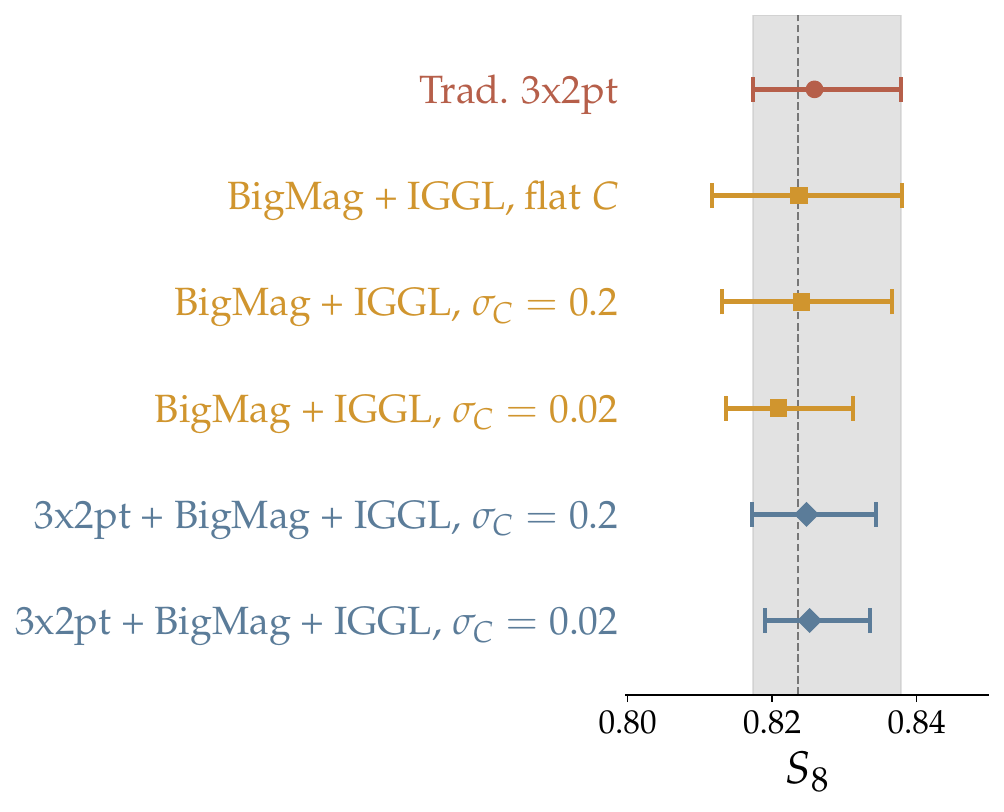}
    \caption{
Marginalized $S_8$ constraints from traditional
$3\times2$pt, BigMag+IGGL, and their joint analysis.
Markers indicate posterior medians, and horizontal
bars show the central $68\%$ credible intervals.
The shaded band reproduces the traditional
$3\times2$pt interval; the dashed vertical line
marks the input value $S_8=0.82355$.
The flat-prior case adopts
$C^{\rm mag}\in[-12,8]$.
The indicated priors apply to all nine position-sample
responses in the magnification analyses, while the
traditional reference retains $\sigma_C=0.2$ for
its three foreground bins.
Shear-calibration and redshift priors are retained
in all cases. The full combination includes shared
measurements once and accounts for their cross-covariance.
}
    \label{fig:s8_constraint_summary}
\end{figure}

Figure~\ref{fig:s8_constraint_summary} summarizes the resulting constraints. The traditional $3\times2$pt analysis gives
\begin{equation}
S_8 = 0.8259^{+0.0120}_{-0.0085},
\end{equation}
corresponding to $\sigma_{68}(S_8)\simeq0.0102$, or approximately $1.2\%$ fractional precision. For context, the DES Y6 $3\times2$pt analysis reports
approximately $1.5\%$ precision on $S_8$
\citep{DESY6_3x2pt}, although its exact data selection and
modeling assumptions differ from those adopted here. With the fiducial $\sigma_C=0.2$ priors, BigMag+IGGL gives
\begin{equation}
S_8 = 0.8241^{+0.0125}_{-0.0110}.
\end{equation}
Its uncertainty is only approximately $15\%$ larger than that of traditional $3\times2$pt. Thus, the optimized magnification observables already provide competitive constraints on the lensing amplitude with the conservative response calibration adopted in the fiducial analysis.

This competitiveness persists when informative external priors on the magnification responses are removed. Assigning broad uniform priors, $C^{\rm mag}\in[-12,8]$, yields
\begin{equation} \label{eq:flat_cmag_s8_constraint}
S_8 = 0.8238^{+0.0142}_{-0.0120},
\qquad \text{BigMag+IGGL, flat } C^{\rm mag}.
\end{equation}
The corresponding fractional uncertainty is approximately $1.6\%$, only $28\%$ larger than the traditional $3\times2$pt reference. Moreover, imposing the fiducial $\sigma_C=0.2$ priors improves this uncertainty by only approximately $11\%$. Much of the cosmological information therefore survives without informative external magnification-response priors.
The six high-redshift responses are constrained well within
the broad prior boundaries, with central $68\%$
credible-interval half-widths of approximately
$0.06$--$0.18$. This demonstrates substantial internal
calibration by the joint observables, conditional on
the other calibration priors and modeling assumptions
retained in the analysis.

The flat-prior configuration is an extremely conservative assumption about magnification-response calibration. The construction of the samples already uses synthetic-source injections to characterize their responses, and this information will be available to inform the cosmological likelihood. The flat case deliberately withholds that external information while retaining the other calibration priors and modeling assumptions. It therefore tests the dependence of the joint cosmological constraint on external knowledge of $C^{\rm mag}$. More precise response calibration can further improve the constraints, as motivated by the injection-statistics study in Appendix~A. Those projections concern statistical precision; achieving a given total calibration uncertainty also requires control of the systematic effects discussed in the response-calibration section.

For the more precise calibration scenario, $\sigma_C=0.02$, BigMag+IGGL gives
\begin{equation}
S_8 = 0.8209^{+0.0103}_{-0.0072}.
\end{equation}
This reduces $\sigma_{68}(S_8)$ by approximately $25\%$ relative to the fiducial $\sigma_C=0.2$ case and produces an uncertainty approximately $14\%$ smaller than the traditional $3\times2$pt reference. For the adopted high-redshift responses, $\sigma_C=0.02$ corresponds to fractional response uncertainties of approximately $0.9$--$1.8\%$. This is comparable in order of magnitude to the assumed shear-calibration precision, although the response parameters and their prior normalizations differ.

\begin{figure}
    \centering
    \includegraphics[width=0.99\linewidth]
    {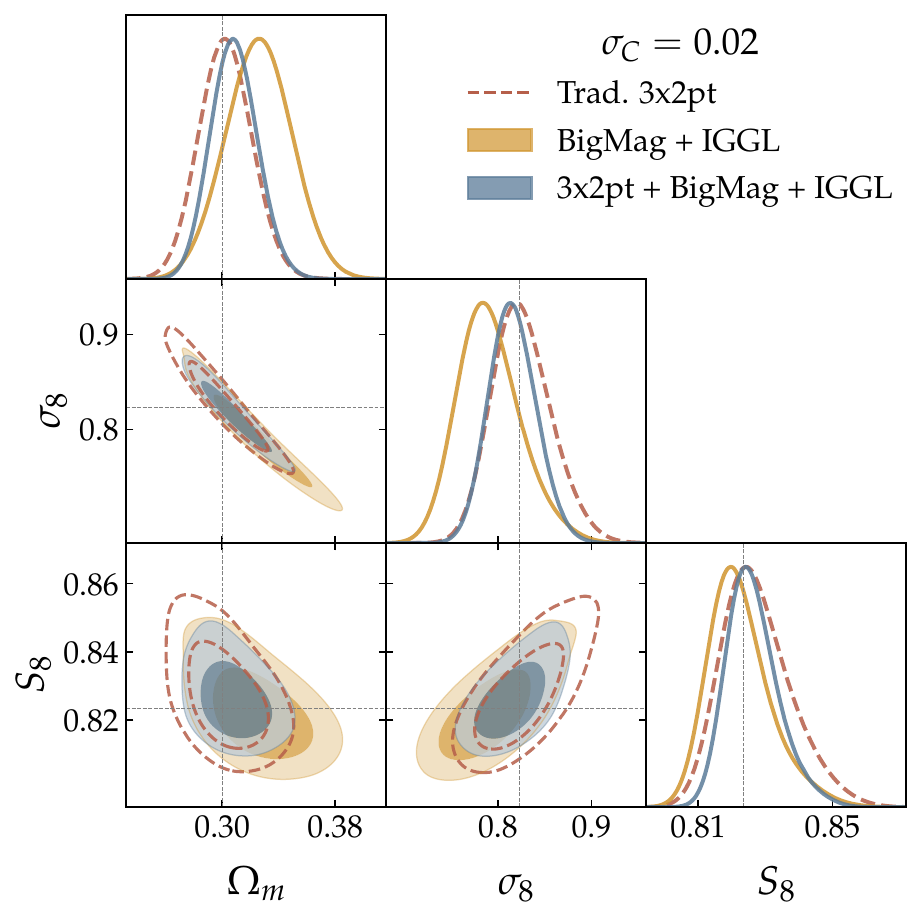}
    \caption{
Cosmological constraints for the tight
magnification-calibration scenario.
Dashed red contours show traditional $3\times2$pt,
gold contours show BigMag+IGGL, and blue contours
show their joint analysis. Contours enclose
$68\%$ and $95\%$ credible regions; diagonal
panels show marginalized one-dimensional posteriors.
The BigMag+IGGL and full combined analyses adopt
independent Gaussian response priors with
$\sigma_C=0.02$ for all nine position samples.
Grey dashed lines indicate the input cosmology.
BigMag+IGGL gives a narrower marginalized $S_8$
interval than traditional $3\times2$pt, while
the traditional analysis constrains
$\Omega_{\rm m}$ and $\sigma_8$ individually
more tightly. Relative to traditional $3\times2$pt, the full
combination reduces the central $68\%$ credible-interval
widths of $\Omega_{\rm m}$, $\sigma_8$, and $S_8$
by approximately $12\%$, $20\%$, and $29\%$, respectively.
}
    \label{fig:bigmag_iggl_3x2pt_tight}
\end{figure}

Figure~\ref{fig:bigmag_iggl_3x2pt_tight} shows the cosmological posteriors for this tight-prior scenario. The different observables constrain overlapping but distinct combinations of cosmology and nuisance parameters. Although BigMag+IGGL provides a narrower marginalized $S_8$ interval, traditional $3\times2$pt retains tighter individual constraints on $\Omega_{\rm m}$ and $\sigma_8$. The full combination improves all three projections, illustrating the additional information obtained by combining the magnification observables with traditional galaxy--galaxy lensing and cosmic shear.

Quantitatively, the full combination yields
\begin{align}
S_8 &= 0.8248^{+0.0096}_{-0.0076},
&& \sigma_C=0.2,\\
S_8 &= 0.8252^{+0.0084}_{-0.0062},
&& \sigma_C=0.02.
\end{align}
These correspond to reductions in $\sigma_{68}(S_8)$ of approximately $16\%$ and $29\%$, respectively, relative to traditional $3\times2$pt. The tight-prior combination reaches approximately $0.9\%$ fractional precision on $S_8$ and also reduces the marginalized interval widths of $\Omega_{\rm m}$ and $\sigma_8$ by approximately $12\%$ and $20\%$. All quoted $S_8$ intervals contain the input value. Because the tight-prior configuration also improves the calibration of the shared foreground responses, its gain relative to the traditional reference includes both the added observables and the changed calibration assumptions. The $\sigma_C=0.2$ comparison holds the response priors on the shared foreground bins fixed.

The improvement reflects the additional lensing information carried by the response-selected high-redshift populations and the different parameter dependences of the observables. BigMag and IGGL constrain their shared magnification responses through different correlations, while traditional galaxy--galaxy lensing and cosmic shear supply further information about the matter distribution and the nuisance parameters entering the joint inference. These measurements share galaxies and cosmological density modes, so their complementarity must be assessed through the joint likelihood rather than by treating their constraints as independent. Within the adopted response and redshift model, the results establish BigMag+IGGL as both a competitive cosmological probe under very conservative external response priors and a useful extension of traditional $3\times2$pt, with larger gains as the magnification calibration improves.

% All chains table
\begin{table*}
\centering
\caption{Summary of the principal chains in
Sections~\ref{sec:results_bigmag}--\ref{sec:results_bigmag_3x2pt}.
$N_{\rm data}$ is the number of retained data-vector elements after scale cuts,
and $(S/N)_{\rm raw}$ is computed from the selected simulated data vector
and its covariance before point-mass or other nuisance-parameter marginalization.
$N_{\rm par}$ counts sampled parameters.
The $S_{8,\eta_*}$ and $\eta_*$ columns give the posterior median and
central $68\%$ credible interval of $S_{8,\eta_*}\equiv\sigma_8(\Omega_{\rm m}/0.3)^{\eta_*}$,
with $\eta_*$ chosen separately for each chain to minimize the interval width.
These optimized combinations describe posterior geometry; precision comparisons
in the text use a common exponent within each comparison.
The final column reports the central $68\%$ credible-interval half-width
$\sigma_{68}(S_8)$ at the common exponent $\eta=0.5$ for every chain,
allowing a direct comparison of marginalized $S_8$ precision.}
\label{tab:chain_summary}
\small
\renewcommand{\arraystretch}{1.18}
\setlength{\tabcolsep}{2.5pt}
\begin{tabular}{llccclccc}
\hline
Chain & Data & $N_{\rm data}$ & $(S/N)_{\rm raw}$ & $N_{\rm par}$
& Priors/model & $S_{8,\eta_*}$ & $\eta_*$ & $\sigma_{68}(S_8)$ \\
\hline
Unsplit samples & $w$ & 140 & 51.055 & 24
& no magnification & -- & -- & $0.2414$ \\
Unsplit samples & $w$ & 140 & 51.064 & 24
& fixed $C^{\rm mag}$ & -- & -- & $0.1995$ \\
BigMag & $w$ & 308 & 91.2 & 33
& $\sigma_C=0.2$ & $0.813_{-0.041}^{+0.044}$ & 0.61 & $0.0442$ \\
BigMag & $w$ & 308 & 91.2 & 33
& $\sigma_C=0.1$ (IS) & $0.822_{-0.027}^{+0.027}$ & 0.65 & $0.0302$ \\
BigMag & $w$ & 308 & 91.2 & 24
& fixed $C^{\rm mag}$ & $0.824_{-0.017}^{+0.018}$ & 0.67 & $0.0245$ \\
BigMag & $w$ & 308 & 91.2 & 42
& NL bias, $\sigma_C=0.2$ & $0.790_{-0.041}^{+0.045}$ & 0.62 & $0.0446$ \\
BigMag & $w$ & 349 & 97.4 & 42
& NL bias + cross, $\sigma_C=0.2$ & $0.794_{-0.041}^{+0.044}$ & 0.65 & $0.0446$ \\
Traditional $2\times2$pt & $w+\gamma_{\rm t}$ & 204 & 138.3 & 25
& $\sigma_C=0.2$ & $0.825_{-0.015}^{+0.016}$ & 0.72 & $0.0243$ \\
\hline
IGGL & $\gamma_{\rm t}$ & 508 & 140.4 & 34
& $\sigma_C=0.2$ & $0.820_{-0.018}^{+0.019}$ & 0.50 & $0.0185$ \\
BigMag + IGGL & $w+\gamma_{\rm t}$ & 857 & 165.4 & 43
& flat $C^{\rm mag}$ & $0.826_{-0.012}^{+0.014}$ & 0.52 & $0.0131$ \\
BigMag + IGGL & $w+\gamma_{\rm t}$ & 857 & 165.4 & 43
& $\sigma_C=0.2$ & $0.827_{-0.011}^{+0.012}$ & 0.53 & $0.0117$ \\
BigMag + IGGL & $w+\gamma_{\rm t}$ & 857 & 165.4 & 43
& $\sigma_C=0.02$ & $0.825_{-0.007}^{+0.009}$ & 0.56 & $0.0087$ \\
Traditional $3\times2$pt & $w+\gamma_{\rm t}+\xi_\pm$ & 610 & 184.7 & 25
& $\sigma_C=0.2$ & $0.827_{-0.008}^{+0.010}$ & 0.59 & $0.0102$ \\
$3\times2$pt + BigMag + IGGL & $w+\gamma_{\rm t}+\xi_\pm$ & 1423 & 212.7 & 43
& $\sigma_C=0.2$ & $0.826_{-0.008}^{+0.009}$ & 0.55 & $0.0086$ \\
$3\times2$pt + BigMag + IGGL & $w+\gamma_{\rm t}+\xi_\pm$ & 1423 & 212.7 & 43
& $\sigma_C=0.02$ & $0.827_{-0.006}^{+0.008}$ & 0.55 & $0.0073$ \\
\hline
\end{tabular}
\vspace{0.4em}

\begin{minipage}{0.98\textwidth}
\footnotesize
$w\equiv w(\theta)$ denotes galaxy-position correlations.
``Unsplit samples'' refers to the parent high-redshift populations;
BigMag uses their positive- and negative-response subsamples.
All models use linear galaxy bias unless marked ``NL bias'', which adds
a second-order bias parameter for each of the nine position samples.
``Cross'' adds the correlations between distinct low-redshift bins,
increasing the clustering block from 308 to 349 elements.
This extended clustering block is also used in BigMag+IGGL and the full
combination; shared measurements are included only once.

The Gaussian response-prior widths $\sigma_C$ apply to all nine position
samples in BigMag and the joint analyses, the six high-redshift position
samples in IGGL alone, and the three foreground bins in traditional
$2\times2$pt and $3\times2$pt.
Flat response priors are uniform over $C^{\rm mag}\in[-12,8]$;
fixed responses are set to their input values.
IS denotes importance reweighting of the fiducial BigMag chain
(weighted ESS $\simeq801$).
All analyses containing $\gamma_{\rm t}$ include independent point-mass
marginalization per retained position--shape pair, vary $A_1$ and baryonic
feedback, and retain shear-calibration and redshift-offset priors of width
$0.01$. Analytically marginalized point-mass amplitudes are excluded
from $N_{\rm par}$.
The two unsplit-sample diagnostics do not provide a useful constrained
amplitude direction; their broad $S_8$ interval widths are sensitive to
the adopted parameter priors, and their raw signal-to-noise is dominated
by the low-redshift autocorrelations.
\end{minipage}
\end{table*}

\newpage

\section{Discussion}
\label{sec:discussion}

The forecasts presented in this work illustrate how
magnification-selected position samples can recover lensing
information lost when populations with opposite responses are
combined. In the configuration adopted here, the
high-redshift samples combine high number densities with
substantial magnification responses. Their cross-correlations
with foreground galaxies, together with foreground
autocorrelations, allow BigMag to break the foreground
galaxy-bias degeneracy, while the same samples provide a
complementary IGGL signal. Combining BigMag and IGGL
self-calibrates the magnification responses of the shared
high-redshift samples and yields competitive constraints on
the growth of structure. Under the adopted sample and
calibration assumptions, BigMag can approach the amplitude
precision of traditional $2\times2$pt, while BigMag+IGGL
provides marginalized $S_8$ constraints comparable to the
traditional $3\times2$pt reference, and tighter in the most
precise magnification-calibration scenario considered here.

These results motivate further development and validation
of response-based galaxy selections. Realizing this potential
in data will require establishing the achievable response
contrast and number densities, independently calibrating
the selected responses, and jointly characterizing their
redshift dependence and the sample redshift distributions.
Additional requirements include controlling intrinsic
clustering from redshift overlap, mitigating observational
effects in the foreground samples and those shared with
the background samples, and validating the theoretical
model on the scales included in the likelihood. In this
section we discuss these requirements, the relation to
previous magnification estimators, and possible extensions
of BigMag.

\subsection{Calibration of the magnification response}
\label{sec:discussion_response_calibration}

Realizing the full constraining power of the analysis requires an accurate absolute calibration of the magnification response of each high-redshift sample, but the observables themselves provide substantial internal calibration. Starting from broad, independent priors of $\sigma(C_{\rm mag})=0.2$, BigMag reduces the marginalized response uncertainties to approximately $0.13$--$0.15$, while IGGL achieves $0.09$--$0.14$ and their combination $0.06$--$0.12$ (Fig.~\ref{fig:bigmag_iggl_cmag}). Even with uniform priors on all responses, BigMag+IGGL yields a fractional precision on $S_8$ of $\simeq 1.6\%$ (Eq.~\ref{eq:flat_cmag_s8_constraint}), only $\simeq 28\%$ larger than that of traditional $3\times2$pt, while constraining the response uncertainties to $\simeq 0.06$--$0.18$. Imposing the fiducial $\sigma_C=0.2$ priors then improves $\sigma_{68}(S_8)$ by only $\simeq 11\%$. This self-calibration arises because each response coefficient must simultaneously describe multiple cross-correlations involving different foreground samples and lensing kernels, while the foreground autocorrelations constrain the galaxy biases entering the BigMag correlation functions. Self-calibration therefore mitigates, but does not eliminate, the need for accurate external response calibration: tightening the response priors substantially improves the cosmological constraints, and in the fixed-response limit the constraining power of BigMag becomes competitive with that of traditional $2\times2$pt for which we assume a $1\%$ relative uncertainty on the multiplicative shear calibration.

Synthetic-source injections provide a direct route to this calibration by propagating a known magnification through detection, measurement, photometric-redshift assignment, and the final sample selection.  Current publicly available injection campaigns, such as DES Y6 \textsc{Balrog}, achieve uncertainties of approximately $\sigma(C_{\rm mag})\sim0.1$--$0.2$ for the samples of this work. Appendix~\ref{app:csample-injection-scaling} shows that this uncertainty follows the expected $N_{\rm inj}^{-1/2}$ scaling. Extrapolating this trend, a campaign covering the full DES footprint at $5$--$10$ times the current injection density, and concentrating injections on the galaxies most likely to contribute to the high-redshift samples, would place the statistical response uncertainties close to $\sim0.01$. The statistical precision assumed in the tight-prior forecasts therefore appears achievable with a larger, dedicated campaign.

Defining the samples places a separate requirement on the injections. Because a sample response averages over many cells, individual cell responses are noisier than the aggregate by roughly the square root of the number of contributing cells. If the same noisy estimates are used both to assign cells to branches and to measure the branch responses, upward fluctuations are preferentially promoted into the positive-response branch and downward fluctuations into the negative-response branch, biasing the apparent contrast high even when each cell estimate is unbiased; this selection bias is not captured by the jackknife uncertainties of \S~\ref{sec:data_sims_des_balrog}. Appendix~\ref{app:csample-heldout-validation} shows that it is large for current DES statistics: on held-out injections, the branch contrast $\Delta C \equiv C^{\rm mag}_{+} - C^{\rm mag}_{-}$ falls from $\simeq 2.7$--$3.5$ to $\simeq 0.9$--$1.7$, with the positive-response branches consistent with the null response. Flagship reproduces this regression at Balrog-like per-cell noise, whereas at full statistics the in-sample and held-out responses agree and a contrast of $\Delta C \simeq 1.6$ survives, so the SOM carries genuine response information that current injections cannot fully exploit. Selections should therefore be defined and calibrated with disjoint injection subsets. The in-sample responses adopted in our forecasts (Table~\ref{tab:data_sims_samples}) should accordingly be read as targets, whose realization requires an injection campaign large enough both to select the branches reliably and to calibrate them, or an alternative method capable of robustly separating the branches.

Increasing the number or area of injections reduces only the statistical component of the uncertainty. A precision analysis must also demonstrate that the injected population accurately represents the data across the joint space of colour, magnitude, size, morphology, and redshift. Because most current source-injection campaigns rely on empirical galaxy samples drawn from relatively small deep fields, additional strategies will be needed to quantify and mitigate the impact of sample variance in these reference populations, potentially using realistic models of galaxy populations \citep[e.g.][]{Alarcon26_diffstarpop, galsbisps,popcosmos}. The systematic accuracy of the injection-based response measurement must also be established, in analogous work to that undertaken by the weak lensing shear community \citep{Y3Imsims, HSC_imsims, KIDS_imsims, Yoon_kids_metacal_imsims_skills, Y6Mdet, Euclid_imsims_I, Euclid_imsims_II}. This includes testing the validity of the finite-difference approximation by comparing responses inferred from positive and negative magnifications, $\pm\Delta\kappa$, and, ideally, from several values of the applied convergence to identify any departures from linearity. In addition, potential residual dependencies of the inferred response on spatially varying observing conditions across the survey footprint should be quantified and, where necessary, incorporated into the calibration uncertainty budget.

The response can also vary within a background tomographic bin \citep[e.g.][]{Wenzl_2024}.  For a redshift distribution $n_i(z_s)$ normalized over source redshift and using $\chi_s\equiv\chi(z_s)$, this dependence enters through the response-weighted magnification kernel
\begin{equation}
W_{{\rm mag},i}(\chi)
=
\frac{3\Omega_{\rm m}H_0^2}{2c^2}
\frac{\chi}{a(\chi)}
\int_{z(\chi)}^\infty {\rm d}z_s\,
n_i(z_s)C_{{\rm mag},i}(z_s)
\frac{\chi_s-\chi}{\chi_s} ,
\label{eq:discussion_redshift_dependent_cmag_kernel}
\end{equation}
such that
$\delta_{g,i}^{\rm mag}(\hat{\boldsymbol n})
=\int {\rm d}\chi\,W_{{\rm mag},i}(\chi)
\delta_{\rm m}(\chi\hat{\boldsymbol n},z(\chi))$.
Appendix~\ref{app:cmag-redshift-dependence} derives this result from the linear selection response of Sec.~\ref{sec:gal_density_mag} and relates it to the sample-averaged response measured with synthetic-source injections. Consequently, the response entering a given cross-correlation is weighted by both the source distribution and the lensing geometry, and its effective value can differ among foreground kernels.  Future analyses should therefore use source injections to characterize any redshift dependence of the magnification response and propagate the joint uncertainty in $n_i(z)C_{{\rm mag},i}(z)$, including the covariance between redshift and response calibration. The partial self-calibration supplied by BigMag and IGGL constrains effective response amplitudes, but cannot by itself protect against unmodeled redshift dependence or selection-response errors that are absorbed by those amplitudes.

\subsection{Redshift distributions and sample optimization}
\label{sec:discussion_redshift_optimization}

Accurate redshift distributions are essential for interpreting the position-position correlations.  For simplicity, our fiducial forecasts assign the positive- and negative-response branches of each high-redshift bin the same Gaussian $n(z)$.  The data-informed SOMPZ distributions shown in \S\ref{sec:data_sims_sample_selection}, however, reveal small differences between these branches, as well as extended low-redshift tails.  A real-data analysis should use a separate redshift distribution for each selected sample and propagate its uncertainty jointly with that of the magnification response. The SOMPZ methodology used here provides a natural route to these branch-specific calibrations.

The low-redshift tails deserve particular attention because galaxies that actually lie at the foreground redshift cluster physically with the lens sample, adding an intrinsic density-density contribution to the measured cross-correlation.  On large scales, its amplitude depends on both the redshift overlap and the biases of the foreground and contaminating populations \citep{Garcia_Fernandez_2018, 2020JCAP...03..044N}. For the three data-informed high-redshift samples, the fractions of their redshift distributions below $z=0.65$ are approximately $0.010$, $0.015$, and $0.020$, with similar values for the positive- and negative-response branches. These numbers describe the total low-redshift tail, not the overlap with each individual lens bin. If the entire low-redshift tail were concentrated within the redshift range of a single lens bin, the resulting contamination would be localized to that cross-correlation; if instead the tail were spread across the full low-redshift range, the overlap with any one lens bin would be correspondingly smaller.

Appendix~\ref{app:redshift-overlap} isolates this effect with a controlled experiment in which a fraction $f_{\rm overlap}$ of each high-redshift distribution is replaced by the Gaussian distribution of a matched lens bin. Thus, $f_{\rm overlap}$ represents a pair-specific contaminating component, rather than the total probability below $z=0.65$.  For the scale cuts and covariance adopted here, $f_{\rm overlap}=0.01$ produces a density-only contribution with a signal-to-noise approximately $10\%$ of that of the magnification signal, with a roughly linear scaling at small overlap. The response split we have adopted here provides an important advantage in this respect, as selecting samples with larger $|C_{\rm mag}|$ strengthens the magnification contribution without producing a corresponding increase in the intrinsic density--density term.  The BigMag selections therefore improve not only the magnification signal-to-noise, but also its contrast with physical-clustering contamination.  Nevertheless, intrinsic clustering contributes positively to both response branches: as the overlap grows, it reinforces the positive-response correlation but suppresses, and can eventually reverse, the negative-response correlation.  Residual overlap must therefore be modelled even when it remains subdominant in the total signal.

These redshift-overlap considerations feed directly into the optimization of the response-selected samples.  The split adopted here, at the null-response boundary ($C_{\rm mag}=0$), removes the leading cancellation between galaxies with opposite magnification responses and increases the contrast of magnification against intrinsic clustering.  It does not, however, necessarily define the optimal selection.  Moving toward more extreme response values can increase $|C_{\rm mag}|$ and make a fixed density-overlap contribution relatively less important, but it also reduces the galaxy number density and may select populations with less secure redshift distributions or response calibrations. Similarly, a denser parent sample can improve the trade-off between number density and magnification response, although deeper selections may also contain larger low-redshift tails. Future optimization should therefore maximize the expected marginalized cosmological information, rather than $|C_{\rm mag}|$ or the raw signal-to-noise alone, while accounting jointly for number density, response uncertainty, redshift overlap, contaminant galaxy bias, and sensitivity to observational systematics.

\subsection{Null-magnification samples}
\label{sec:discussion_null_mag}

A useful by-product of the framework developed here is the ability to
construct samples that suppress magnification. Once the parent population
has been characterized and divided according to its magnification
response, the same information can be used both to select samples with
large $|C_{\rm mag}|$ and to identify selections with
$C_{\rm mag}\simeq0$. The latter could be obtained by selecting regions
of galaxy-property space with small responses, or by combining
opposite-response populations with suitable weights. Their construction
therefore follows naturally from the calibration and sample-selection
machinery already required for BigMag, without a separate calibration
programme. Such samples would be useful for the galaxy-position component
of traditional $3\times2$pt analyses: suppressing their magnification
response would reduce the additional contributions to galaxy clustering
and galaxy--galaxy lensing while retaining the intrinsic clustering and
its correlation with matter. This would simplify the interpretation of
these observables and reduce their sensitivity to uncertainties in the
magnification coefficients, whose importance has been demonstrated for
current and future surveys
\citep{Euclidmagnification2021,y3-2x2ptmagnification}. The statistical
benefit would depend on the number density, galaxy bias, and redshift
distribution of the resulting selections, so a sample optimized for
minimal magnification need not maximize the total cosmological
information.

Null-response lens samples would be particularly valuable for
lensing-ratio measurements
\citep{Jain2003,DES_spt_lensingratio,Unruh_2019, Prat_2022,2024OJAp....7E..97E,desy6-ggl, 2026PhRvD.113f3534R}. In the narrow-lens-bin limit,
ratios of lensing signals measured around the same foreground galaxies
can cancel the common dependence on galaxy bias and the
galaxy--matter power spectrum, leaving ratios of lensing-distance
factors. Magnification of the lens population introduces an additional
correlation between foreground matter and the measured background
lensing signal. This contribution involves a different line-of-sight
kernel and does not generally cancel between the numerator and
denominator, introducing dependence on the matter distribution and
potentially on angular scale
\citep{Unruh_2019,BermejoCliment2020}. Selecting lenses with
$C_{\rm mag}\simeq0$ would suppress this contribution and help preserve
the geometric interpretation that constitutes the main motivation behind these measurements.
Realizing this benefit requires controlling the residual response
across the lens redshift distribution, alongside the usual treatment
of finite bin widths and other observational effects. Although
optimizing and validating such samples is beyond the scope of this
paper, the ability to construct both magnification-enhanced and
near-null samples from the same calibrated parent population broadens
the applications of the proposed selection framework.

\subsection{Observational systematics}
\label{sec:discussion_observational_systematics}

The BigMag background samples are dense and extend to faint galaxies close to the detection limit, making them particularly sensitive to spatial variations in observing conditions.  As discussed in \S\ref{sec:systematics}, this challenge is partly reduced because we do not use the autocorrelations of these samples, but only their cross-correlations with brighter foreground galaxies.  The foreground samples retain the usual systematics requirements of a clustering analysis, since their autocorrelations enter the data vector directly.  Once the foreground density field has been properly corrected, fluctuations affecting only the background selection do not bias the mean cross-correlation, provided that they are uncorrelated with the corrected foreground field.  They can nevertheless increase the covariance and reduce the achievable signal-to-noise.  Shared or correlated residual systematics can still bias the measurement. In an optimal analysis both samples should be carefully characterized and corrected, following the systematics-mitigation strategies developed for modern clustering analyses using survey-property maps of depth, seeing, extinction or stellar density \citep{Noah_weights,Berlfein2024_systematics,DESY6_systematics, highz}.  

Observing conditions can also change the composition of a selected sample across the footprint, producing spatial variations in its redshift distribution, galaxy bias, or magnification response.  These subsample effects should be propagated into the theoretical prediction or corrected, since otherwise they may bias the inferred cosmological signal \citep[e.g.][]{Kong_subsample_sys,Zhang_clustering_variablenz}.  A real-data analysis must additionally test whether the background selection is independent of the foreground density itself.  This assumption can fail through effects local to the foreground galaxies, including blending and deblending failures, local background-subtraction errors, and dust extinction, obscuration or chromatic effects induced by nearby galaxies or intracluster light from the foreground sample.  Such effects can generate foreground--background correlations even after large-scale observing conditions have been corrected. They have previously been investigated in flux-magnification analyses \citep{2010MNRAS.405.1025M,2016MNRAS.455.3943H,2021MNRAS.503.4964G}, and their impact on observed number counts can be further characterized and mitigated using image-level source-injection campaigns such as \textsc{Balrog}.

\subsection{Theoretical modeling, foreground magnification and small-scale extensions}
\label{sec:discussion_theory_small_scales}

The forecasted constraining power depends on the adopted nonlinear model and scale cuts.  We use a minimum transverse separation of $4\,h^{-1}\mathrm{Mpc}$ for the clustering measurements entering BigMag and for traditional galaxy-galaxy lensing.  As shown in \S\ref{sec:results_bigmag}, adding a second-order galaxy-bias parameter produces only a small loss of precision, which is encouraging, but does not by itself demonstrate that the model will remain unbiased when applied to data.  A real data analysis should validate the scale cuts using realistic simulations or contaminated synthetic data, including scale-dependent galaxy bias or baryonic effects absent from the fiducial model.  The adopted choices should only be accepted after showing that they recover unbiased cosmological constraints, following tests similar to those developed in usual lensing analyses \citep[e.g.][]{desy6methods}.

The position-position magnification signal has a different radial dependence
from traditional galaxy--galaxy lensing.  Schematically,
\begin{align}
    \gamma_{\rm t}(R)
    &\propto
    \Delta\Sigma(R)
    =
    \overline{\Sigma}(<R)-\Sigma(R),
    \\
    w_{g\mu}(R)&\propto\Sigma(R),
\end{align}
up to the corresponding redshift kernels and response factors.  The galaxy-galaxy lensing signal at a retained separation therefore depends on the matter distribution at all smaller separations.  As discussed in \S\ref{sec:results_bigmag_calibration}, uncertainty in the mass below the minimum modeled scale can leak into the retained measurements and is commonly mitigated through point-mass marginalization or an equivalent localization procedure \citep{MacCrann_2019,Prat_2023}.  The leading BigMag signal does not contain this same enclosed-mass contribution, so its scale cut acts more locally in projected separation.  This does not make BigMag insensitive to nonlinear physics on the retained scales, but it may make the separation between modeled and excluded scales cleaner than in traditional galaxy-galaxy lensing.

In our forecasts we conservatively apply point-mass marginalization to every retained position-shear correlation, including those dominated by IGGL. The dominant IGGL contribution, $\langle\kappa\gamma\rangle$, and the residual traditional-lensing contribution, $\langle\delta_g\gamma\rangle$, have different physical kernels, so the nuisance treatment inherited from traditional galaxy-galaxy lensing might be conservative or at least non-optimal.  A future analysis could model these terms separately and associate unresolved-scale nuisance parameters only with the contributions that require them, after validating the procedure with simulations.  We have also simplified the forecasts by setting the magnification response of the foreground samples to zero.  In real data this response will generally be nonzero, introducing additional density-magnification terms in the foreground autocorrelations and a smaller $\langle\kappa\kappa\rangle$ contribution to the foreground--background cross-correlations.  These terms should be included in the model and may themselves carry useful cosmological information, although the magnification contribution to clustering autocorrelations is small compared to intrinsic clustering.

The different and more local projected-scale dependence of BigMag also makes it an interesting probe of the galaxy-halo connection on smaller scales \citep{Wechsler_2018}. A more complete small-scale analysis could combine BigMag with galaxy-galaxy lensing and clustering using halo models \citep{Zacharegkas_des_stellarmass,Dvornik2023_kids_galhalo, Chaurasiya2024_HSC_galhalocon}, subhalo abundance matching \citep{Mahony2026_shame}, or differentiable galaxy--halo models based on parametrized physical prescriptions for galaxy star formation history \citep{hearin_etal21_diffmah, alarcon_etal23_diffstar, Alarcon26_diffstarpop} and stellar population synthesis \citep{hearin_etal23_dsps}. The complementary radial weighting of these observables could help separate galaxy-halo physics from cosmology.  Exploiting this information will require accurate modeling of nonlinear bias, baryonic effects, higher-order magnification, and the covariance with the large-scale data vector.

Complementing these model-dependent small-scale extensions, Appendix~\ref{app:magnification-ratios} considers a more agnostic use of the excluded BigMag measurements through magnification ratios. These ratios partially cancel the unknown small-scale galaxy--matter correlation while retaining information about relative magnification responses and, when different source redshifts are compared, lensing geometry, similarly to conventional lensing ratios \citep{DES_spt_lensingratio}. For the positive- and negative-response branches of the same tomographic bin, the common foreground mass profile cancels to leading order, such that their correlation ratio directly constrains $C_{\rm mag}^{-}/C_{\rm mag}^{+}$ and hence the anti-correlated combination of the two responses. However, a common rescaling of both coefficients leaves this ratio unchanged and can be absorbed by the BigMag lensing amplitude, yielding little improvement in the cosmological constraints.  Ratios across source-redshift bins additionally retain a lensing-efficiency ratio, although this geometric information is weak for our sample configuration and remains degenerate with response and redshift-distribution uncertainties. These measurements can therefore provide useful internal calibration and consistency tests, but they do not determine the absolute response normalization nor add significant cosmological information for the configuration considered here.

\subsection{Relation to previous magnification estimators}
\label{sec:discussion_previous_magnification}

The analysis presented here builds on a broad literature investigating magnification as both a cosmological signal and a contribution to standard large-scale-structure observables. Early studies developed quasar--galaxy density correlations and related higher-order statistics \citep{2002A&A...386..784M,2003A&A...409..411M}. They showed that, because faint sources with $C_{\rm mag}<0$ are anti-correlated with foreground mass, an unweighted sample can suppress or even reverse the signal. It also derived optimal weights under which every magnitude bin contributes constructively. Subsequent work examined the complementarity of number-count magnification with shear \citep{2010MNRAS.401.2093V,2014MNRAS.437.2471D} and of size magnification with shape \citep{2013MNRAS.433L...6H}; its effects on source counts and fluxes \citep{2011MNRAS.411.2113J}; and estimators that null, remove or cancel the intrinsic clustering signal \citep{2005PhRvL..95x1302Z,2006MNRAS.367..169Z,2011MNRAS.415.1681H,2021RAA....21..247H,2024MNRAS.527.7547M}. The sensitivity of inverse galaxy--galaxy lensing to magnification has also been explored \citep{HSTIGGL,IGGL}. In the context of modern surveys, several studies have shown that neglecting magnification can bias clustering, lensing, and $3\times2$pt analyses \citep{2014MNRAS.437.2471D,2016PhRvD..94d3007C,2018JCAP...04..033V,2018PhRvD..97b3537L,2020MNRAS.491.1746T,2020A&A...636A..95D,2020A&A...638A..96U,2022MNRAS.515.1130D,Euclidmagnification2021,2024A&A...685A.167E}, while recent work has assessed its additional constraining power in LSST-like and submillimetre surveys \citep{2022MNRAS.513.1210M,2025Galax..13...89C,2025arXiv251213894B}. Together, these studies establish both the need to model magnification accurately and its potential value as a deliberately constructed cosmological observable.

The measurements most closely related to BigMag use magnification bias in galaxy number density. In this approach, foreground positions are correlated with the angular density of a more distant population, producing a position--position lensing statistic when the samples have negligible redshift overlap. Such measurements have been performed using quasars \citep{2003ApJ...589...82G,2005MNRAS.359..741M,2005ApJ...633..589S}, Lyman-break galaxies \citep{2009A&A...507..683H,2012MNRAS.426.2489M}, and photometrically selected galaxy samples \citep{Garcia_Fernandez_2018,2020JCAP...03..044N}. Several of these analyses measured the correlation in magnitude bins, observed the sign change predicted by $C_{\rm mag}$ between bright and faint sources, and combined bins with weights \citep{2005ApJ...633..589S, 2009A&A...507..683H, 2012MNRAS.426.2489M, Garcia_Fernandez_2018}. A sustained line of work has used high-redshift submillimetre galaxies, whose steep number counts and strong redshift separation can produce a large magnification-bias signal \citep{2010MNRAS.406.2352L,2011MNRAS.414..596W,2013MNRAS.429.3230H,2014MNRAS.442.2680G,2017JCAP...10..024G,2021A&A...646A.152G,2026ExA....61...18G,2019JCAP...09..021B,2020A&A...639A.128B,2021A&A...656A..99B,2024A&A...686A.191B,2021A&A...645A.126C,2022A&A...662A..44C,2024A&A...686A.190C,2022A&A...667A.146C,2025arXiv251023582F}. These applications range from cross-correlation detections and tomography to constraints on cosmology, halo models, and the halo mass function.

Related count-based measurements have used magnification bias as a mass probe for galaxy groups and clusters \citep{2010ApJ...723L..13V,2011ApJ...733L..30H,2012ApJ...754..143F,2013ApJ...769...13U,2013ApJ...772...65C,2014MNRAS.439.3755F,2016MNRAS.457.3050C,2020MNRAS.495..428C,2017A&A...608A.141T,2025A&A...697A.141M}. Flux-dependent combinations of galaxy-count maps have also been used to reconstruct convergence from DECaLS and DES Y3 data by cancelling deterministic galaxy bias \citep{2024PhRvD.110j3538Q,2025PhRvD.112d3522Q,2025PhRvD.112j3514Q}. These methods demonstrate the range of information available from source counts, but also emphasize their sensitivity to the response of the selected population and to contaminants such as redshift overlap, dust, blending, and obscuration.

Other estimators use lensing-induced changes in galaxy flux, magnitude, or size more directly, with the unlensed reference supplied by a population distribution or an empirical scaling relation. Applications include correlated shifts in ensemble observables \citep{2010MNRAS.405.1025M,2012ApJ...744L..22S,2014MNRAS.440.3701B,2024ApJ...973..102X}, Fundamental-Plane residuals used as noisy galaxy-level convergence proxies \citep{2006ApJ...648L..17B,2011A&A...532A..37S,2014ApJ...780L..16H,2020MNRAS.496.2998F}, and Bayesian convergence likelihoods based on the joint intrinsic distribution of galaxy size, magnitude, and redshift \citep{2015MNRAS.452.1202A}. A related population-inference principle underlies Bayesian Fourier Domain methods, which operate on PSF-corrected image moments and could in principle infer magnification jointly with shear, although published demonstrations have focused on shear \citep{2014MNRAS.438.1880B,2016MNRAS.459.4467B}. These direct estimators require careful calibration of the unlensed population and control of photometric noise, dust, blending, obscuration, and selection effects \citep{2016MNRAS.455.3943H,2021MNRAS.503.4964G}. Fundamental-Plane residuals present the additional complication that they correlate intrinsically with large-scale environment and can therefore mimic lensing \citep{2015MNRAS.454..478J}.

BigMag belongs to the established class of position--position magnification estimators. It shares with the $C_{\rm mag}$-weighted estimators and flux-bin estimators mentioned before the principle that sources with opposite responses should not be combined with a common sign. It differs from them in three respects. First, the response of each candidate selection is calibrated end-to-end with image-level injections, which include detection, size, blending and SOM/photometric-redshift assignment, rather than being inferred from how the luminosity function translates into an $\alpha(m)$ relation. Second, these same calibrated responses are used to define the background samples in multi-dimensional color-magnitude space using the SOM. Third, the resulting branches enter as separate tracers in a joint likelihood with foreground clustering and with IGGL, rather than being compressed into a single weighted correlation or convergence map. Keeping these populations separate recovers signal that would otherwise be lost through cancellation and provides internal tests using positive, negative, and near-zero-response samples. In addition, the different positive and negative branches trace a common nuisance field, the intrinsic clustering density field of the background sources, acting as a sort of multi-tracer field included through the joint covariance. This is the principle behind flux-bin magnification reconstructions \citep{2005PhRvL..95x1302Z,  2024PhRvD.110j3538Q, 2025PhRvD.112d3522Q, 2025PhRvD.112j3514Q}, and is closely related to multi-tracer cancellation of sample variance \citep{2009PhRvL.102b1302S}.

A natural extension of BigMag would be to use the same background populations to construct a complementary flux-magnification observable by cross-correlating foreground galaxy positions with fluctuations in the mean flux or magnitude of the background galaxies
\citep{2010MNRAS.405.1025M,2011MNRAS.411.2113J,2014MNRAS.440.3701B}. With negligible redshift overlap, this observable depends on the same foreground galaxy bias and matter--convergence correlation as the position-position magnification signal, but its amplitude is set by a distinct flux-response coefficient. The effective response of the measured ensemble flux, including the effects of detection, measurement noise, blending, non-detections, and sample selection, could in principle be calibrated using the same matched magnified and unmagnified synthetic-source injections employed for the number-density response
\citep{2016MNRAS.455.3943H,y3-2x2ptmagnification,Y6Balrog,Legnani2026}. The samples optimized for number-density magnification need not be optimal for the flux observable, but their joint analysis could recover additional information and provide further consistency tests. Such an analysis would require careful treatment of dust extinction, obscuration, blending, and spatial photometric-calibration errors
\citep{2010MNRAS.405.1025M,2013MNRAS.429.3230H,2016MNRAS.455.3943H,2021MNRAS.503.4964G}.
Multi-band flux and colour information could help separate these effects because gravitational magnification is achromatic, whereas dust extinction is generally wavelength dependent. We leave the development of this joint number-density and flux-magnification analysis to future work.

\section{Conclusion}
\label{sec:Conclusion}

In this paper, we have developed an injection-calibrated strategy for selecting galaxy
subpopulations according to their magnification response, using observed
photometric properties and self-organizing maps.  We call the resulting
position-only analogue of a traditional $2\times2$pt analysis
BigMag: it combines the autocorrelations of foreground position
samples with lensing-dominated position--position cross-correlations between
foreground and background galaxies.  We assessed its cosmological potential
using DES-informed redshift distributions and number densities together with
simulated data vectors.  The cosmological forecasts adopt a possible but optimistic 
response-separation scenario motivated by the DES measurements and by the
Flagship validation tests.  Our main conclusions are:

\begin{itemize}

\item BigMag extracts weak-lensing information using galaxy positions alone.  Separating background galaxies with different magnification responses prevents positive and negative contributions to the foreground--background cross-correlations from cancelling.  Combining these cross-correlations with foreground galaxy clustering helps break the degeneracy between foreground galaxy bias and the amplitude of matter fluctuations.

\item Using DES Year 6 photometry, SOMPZ redshift calibration, and \textsc{Balrog} synthetic-source injections, we constructed dense high-redshift candidate samples and found direct evidence for substantial magnification-response cancellation.  Complementary magnitude partitions produce order-unity response differences, with positive- and negative-response populations in the first two high-redshift bins and a predominantly positive response in the third.  Validation and simulation tests show, however, that the current \textsc{Balrog} statistics are insufficient to optimally achieve the full magnification response separation in the data with a SOM. DES-like Flagship simulations demonstrate that a SOM partition can recover a large and stable held-out separation when the per-cell responses are measured with sufficiently high signal-to-noise.

\item In the adopted response-separation scenario, splitting the background samples increases the raw signal-to-noise
of the foreground--background position cross-correlations,
before analysis scale cuts, from approximately 64 to 194.  The resulting correlations break the foreground-bias--matter-amplitude degeneracy. When the magnification coefficients are fixed, the marginalized $S'_8$ uncertainty from BigMag is approximately $17\%$ larger than that from the reference traditional $2\times2$pt analysis, which retains its fiducial small shear-calibration uncertainty.

\item The same response-selected position samples generate a high signal-to-noise inverse galaxy-galaxy lensing (IGGL) observable with complementary dependencies on cosmology and nuisance parameters.  Combining BigMag and IGGL reduces the marginalized uncertainties on the six high-redshift magnification responses from prior widths of $0.2$ to approximately $0.055$--$0.12$.  With the fiducial response priors, their joint analysis yields $\sigma_{68}(S_8)=0.0117$, approximately $15\%$ larger than the traditional $3\times2$pt reference.  For $\sigma(C_{\rm mag})=0.02$, the joint magnification constraint becomes approximately $14\%$ tighter than that reference.

\item Adding BigMag and IGGL to traditional $3\times2$pt reduces the forecast $S_8$ uncertainty by approximately $16\%$ for the fiducial response priors and by $29\%$ for $\sigma(C_{\rm mag})=0.02$.  These gains are conditional on the adopted response amplitudes and modeling assumptions. Realizing them in data will require independently validated magnification responses, accurate redshift calibration, control of redshift overlap and shared observational systematics, and validated nonlinear modeling.  The measured injection-statistics scaling from Appendix~\ref{app:csample-injection-scaling} suggests that statistical response uncertainties near $0.01$ could be reached with a full-footprint injection campaign at approximately $5$--$10$ times the current injection density, although systematic calibration uncertainties must be controlled separately.

\item The results in this paper motivate further development and validation
of galaxy selections that maximize the cosmological information
available from magnification. Applying the strategies explored
here, and investigating alternative selections and galaxy
populations, will establish the performance achievable in real
data. The resulting samples can then be used to constrain
cosmology through BigMag and IGGL, both jointly and in
combination with traditional $3\times2$pt analyses. The same framework can also be used to construct samples with
near-zero magnification response, suppressing lens-magnification
contributions to traditional $3\times2$pt analyses and helping
preserve the geometric interpretation of lensing-ratio
measurements.

\end{itemize}

\section*{Acknowledgments}

AA acknowledges support from projects PID2023-153229NA-I00 and PID2025-172756NB-C31 funded by MICIU/AEI/10.13039/501100011033 and FEDER, UE, and grant RYC2024-050321-I funded by MICIU/AEI/10.13039/501100011033 and by ESF+. The project that gave rise to these results received the support of a fellowship from "la Caixa" Foundation (ID 100010434). The fellowship code is LCF/BQ/PI23/11970028. AA would also like to thank Laura Casas for useful discussions during the preparation of this paper.

CS acknowledges support from projects PID2022-138123NA-I00, PID2025-172756NB-C32 and CNS2025-166489 funded by MICIU/AEI/10.13039/501100011033 and FEDER, UE, and 
Ramón y Cajal project RYC2021-031194-I, funded by
MCIN/AEI/10.13039/501100011033 and by the “European Union NextGenerationEU/PRTR”.

We thank the developers of {\tt NumPy} \citep{numpy_ndarray}, {\tt SciPy} \citep{scipy}, Jupyter \citep{jupyter}, IPython \citep{ipython}, and Matplotlib \citep{matplotlib} for their extremely useful free software. While writing this paper we made extensive use of the Astrophysics Data Service (ADS) and {\tt arXiv} preprint repository.

This work has made use of CosmoHub \citep{2017ehep.confE.488C,TALLADA2020100391}, developed by PIC (maintained by IFAE and CIEMAT) in collaboration with ICE-CSIC. It received funding from the Spanish government (grant EQC2021-007479-P funded by MCIN/AEI/10.13039/501100011033), the EU NextGeneration/PRTR (PRTR-C17.I1), and the Generalitat de Catalunya.

This project used public archival data from the Dark Energy Survey (DES). Funding for the DES Projects has been provided by the U.S. Department of Energy, the U.S. National Science Foundation, the Ministry of Science and Education of Spain, the Science and Technology FacilitiesCouncil of the United Kingdom, the Higher Education Funding Council for England, the National Center for Supercomputing Applications at the University of Illinois at Urbana-Champaign, the Kavli Institute of Cosmological Physics at the University of Chicago, the Center for Cosmology and Astro-Particle Physics at the Ohio State University, the Mitchell Institute for Fundamental Physics and Astronomy at Texas A\&M University, Financiadora de Estudos e Projetos, Funda{\c c}{\~a}o Carlos Chagas Filho de Amparo {\`a} Pesquisa do Estado do Rio de Janeiro, Conselho Nacional de Desenvolvimento Cient{\'i}fico e Tecnol{\'o}gico and the Minist{\'e}rio da Ci{\^e}ncia, Tecnologia e Inova{\c c}{\~a}o, the Deutsche Forschungsgemeinschaft, and the Collaborating Institutions in the Dark Energy Survey.
The Collaborating Institutions are Argonne National Laboratory, the University of California at Santa Cruz, the University of Cambridge, Centro de Investigaciones Energ{\'e}ticas, Medioambientales y Tecnol{\'o}gicas-Madrid, the University of Chicago, University College London, the DES-Brazil Consortium, the University of Edinburgh, the Eidgen{\"o}ssische Technische Hochschule (ETH) Z{\"u}rich,  Fermi National Accelerator Laboratory, the University of Illinois at Urbana-Champaign, the Institut de Ci{\`e}ncies de l'Espai (IEEC/CSIC), the Institut de F{\'i}sica d'Altes Energies, Lawrence Berkeley National Laboratory, the Ludwig-Maximilians Universit{\"a}t M{\"u}nchen and the associated Excellence Cluster Universe, the University of Michigan, the National Optical Astronomy Observatory, the University of Nottingham, The Ohio State University, the OzDES Membership Consortium, the University of Pennsylvania, the University of Portsmouth, SLAC National Accelerator Laboratory, Stanford University, the University of Sussex, and Texas A\&M University.
Based in part on observations at Cerro Tololo Inter-American Observatory, National Optical Astronomy Observatory, which is operated by the Association of Universities for Research in Astronomy (AURA) under a cooperative agreement with the National Science Foundation.

\section*{Statement on AI use}

Most of the central ideas underlying this paper were developed approximately a year ago, without the use of LLMs, through consideration of the physics of magnification and exploratory work with the publicly available \textsc{Balrog} products. This led to the preparation of the configuration files used to run the inference chains with \textsc{CosmoSIS} in successive stages, with only minimal assistance from LLMs. After the main results had been obtained, Codex was used more extensively to accelerate the production of plots, which were subsequently reviewed and manually edited. Codex was also used to assist with drafting and editing portions of the manuscript, as well as with the full implementation and partial drafting of the four appendices. Claude was additionally used to provide referee-style feedback on the manuscript. Both Claude and Codex were also used to comment on whether literature citations appropriately supported the corresponding statements and as search tools to identify related ideas and approaches in the literature. Any references identified in this way were subsequently checked manually, and, to the best of our ability, all citations in the manuscript were reviewed without relying solely on information provided by LLMs. In the interest of full disclosure, this paragraph was itself edited with the assistance of Codex.

\bibliographystyle{aasjournal} 
\bibliography{Paper, intro}% Produces the bibliography via BibTeX.

\appendix

\section{Magnification response}

\subsection{DES Y6-like sample in Flagship}
\label{app:csample-flagship}

To aid the interpretation of the trends measured in DES Y6, we construct a DES-like galaxy sample from the Flagship simulation \citep{flagship_paper}.  Flagship provides noiseless attenuated DECam $griz$ fluxes, but not a corresponding DES noise model.  We therefore use the band-matched LSST flux uncertainties and noise realizations supplied with Flagship, scaling both by a factor of four to approximate the depth of the DES measurements.  The fiducial observed flux in each band is constructed by adding this scaled noise realization to the DECam flux.  For the magnified realization, we multiply the noiseless flux by $\mu=(1-\Delta\kappa)^{-2}$, with $\Delta\kappa=0.01$, before adding the same noise realization.  Keeping the noise fixed ensures that changes in sample membership are produced by the applied magnification rather than by a second noise draw.

Our parent catalog is the union of galaxies that satisfy an observed $i$-band signal-to-noise threshold of five in either the fiducial or magnified realization.  Including objects selected only after magnification is essential: otherwise, the response from galaxies entering the catalog through the detection boundary would be omitted.  We train a $32\times32$ SOM using the fiducial $griz$ photometry and its uncertainties, and apply the same trained map to the fiducial and magnified realizations.  Their matched count difference then gives $C_{\rm sample}$ through Eq.~\eqref{eq:csample_estimator}.  This is a catalog-level validation rather than a fully image-level simulation: it captures the photometric selection and SOM reassignment caused by magnification, but not all of the blending and measurement effects present in the DES \textsc{Balrog} analysis.

\begin{figure}
  \centering
  \includegraphics[width=0.99\linewidth]
  {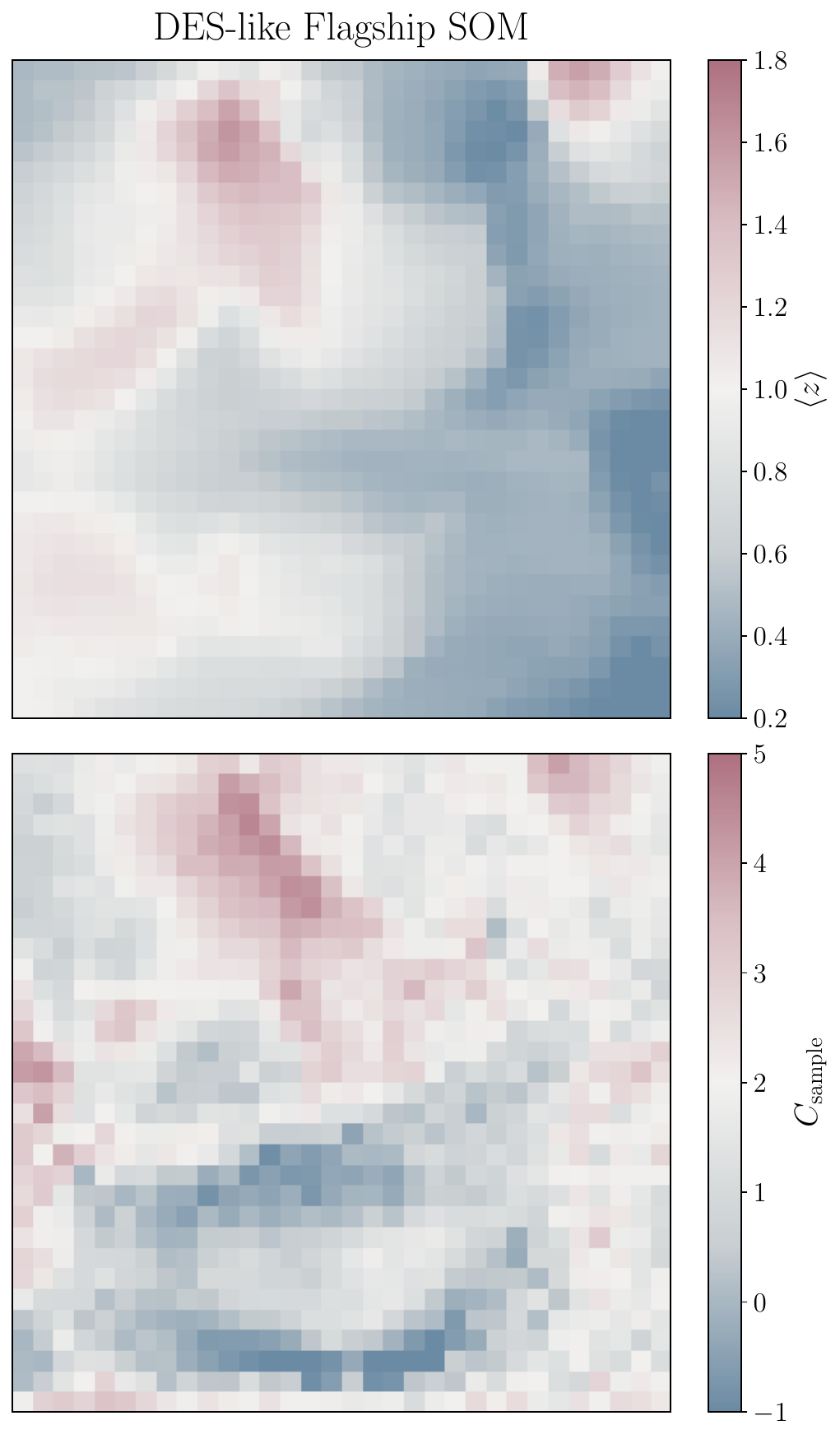}
  \caption{Properties of the $32\times32$ SOM trained on the DES-like Flagship sample.  The upper panel shows the mean true redshift of the galaxies assigned to each cell.  The lower panel shows the corresponding sample response $C_{\rm sample}$, measured by comparing the matched fiducial and $\Delta\kappa=0.01$ realizations.  The response map uses all Flagship SOM assignments available for this analysis.  The null magnification-selection response is $C_{\rm sample}=2$.}
  \label{fig:flagship-des-som-response}
\end{figure}

Figure~\ref{fig:flagship-des-som-response} shows how the mean redshift and $C_{\rm sample}$ vary across the Flagship SOM.  As in the data, the SOM organizes galaxies into extended regions with smoothly varying redshift, while the response contains additional structure within and across those regions. Cells at similar mean redshift can therefore have substantially different magnification responses.  This is the behavior needed for the SOM to provide a higher-dimensional response selection rather than acting only as a redshift estimator.

\begin{figure}
  \centering
  \includegraphics[width=0.88\linewidth]
  {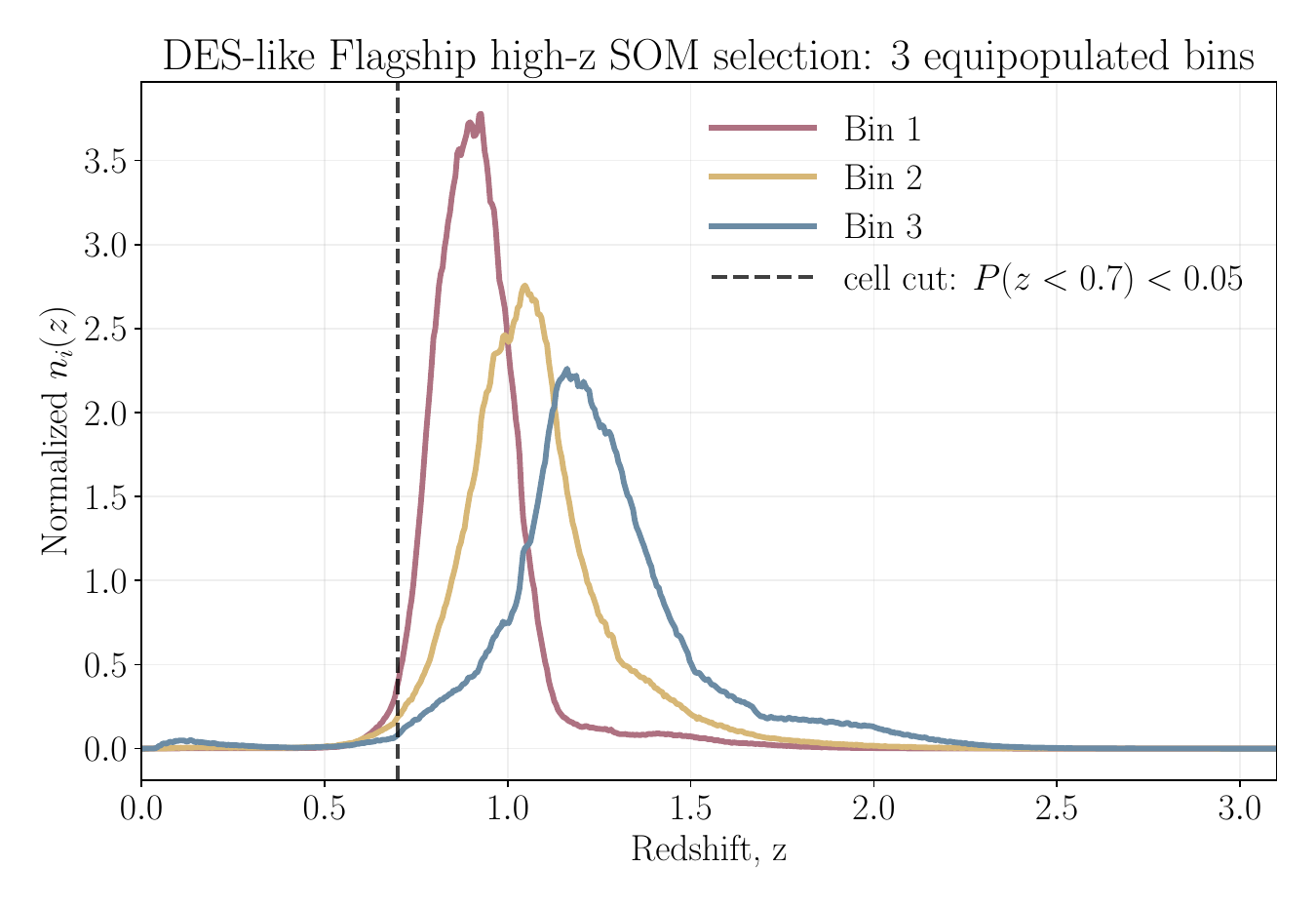}
  \caption{Normalized redshift distributions of the three DES-like Flagship high-redshift samples.  SOM cells are required to satisfy $P(z<0.7\mid\mathrm{cell})<0.05$ and are then divided into three approximately equipopulated samples after ordering by their mean redshift. The vertical dashed line marks $z=0.7$.}
  \label{fig:flagship-des-highz-nz}
\end{figure}

We form three high-redshift samples using the same cell-based construction as for the DES selections.  We first retain cells for which less than five per cent of the assigned population lies below $z=0.7$.  We then order the retained cells by their mean redshift and divide them into three approximately equipopulated bins.  Their normalized redshift distributions are shown in Fig.~\ref{fig:flagship-des-highz-nz}.  The bins have mean redshifts of approximately $0.92$, $1.07$, and $1.25$, and reproduce the broad ordering and overlap of the DES high-redshift samples sufficiently well for the qualitative tests below.  Differences in the detailed redshift distributions and number densities mean that the comparison should not be interpreted as a precise simulation-based prediction for DES.

As a first test of response cancellation, we repeat the complementary magnitude partitions used for the DES data in Fig.~\ref{fig:y6_magnitude_fraction_splits}.  For every threshold, the brighter and fainter branches form an exhaustive partition of the parent tomographic sample.  Magnification moves galaxies through their common magnitude boundary, increasing the abundance of one branch while decreasing that of the other. This internal transfer cancels when the two branches are recombined, so a small response of the parent sample can conceal two substantially larger and oppositely signed contributions.

\begin{figure*}
  \centering
  \includegraphics[width=0.99\linewidth]
  {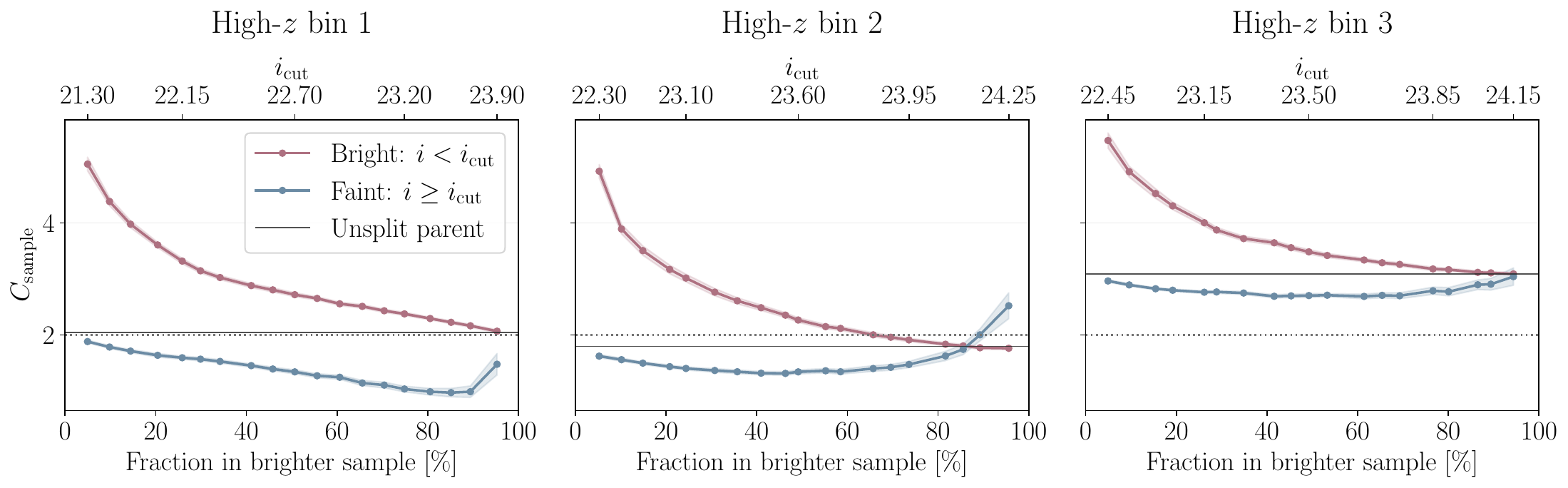}
  \caption{Sample response of complementary magnitude partitions in the three DES-like Flagship high-redshift bins, analogous to Fig.~\ref{fig:y6_magnitude_fraction_splits}.  The horizontal axis gives the fraction of the fiducial parent sample in the brighter branch; the upper axis gives the corresponding $i$-band threshold.  Rose and blue curves show the bright and faint branches, respectively, and the dotted line marks the null selection response, $C_{\rm sample}=2$ or $C_{\rm mag}=0$.  Shaded regions show the uncertainty on each fixed partition.  The first two bins contain branches on opposite sides of the null response over a broad range of thresholds, whereas both branches in the third bin are predominantly positive.}
  \label{fig:flagship-des-magnitude-splits}
\end{figure*}

Figure~\ref{fig:flagship-des-magnitude-splits} shows that the Flagship samples display the same qualitative behavior as the data.  At the illustrative 25/75 bright/faint split, the Flagship branch responses in $C_{\rm mag}$ units are approximately $(+1.32,-0.41)$, $(+1.02,-0.60)$, and $(+2.01,+0.76)$ for bins 1--3, respectively.  The precise amplitudes differ from those measured in DES, as expected from the differences in photometry, redshift distributions, number densities, and measurement model.  Nevertheless, both data sets exhibit order-unity branch responses: the first two bins contain clear cancellation between positive- and negative-response populations, while the response of the third bin is predominantly positive.  The agreement in these broad trends supports the physical interpretation of the DES magnitude partitions and establishes that significant response separation is possible without selecting on noisy per-cell response estimates.

\subsection{Held-out validation of magnification response}
\label{app:csample-heldout-validation}

The magnitude-partition test demonstrates that the parent samples contain populations with different responses, but it does not establish how accurately those populations can be identified from the current SOM-cell response estimates.  If the same finite injection sample is used both to rank cells and to evaluate the resulting branches, statistical fluctuations preferentially promote cells with upward fluctuations into the higher-response branch and cells with downward fluctuations into the lower-response branch.  The apparent contrast is then biased high by this selection, even when the response estimate of every individual cell is unbiased before selection.

We test this effect with a held-out construction.  For the DES Y6 \textsc{Balrog} catalogs, we divide the parent deep-galaxy IDs into two independent halves, keeping all repeated injections of a given deep object in the same half.  One half is used to estimate the response of each SOM cell and to define branches with $\widehat C_{\rm cell}>2$ and $\widehat C_{\rm cell}<2$.  The aggregate responses of these fixed branches are then evaluated using only the other half.  We repeat the procedure over 20 random divisions and exchange the roles of the two halves.  The Flagship tests use the analogous train/held-out split of the matched simulated galaxies.

\begin{figure*}
  \centering
  \includegraphics[width=0.99\linewidth]
  {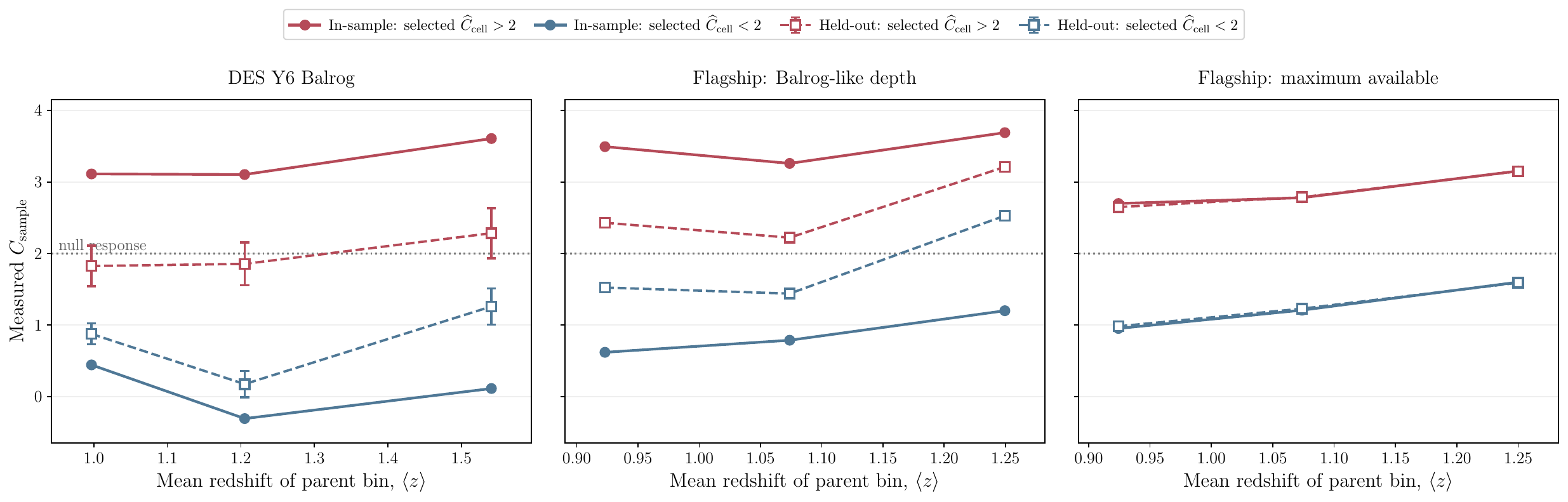}
  \caption{Held-out validation of SOM-cell response selections in DES Y6 and the DES-like Flagship sample.  Filled circles and solid lines show in-sample responses for branches selected above (red) and below (blue) the estimated cell-level null response.  Open squares and dashed lines show the responses obtained when the cells are selected using one half of the objects and the fixed branches are evaluated using the independent half.  The left panel shows the DES Y6 \textsc{Balrog} result, with error bars giving the scatter across the held-out realizations.  The middle panel shows Flagship after culling it to approximately one per cent of the eventual catalog, producing a Balrog-like per-cell noise level.  The right panel uses the maximum available Flagship sample.  The contraction of the held-out contrast at low statistics, and its disappearance in the high-statistics Flagship result, isolate the selection bias caused by noisy per-cell response estimates.}
  \label{fig:csample-heldout-real-flagship}
\end{figure*}

Figure~\ref{fig:csample-heldout-real-flagship} compares the resulting in-sample and held-out measurements.  In the DES panel, the held-out branch contrast is substantially smaller than the in-sample contrast.  The response ordering still contains predictive information, but the current \textsc{Balrog} statistics are insufficient to validate the full apparent separation obtained when the same injections are used for both cell selection and response measurement. The middle panel repeats the test after reducing the Flagship catalog to one per cent of the eventual simulated sample, chosen to give a Balrog-like per-cell noise level.  It exhibits the same regression of the held-out values toward the null response, demonstrating that finite per-cell statistics alone can reproduce the qualitative behavior seen in the data.

The right panel instead uses the maximum available Flagship sample.  In this high-statistics limit, the in-sample and held-out measurements agree closely in all three tomographic bins, and the fixed SOM split recovers a large response contrast on unseen galaxies.  The comparison between the middle and right panels is particularly informative: both use the same underlying galaxy population and SOM, and differ primarily in the number of objects available to calibrate each cell.  The reduction seen in the middle panel is therefore not evidence that the SOM variables lack response information; it is the expected winner's-curse effect from optimizing on noisy cell estimates.

Based on these tests, the current SOM-selected DES branches identify promising directions in galaxy-property space, but their in-sample amplitudes should be regarded as optimistic targets rather than as independently validated response calibrations.  At the same time, the magnitude partitions establish directly in the data that substantial cancellation is present, and the high-statistics Flagship test shows that a SOM-based selection can recover a strong and stable held-out separation once the cell responses are measured precisely.  A larger image-injection campaign, with disjoint optimization and validation samples, is therefore the natural route to determining how much of this potential can be realized in DES-like data or in future surveys. The following subsection quantifies how the statistical precision of the response calibration scales with the size of such an injection campaign.

\subsection{Statistical precision of the magnification-response calibration}
\label{app:csample-injection-scaling}

The cosmological information in BigMag depends directly on the calibration of the magnification response of each high-redshift sample.  In the notation of Sec.~\ref{sec:data_sims_des_balrog}, the selection response is estimated from the weighted counts in the fiducial and magnified \textsc{Balrog} runs,
\begin{align}
  \widehat C_{\rm sample}(S)
  &=
  \frac{\widehat N_{\rm mag}(S)-\widehat N_{\rm fid}(S)}
       {\widehat N_{\rm fid}(S)\,\Delta\kappa},
  \\
  \widehat N_r(S)
  &=
  \sum_{j\in D_r(S)}\frac{1}{N_{{\rm inj},j}} .
  \label{eq:app-csample-scaling-estimator}
\end{align}
Here $N_{{\rm inj},j}$ is the number of total attempted injections of the deep-field object underlying detection $j$.  Since $C_{\rm mag}=C_{\rm sample}-2$, the statistical uncertainty discussed below is the same for $C_{\rm sample}$ and $C_{\rm mag}$.  This appendix uses the existing injection catalog to measure how that uncertainty changes with the size of the injection campaign, and to assess the scale of future campaigns required for more precise calibration.

\begin{figure}
  \centering
  \includegraphics[width=0.99\linewidth]
  {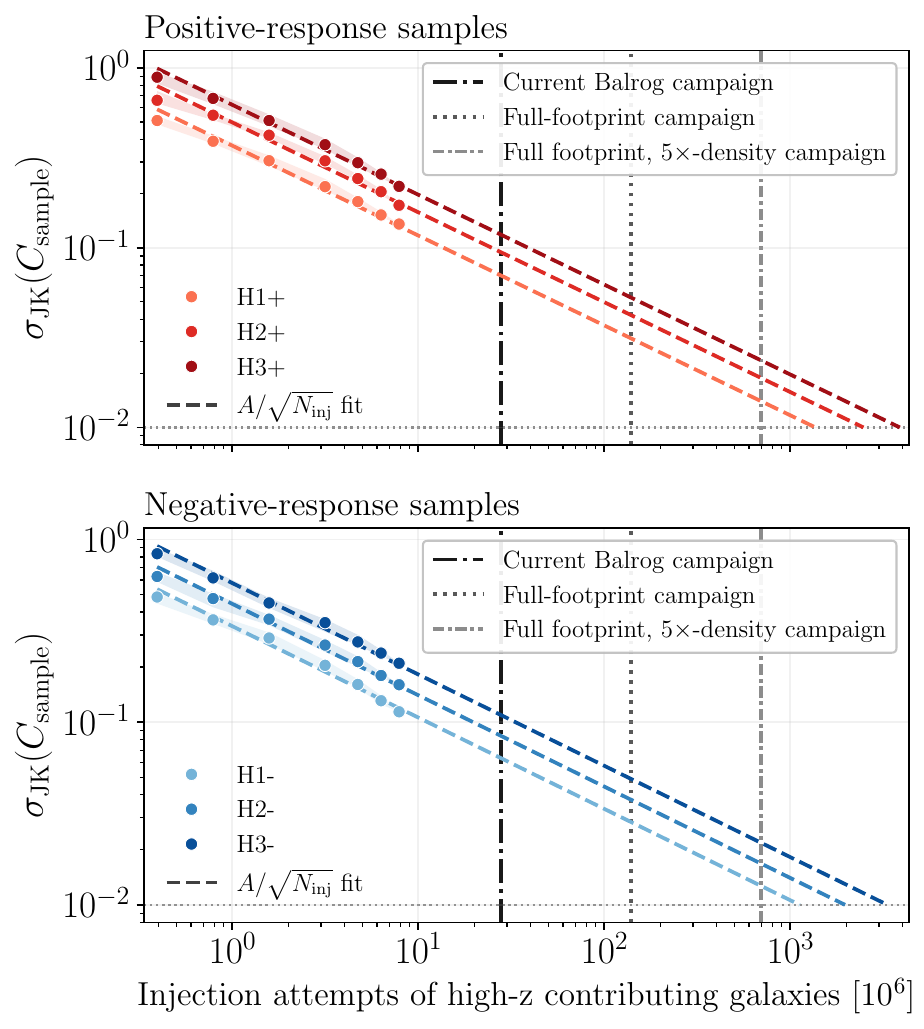}
  \caption{
    Estimated jackknife uncertainty on $C_{\rm sample}$ as a function of the number of attempted injections associated with deep IDs that contribute to the six high-$z$ samples.  The upper and lower panels show the positive- and negative-response samples, respectively.  Points give the median over 20 fixed-footprint thinning realizations and shaded regions enclose the 16th--84th percentile range.  Dashed curves show independent logarithmic-space fits of Eq.~\eqref{eq:app-csample-sqrt-scaling}, with the slope fixed to $-1/2$; their continuation beyond the measured range is an extrapolation.  The horizontal dotted line marks $\sigma(C_{\rm sample})=0.01$.  The vertical lines mark the number of injections in the current Balrog magnification campaign, which covers one fifth of the survey footprint, as well as the corresponding numbers expected for injections over the full footprint and for five times the injection density over the full footprint.
  }
  \label{fig:app-csample-injection-scaling}
\end{figure}

Figure~\ref{fig:app-csample-injection-scaling} shows how the uncertainty on $C_{\rm sample}$ evolves with the number of attempted injections of deep-field objects that enter any of the high-$z$ samples considered in this work. We consider retained fractions of $f_{\rm inj}={0.05,0.10,0.20,0.40,0.60,0.80,1.00}$. Uncertainties are estimated using 100 jackknife regions in order to preserve survey-area effects. To avoid introducing differences between the fiducial and magnified runs, however, we remove the same blocks of (deep-field ID, patch) from both. For each retained fraction, we analyze 20 random realizations, recomputing both the inverse-injection weights and the full delete-one jackknife uncertainty for each realization. We report the median uncertainty together with its 16th--84th percentile range.

For each sample $s$, we describe the injection-limited trend by
\begin{equation}
  \sigma_{\rm JK}\!\left(C_{{\rm sample},s};N_{{\rm inj},H}\right)
  =
  \frac{A_s}{\sqrt{N_{{\rm inj},H}}} .
  \label{eq:app-csample-sqrt-scaling}
\end{equation}
We fix the logarithmic slope to $-1/2$ and fit $A_s$ in log space, giving comparable weight to fractional deviations across the tested range.  The origin of this scaling is discussed in \S\ref{app:binomial_csample}. In Figure~\ref{fig:app-csample-injection-scaling}, we overlay the resulting best-fit $N_{{\rm inj},H}^{-1/2}$ scaling for each sample as a dashed line of the corresponding color.

We extrapolate these trends to estimate the number of injections required to reach an uncertainty of $\sigma(C_{\rm sample})\sim0.01$, indicated by the horizontal dotted line. For reference, vertical lines mark the number of injections in the entire current Balrog magnification campaign, which covers one fifth of the survey footprint, as well as the corresponding numbers expected for injections over the full footprint and for five times the injection density over the full footprint. The extrapolation suggests that a campaign with approximately $5$--$10\times$ the number of injections in a full-footprint Balrog realization would be sufficient to achieve excellent statistical precision. This requirement could likely be reduced further by tailoring the injection strategy to the target high-$z$ populations. We leave a detailed investigation of systematic uncertainties, together with validation of these statistical-uncertainty projections, to future work.

\subsubsection{Origin of the inverse-square-root scaling}
\label{app:binomial_csample}

The inverse-square-root scaling can be understood using an idealized Bernoulli model in which all injection attempts share the same selection probabilities within the fiducial/magnified campaigns, and different attempts are statistically independent.  Consider $n$ independent injection attempts and let $p_0$ and $p_\kappa$ be the probabilities that an injected object enters the sample in the fiducial and magnified runs.  If the two selection indicators for a paired attempt have covariance $c$, then $\operatorname{Var}(\widehat p_r)=p_r(1-p_r)/n$ and $\operatorname{Cov}(\widehat p_\kappa,\widehat p_0)=c/n$.  

Expanding the estimator
\begin{equation}
  \widehat C_{\rm sample}
  =
  \frac{\widehat p_\kappa-\widehat p_0}
       {\widehat p_0\,\Delta\kappa}
  \label{eq:app-csample-probability-estimator}
\end{equation}
to first order in the fluctuations of
$\widehat p_0$ and $\widehat p_\kappa$ about their expectation values, and propagating their variances and covariance, gives
\begin{equation}
  \operatorname{Var}(\widehat C_{\rm sample})
  \simeq
  \frac{1}{n\,\Delta\kappa^2}
  \left[
    \frac{p_\kappa(1-p_\kappa)}{p_0^2}
    +
    \frac{p_\kappa^2(1-p_0)}{p_0^3}
    -
    \frac{2p_\kappa c}{p_0^3}
  \right].
  \label{eq:app-csample-binomial-variance}
\end{equation}

The bracket determines the normalization, while the explicit $1/n$ factor gives $\sigma(C_{\rm sample})\propto n^{-1/2}$.  If
$p_\kappa\simeq p_0=p$ (as is typically the case, since $\lvert\Delta\kappa\,C_{\rm sample}\rvert =\mathcal{O}(10^{-2})$) and the paired indicators have correlation coefficient $\rho$, this reduces to
\begin{equation}
  \sigma(C_{\rm sample})
  \simeq
  \frac{1}{|\Delta\kappa|}
  \sqrt{\frac{2(1-\rho)(1-p)}{np}} .
  \label{eq:app-csample-binomial-limit}
\end{equation}
Thus rare selections have larger calibration noise, whereas positive covariance between paired fiducial and magnified outcomes reduces it.

The actual \textsc{Balrog} estimator is weighted and combines galaxies with heterogeneous selection probabilities, so it is not exactly binomial.  For a fixed heterogeneous population, its variance instead contains weighted sums of the individual $p_i(1-p_i)$ contributions.  Provided the injected population and distribution of observing conditions remain fixed, these terms still decrease inversely with the effective number of independent attempts. The fitted amplitude $A_s$ absorbs the selection efficiencies, weights, pairing covariance, and residual spatial correlations.  A fixed-footprint jackknife may also contain an area-dependent component that does not continue to decrease as $N_{\rm inj}^{-1/2}$, so the long-range extrapolations should not be interpreted as a complete error model. We leave these explorations to future work, but this back-of-the-envelope calculation already demonstrates that exquisite statistical precision in the calibration of the magnification response is achievable with a dedicated injection campaign. 

\section{Robustness to redshift overlap}
\label{app:redshift-overlap}

The fiducial BigMag simulations have little overlap between the
low-redshift foreground samples and the high-redshift
magnification-selected samples. Their cross-correlations are therefore
dominated by magnification. In real data, low-redshift tails in the
background distributions also generate intrinsic galaxy clustering.
Here we quantify how this contribution changes the signal and its
positive/negative response pattern, and compare the controlled tests
with the data-informed redshift distributions of
Sec.~\ref{sec:data_sims}.

\subsection{Controlled overlap tests}

We introduce a low-redshift component into each normalized background
distribution according to
\begin{equation}
\begin{split}
         &n'_{H_j^\epsilon}(z)
 = (1-f_{\rm overlap})n_{H_j^\epsilon}(z)
 + f_{\rm overlap}n_{L_j}(z),
 \\ & \text{with } \epsilon\in\{+,-\},
 \label{eq:overlap-mixture}
 \end{split}
\end{equation}
with $H_1^\pm$ matched to $L_1$, $H_2^\pm$ to $L_2$, and
$H_3^\pm$ to $L_3$. This assigns the same contaminating fraction
to both response branches within each high-redshift bin. The mapping
defines a controlled tail shape; it does not imply that contaminants
in the observed samples follow these particular foreground
distributions. All foreground--background pairs are retained, including
correlations with foreground bins other than the matched one.
We consider
\begin{equation}
 f_{\rm overlap}=\{0,\,0.002,\,0.005,\,0.01,\,0.02,\,0.05,\,0.10\}.
\end{equation}
The cosmology, sample number densities, galaxy biases, and nominal
magnification responses are held fixed. In particular, the
contaminating component inherits the bias and response assigned to
its background sample; these tests do not independently vary the
properties of the contaminants.

For each modified set of redshift distributions, we evaluate the
correlations with the nominal responses and with all position-sample
responses set to $C_{\rm mag}=0$. The latter calculation isolates
the intrinsic density contribution, allowing the decomposition
\begin{equation}
 \boldsymbol w_{\rm total}
 =\boldsymbol w_{\rm dens}+\boldsymbol w_{\rm mag},
 \qquad
 \boldsymbol w_{\rm mag}
 \equiv\boldsymbol w_{\rm total}-\boldsymbol w_{\rm dens}.
 \label{eq:overlap-decomposition}
\end{equation}
The magnification component is thus evaluated using the modified
redshift distributions, rather than fixed to its no-overlap value.

We apply the same $4\,h^{-1}\mathrm{Mpc}$ scale cuts as in the
fiducial BigMag analysis. For foreground bins $L_1$, $L_2$, and
$L_3$, the corresponding minimum angular separations are
$23.96'$, $14.65'$, and $11.63'$, respectively. The selected vector
contains 264 measurements from the 18 foreground--background
correlations. For each component $X$, we compute
\begin{equation}
 (S/N)_X
 =\left[\boldsymbol w_X^{\mathsf T}
 \boldsymbol{\mathsf C}_{\rm fid}^{-1}
 \boldsymbol w_X\right]^{1/2},
 \label{eq:overlap-snr}
\end{equation}
using the same fiducial covariance submatrix, including correlations
between all retained measurements. These are signal-to-noise measures
at fixed parameters, before nuisance-parameter marginalization. Their
ratio compares the covariance-weighted sizes of the two components;
it is not a fractional contamination in every angular bin or a
prediction of the cosmological parameter bias.

\begin{table}[tb]
\centering
\caption{Signal-to-noise of the foreground--background correlations
after the fiducial BigMag scale cuts, as a function of the injected
fraction in Eq.~(\ref{eq:overlap-mixture}). All components use the
same fiducial covariance. The density contribution is obtained by
setting all position-sample magnification responses to zero.}
\label{tab:overlap-snr}
\small
\setlength{\tabcolsep}{4pt}
\begin{tabular}{ccccc}
\hline
$f_{\rm overlap}$ & $(S/N)_{\rm total}$ & $(S/N)_{\rm dens}$ &
$(S/N)_{\rm mag}$ & $\dfrac{(S/N)_{\rm dens}}{(S/N)_{\rm mag}}$ \\
\hline
0.000 & 41.68 &  0.09 & 41.69 & 0.002 \\
0.002 & 41.52 &  0.81 & 41.61 & 0.019 \\
0.005 & 41.31 &  2.00 & 41.50 & 0.048 \\
0.010 & 41.04 &  4.00 & 41.32 & 0.097 \\
0.020 & 40.79 &  8.00 & 40.96 & 0.195 \\
0.050 & 42.42 & 20.00 & 39.87 & 0.502 \\
0.100 & 51.78 & 39.99 & 38.07 & 1.050 \\
\hline
\end{tabular}
\end{table}

Table~\ref{tab:overlap-snr} shows that the intrinsic clustering
contribution grows approximately linearly with the injected fraction.
Its signal-to-noise is about $10\%$ of that of magnification at
$f_{\rm overlap}=0.01$, about $20\%$ at $0.02$, and comparable
to it at $0.10$. The magnification signal decreases more slowly,
as some of the background population is moved to lower redshift.
The component signal-to-noise values do not add in quadrature,
because the density and magnification vectors have a nonzero
covariance-weighted inner product. In particular, the increase in
the total signal-to-noise at large overlap reflects additional
intrinsic clustering, rather than increased magnification precision.

As a check, we repeat the calculation at $f_{\rm overlap}=0.02$
using the regenerated covariance of the total signal for all three
components. This gives
\begin{equation}
 \bigl((S/N)_{\rm total},(S/N)_{\rm dens},(S/N)_{\rm mag}\bigr)
 =(41.44,\,8.17,\,41.97),
\end{equation}
and $(S/N)_{\rm dens}/(S/N)_{\rm mag}=0.195$.
The individual values differ from the fixed-covariance results by
less than $3\%$, while their ratio is nearly unchanged. Thus, at
this representative overlap fraction, the component comparison is
insensitive to the covariance update.

We also examine the expected sign reversal between the positive-
and negative-response branches. For each of the nine $(L_i,H_j)$
combinations, we compute the arithmetic mean of
$\theta w_{ij}^{\pm}(\theta)$ over the retained angular bins,
and check whether the positive branch remains positive and the
negative branch remains negative. All nine combinations retain
these signs for $f_{\rm overlap}\leq0.01$. At $0.02$, seven
remain: the negative branches of $L_1\times H_1$ and
$L_2\times H_2$ change sign. Six combinations retain the
expected signs at $0.05$, and three at $0.10$. These counts
describe the mean theoretical signals, not the statistical
significance of a sign-reversal detection. They illustrate how
intrinsic clustering can erase the negative-response signal even
while the magnification contribution remains substantial.

\subsection{Comparison with the data-informed distributions}

We next examine the SOMPZ redshift distributions used to characterize
the selections in Sec.~\ref{sec:data_sims}. The high-redshift
population is first divided into three tomographic bins, and each
bin is then split by magnification response. We evaluate the
low-redshift tail separately for each parent bin and each response
branch, using their respective redshift distributions. Specifically,
\begin{equation}
 F_{H_j^\epsilon}(z<0.65)
 =\int_0^{0.65} dz\,\hat n_{H_j^\epsilon}(z),
 \label{eq:overlap-tail-fraction}
\end{equation}
where $\hat n$ has unit integral. For the tabulated SOMPZ
distributions, this integral is evaluated by summing the normalized
redshift-bin probabilities up to the bin edge at $z=0.65$.

\begin{table}[tb]
\centering
\caption{Fractions below $z=0.65$ in the data-informed SOMPZ
redshift distributions. Parent denotes the unsplit population
within each high-redshift tomographic bin; the other columns
refer to its positive- and negative-response selections. These
are total low-redshift tail fractions, not injected fractions or
overlaps with an individual foreground bin.}
\label{tab:data-nz-overlap}
\begin{tabular}{lccc}
\hline
Bin & Parent & Positive response & Negative response \\
\hline
$H_1$ & 0.0123 & 0.0109 & 0.0128 \\
$H_2$ & 0.0179 & 0.0129 & 0.0199 \\
$H_3$ & 0.0232 & 0.0207 & 0.0253 \\
\hline
\end{tabular}
\end{table}

The low-redshift fractions are therefore of order $1$--$3\%$,
with differences between the response branches that should be
retained in a real-data model. However, these fractions cannot
be identified directly with $f_{\rm overlap}$ in the controlled
scan. Equation~(\ref{eq:overlap-mixture}) assigns the contaminating
population a specific foreground-bin distribution, whereas the
SOMPZ tails can be spread across the full low-redshift range.
Two samples with the same $F_H(z<0.65)$ can consequently have
different overlaps with a given foreground bin and different
intrinsic clustering signals.

Predicting the intrinsic clustering contribution requires the full
redshift distributions of both samples, together with their galaxy
biases and the matter correlations. We therefore do not infer
contamination signal-to-noise values for the data-informed samples
by matching their total low-redshift tail fractions to the
controlled scan.

These tests show that percent-level overlap will not erase or wash out the
magnification signal, but its impact must be assessed using the
final low- and high-redshift data samples, including the separate
positive- and negative-response selections. Minimizing their
physical redshift overlap should be an explicit goal
of sample optimization, alongside number density and magnification
response. Any residual intrinsic clustering that cannot be shown
to be negligible at the required precision should be included
in the model, with uncertainties in the redshift tails and
contaminant galaxy biases propagated into the cosmological
constraints. For the tail shapes and biases considered here,
$f_{\rm overlap}\leq0.01$ preserves all nine sign reversals,
but this is not a universal tolerance: even at this overlap
fraction, the density contribution has a signal-to-noise of
approximately four. Also, preserving the sign reversal therefore does
not by itself establish negligible cosmological bias.

IGGL provides a complementary observable with a different
sensitivity to these tails. Because it correlates the high-redshift
galaxy positions with source shapes, it has no direct
density--density contribution of the kind considered above.
Its dominant magnification--shear signal also does not depend
on the galaxy bias of the position sample. However, low-redshift
contaminants in that sample can lie in front of the shape sources
and generate a conventional galaxy--galaxy lensing contribution,
while overlap with the shape-source distribution can introduce
density--intrinsic-alignment correlations \citep{IGGL}.
The relevant redshift ordering must therefore be characterized
using both the position and shape samples. Given the strong
constraining power of IGGL in our forecasts, assessing these
contributions for the final selections is particularly important:
combining BigMag and IGGL offers complementary sensitivity to
redshift-tail contamination, but does not remove the need to
characterize and model it.

\section{Small-scale magnification ratios}
\label{app:magnification-ratios}

The fiducial BigMag analysis excludes projected separations below
$4\,h^{-1}\mathrm{Mpc}$, where predicting the absolute clustering
signal requires more detailed modeling of nonlinear galaxy bias
and baryonic effects. Here we explore whether these measurements
can instead constrain ratios of magnification responses or lensing
efficiencies by cancelling the unknown small-scale
galaxy--matter correlation.

We write $w_{ij}^{\epsilon}(\theta)
\equiv w^{L_iH_j^{\epsilon}}(\theta)$ and
$C_j^{\epsilon}\equiv C^{\rm mag}(H_j^{\epsilon})$, with
$\epsilon\in\{+,-\}$. For negligible foreground--background
redshift overlap, the leading magnification contribution is
\begin{equation}
  w_{ij}^{\epsilon}(\theta)
  \simeq C_j^{\epsilon}
  \left\langle\delta_{g,i}\kappa_j\right\rangle(\theta).
  \label{eq:mag-ratio-leading}
\end{equation}
Rather than dividing noisy measurements directly, we fit the
original correlation functions with their joint covariance,
profiling over nuisance amplitudes describing the unknown
small-scale signal. These amplitudes are shared between the
correlations whose common dependence is assumed to cancel.

For positive- and negative-response samples with the same true
redshift distribution, this construction measures
\begin{equation}
  R_j^{-/+}
  \equiv
  \frac{w_{ij}^{-}(\theta)}{w_{ij}^{+}(\theta)}
  \simeq \frac{C_j^-}{C_j^+}.
  \label{eq:mag-ratio-cross-sign}
\end{equation}
The common galaxy--matter correlation and lensing kernel cancel
to leading order, so different foreground bins and angular scales
provide repeated measurements of the same response ratio.
However, multiplying both coefficients by a common factor leaves
the ratio unchanged. These measurements therefore constrain the
relative responses, but cannot determine their absolute
normalization.

Comparing samples of the same response sign at different source
redshifts instead gives, in the narrow-foreground-bin limit,
\begin{equation}
  R_{i,jk}^{\epsilon}
  \equiv
  \frac{w_{ij}^{\epsilon}(\theta)}
       {w_{ik}^{\epsilon}(\theta)}
  \simeq
  \frac{C_j^{\epsilon}}{C_k^{\epsilon}}
  \frac{\beta_{ij}}{\beta_{ik}},
  \label{eq:mag-ratio-cross-redshift}
\end{equation}
where $\beta_{ij}$ is the lensing efficiency averaged over the
source redshift distribution. This ratio retains geometric
information, analogously to conventional lensing ratios
\citep{DES_spt_lensingratio}, but that information is degenerate with the
relative responses and source redshift distributions. For our
sample configuration, changing $\Omega_{\rm m}$ from $0.3$ to
$0.2$ or $0.4$ changes the relevant distance ratios by only
approximately $0.03$--$0.25\%$, compared with statistical ratio
uncertainties of approximately $3$--$8\%$. The geometric
sensitivity is therefore weak.

We explored both constructions using the BigMag-only posterior
with the fiducial response priors $\sigma_C=0.2$.
A Gaussian covariance update predicts a reduction of less than
$1\%$ in the marginalized fluctuation-amplitude uncertainty for
either construction. Direct importance reweighting yields a low
effective sample size because the ratio likelihoods select narrow
directions in response space; apparent changes of several per
cent in those reweighted constraints are consequently not robust.
These exploratory results indicate no significant improvement
in cosmological precision for the configuration considered here.

The limited gain follows from the remaining absolute-response
degeneracy. Even combining both constructions leaves all ratios
unchanged under a global rescaling
$C_j^{\epsilon}\rightarrow\lambda C_j^{\epsilon}$.
This normalization remains degenerate with the cosmological
lensing amplitude in the large-scale BigMag correlations.
An absolute response calibration, for example from synthetic
source injections, or complementary information from other
observables is therefore still needed to constrain this mode.

Small-scale ratios can thus complement absolute calibration by
constraining relative responses and providing internal consistency
tests. Their interpretation assumes sufficiently well-matched
lensing kernels and control of residual redshift overlap and
observational contamination. Because the two constructions use
the same underlying correlations, a combined analysis must also
retain their covariance, together with the covariance with the
large-scale measurements.

\section{Redshift dependence of the magnification response}
\label{app:cmag-redshift-dependence}

Section~\ref{sec:gal_density_mag} derives the linear response of the
projected number density of a selected galaxy sample to a weak convergence,
\begin{equation}
  \delta_g^{\rm mag}(\hat{\mathbf n})
  = C_{\rm mag}\,\kappa(\hat{\mathbf n}),
  \qquad
  C_{\rm mag}=-2+C_{\rm sample}.
  \label{eq:app-cmag-scalar-response}
\end{equation}
This expression treats the response and convergence as effective quantities for the complete source sample.  
A redshift-dependent number-count slope inside the magnification kernel is standard in models of galaxy number counts, which follow from the evolution of the luminosity function \citep{2016PhRvD..94d3007C,2018PhRvD..97b3537L,2018JCAP...04..033V,Euclidmagnification2021,2022MNRAS.515.1130D}, while \citet{Wenzl_2024} found evidence for redshift evolution within BOSS CMASS sample and showed that its effective value then differs between observables. Here we generalize this to an arbitrary selection response and relate it to the quantity measured by synthetic-source injections.  

Consider a galaxy selection $i$, described by a selection function
$S_i(\mathbf F')$ applied to observed properties $\mathbf F'$.  The selection
may include detection, catalogue weights, colour and magnitude cuts, and
photometric-redshift bin assignment.  Let
$\overline N(\mathbf F,z_s)\,d\mathbf F\,dz_s$ be the ensemble-mean number of
galaxies with intrinsic properties $\mathbf F$ and true redshift $z_s$ in an
unlensed angular element $\Delta\Omega(0)$.  The unlensed differential surface
density of the selected sample is
\begin{equation}
  \overline n_i^0(z_s)
  =
  \frac{1}{\Delta\Omega(0)}
  \int d\mathbf F\,
  S_i(\mathbf F)\,\overline N(\mathbf F,z_s).
  \label{eq:app-unlensed-differential-counts}
\end{equation}
A galaxy at $z_s$ experiences
$\kappa_s\equiv\kappa(\hat{\mathbf n},z_s)$.  Repeating the expansion of
Sec.~\ref{sec:gal_density_mag} at fixed $z_s$ gives
\begin{align}
  \overline n_i(z_s,\kappa_s)
  & =
  \frac{1-2\kappa_s}{\Delta\Omega(0)}
  \int d\mathbf F\,
  S_i\!\left[\mathbf F'(\mathbf F,\kappa_s)\right]
  \overline N(\mathbf F,z_s)
  \nonumber\\
  & \simeq
  \overline n_i^0(z_s)
  \left[1+C_{{\rm mag},i}(z_s)\kappa_s\right],
  \label{eq:app-redshift-dependent-count-response}
\end{align}
where
\begin{align}
  C_{{\rm mag},i}(z_s)
  &=-2+C_{{\rm sample},i}(z_s),
  \label{eq:app-redshift-dependent-cmag}\\
  C_{{\rm sample},i}(z_s)
  &=
  \frac{
    \displaystyle
    \int d\mathbf F\,
    \left.\frac{\partial S_i}{\partial\kappa}\right|_{\kappa=0}
    \overline N(\mathbf F,z_s)
  }{
    \displaystyle
    \int d\mathbf F\,
    S_i(\mathbf F)\,\overline N(\mathbf F,z_s)
  }.
  \label{eq:app-redshift-dependent-csample}
\end{align}
Thus $C_{{\rm mag},i}(z_s)$ is the response derived in
Sec.~\ref{sec:gal_density_mag}, conditioned on true source redshift.  In
particular, lensing-induced migration across any boundary included in $S_i$ is
part of $C_{{\rm sample},i}(z_s)$.

To include intrinsic clustering, write the local galaxy distribution as its
ensemble mean plus a fluctuation and define
$\delta_{g,i}^{\rm int}(\hat{\mathbf n},z_s)$ as the corresponding
selection-weighted intrinsic density contrast.  Keeping terms linear in
$\delta_g$ and $\kappa$ gives
\begin{equation}
  n_i^{\rm obs}(\hat{\mathbf n},z_s)
  =
  \overline n_i^0(z_s)
  \left[
    1+\delta_{g,i}^{\rm int}(\hat{\mathbf n},z_s)
    +C_{{\rm mag},i}(z_s)\kappa(\hat{\mathbf n},z_s)
  \right].
  \label{eq:app-redshift-resolved-observed-density}
\end{equation}
Products such as $\kappa\delta_g$, including fluctuations in the population
mix that modulates the local response, enter beyond this linear-order
expression.

Projecting Eq.~\eqref{eq:app-redshift-resolved-observed-density} over source
redshift then yields
\begin{equation}
  \delta_{g,i}^{\rm obs}(\hat{\mathbf n})
  =
  \int dz_s\,p_i(z_s)
  \left[
    \delta_{g,i}^{\rm int}(\hat{\mathbf n},z_s)
    +C_{{\rm mag},i}(z_s)\kappa(\hat{\mathbf n},z_s)
  \right],
  \label{eq:app-projected-observed-density}
\end{equation}
\begin{equation}  
  \delta_{g,i}^{\rm mag}(\hat{\mathbf n})
  =
  \int dz_s\,p_i(z_s)C_{{\rm mag},i}(z_s)
  \kappa(\hat{\mathbf n},z_s).
  \label{eq:app-projected-magnification-density}
\end{equation}
The source quantity relevant to the magnification observable is therefore the
response-weighted redshift distribution
$p_i(z_s)C_{{\rm mag},i}(z_s)$, rather than an independently averaged
response and redshift distribution.

For the flat geometry assumed in this work, the convergence to a source at
$\chi_s\equiv\chi(z_s)$ is
\begin{equation}
  \kappa(\hat{\mathbf n},z_s)
  =
  \frac{3\Omega_{\rm m}H_0^2}{2c^2}
  \int_0^{\chi_s}d\chi\,
  \frac{\chi}{a(\chi)}
  \frac{\chi_s-\chi}{\chi_s}
  \delta_{\rm m}(\chi\hat{\mathbf n},z(\chi)).
  \label{eq:app-convergence-source-plane}
\end{equation}
Substituting this relation into
Eq.~\eqref{eq:app-projected-magnification-density} and exchanging the order of
integration gives
\begin{equation}
  \delta_{g,i}^{\rm mag}(\hat{\mathbf n})
  =
  \int d\chi\,W_{{\rm mag},i}(\chi)
  \delta_{\rm m}(\chi\hat{\mathbf n},z(\chi)),
  \label{eq:app-projected-magnification-kernel}
\end{equation}
with
\begin{equation}
  W_{{\rm mag},i}(\chi)
  =
  \frac{3\Omega_{\rm m}H_0^2}{2c^2}
  \frac{\chi}{a(\chi)}
  \int_{z(\chi)}^\infty dz_s\,
  p_i(z_s)C_{{\rm mag},i}(z_s)
  \frac{\chi_s-\chi}{\chi_s}.
  \label{eq:app-redshift-dependent-cmag-kernel}
\end{equation}

where $\chi_s\equiv\chi(z_s)$.
Equation~\eqref{eq:app-redshift-dependent-cmag-kernel} is the generalization of the usual magnification kernel for a redshift-dependent selection response.

An injection experiment that applies the same artificial convergence to all sources measures the response of the total selected counts.  In the linear-response limit this is
\begin{equation}
  \overline C_{{\rm mag},i}^{\rm inj}
  =
  \int dz_s\,p_i(z_s)C_{{\rm mag},i}(z_s).
  \label{eq:app-injection-averaged-cmag}
\end{equation}
The area contribution is not generated by magnifying sources at fixed
injection positions; in practice the injection measurement supplies
$\overline C_{{\rm sample},i}^{\rm inj}$ and the known area response is added
as $\overline C_{{\rm mag},i}^{\rm inj}=-2+
\overline C_{{\rm sample},i}^{\rm inj}$.  Equation \eqref{eq:app-injection-averaged-cmag} is an unweighted average over the
unlensed source distribution.  By contrast, cosmological magnification weights
each source by its lensing efficiency.  Consequently,
\begin{equation}
  \int dz_s\,p_i(z_s)C_{{\rm mag},i}(z_s)\kappa(z_s)
  \ne
  \overline C_{{\rm mag},i}^{\rm inj}
  \int dz_s\,p_i(z_s)\kappa(z_s)
  \label{eq:app-cmag-average-noncommutativity}
\end{equation}
in general.  The scalar treatment in Eq.~\eqref{eq:app-cmag-scalar-response}
is recovered when the source bin is narrow or when $C_{{\rm mag},i}(z_s)$ is
approximately constant over the redshift range carrying appreciable lensing
weight.

In future work we will use the fiducial and magnified injection
catalogues to estimate the response of the differential counts
directly,
\begin{equation}
  C_{{\rm sample},i}(z)
  \simeq
  \frac{
    \overline n_i(z,\Delta\kappa)
    -\overline n_i(z,0)
  }{
    \Delta\kappa\,\overline n_i(z,0)
  },
  \label{eq:app-finite-difference-redshift-response}
\end{equation}
using a representative redshift calibration of the injected population.

\vspace{1em}

\noindent\rule{\linewidth}{0.4pt}

\vspace{0.1em}

\end{document}